\documentclass[smallextended]{svjour3}       % onecolumn (second format)

\usepackage{amsfonts}
\usepackage{amssymb}
\usepackage{amsmath}
\usepackage{hyperref}
\usepackage{graphicx,subfigure}
\usepackage{latexsym}
\usepackage{mathptmx}      % use Times fonts if available on your TeX system

\journalname{Quantum Information Processing}

\newcommand{\calT}{\mathcal{T}}
\newcommand{\calG}{\mathcal{G}}
\newcommand{\calH}{\mathcal{H}}
\newcommand{\calO}{\mathcal{O}}

\newenvironment{coi}{\noindent}{}

\begin{document}

\title{Generalized Ramsey numbers through adiabatic quantum optimization}

\author{Mani Ranjbar, William G. Macready,\\ Lane Clark, Frank Gaitan}

\institute{Frank Gaitan \at
                Laboratory for Physical Sciences, 8050 Greenmead Dr, 
                College Park, MD 20740\\
                Tel.: 301-935-6531\\
                Fax: 301-935-6723\\
                \email{fgaitan@lps.umd.edu}
                  \and
                Mani Ranjbar, William G. Macready \at
                D-Wave Systems, Inc., 3033 Beta Avenue, Burnaby, British
                Columbia V5G 4M9, Canada
                  \and
               Lane Clark \at
               Department of Mathematics, Southern Illinois University, 
               Carbondale, IL 62901-4401
               }

\date{\today}

\maketitle

\begin{abstract}
Ramsey theory is an active research area in combinatorics whose central 
theme is the emergence of order in large disordered structures, with Ramsey 
numbers marking the threshold at which this order first appears. For generalized
Ramsey numbers $r(G,H)$, the emergent order is characterized by graphs $G$ and 
$H$. In this paper we: (i)~present a quantum algorithm for computing generalized
Ramsey numbers by reformulating the computation as a combinatorial optimization 
problem which is solved using adiabatic quantum optimization; and (ii)~determine
the Ramsey numbers $r(\calT_{m},\calT_{n})$ for trees of order $m,n = 6,7,8$, 
most of which were previously unknown.


\end{abstract}

\section{Introduction\label{sec1}}

To get a taste of the type of problem considered in Ramsey theory, consider an 
arbitrary gathering of $N$ people. One might wonder whether there is a group of
$m$ people at the party who are all mutual acquaintances, or a group of $n$ 
people who are all mutual strangers. Using Ramsey theory \cite{GRS,JNes} it can 
be shown that once the party size $N$ reaches a threshold size $r(m,n)$,
\textit{every\/} party with $N\geq r(m,n)$ people \textit{must\/} contain
either $m$ mutual acquaintances or $n$ mutual strangers. The unforced
and guaranteed emergence of order (viz.~a cluster of $m$ mutual friends or 
$n$ mutual strangers) upon reaching the threshold size is an essential 
characteristic of problems in Ramsey theory. The threshold $r(m,n)$ is an 
example of a two-color Ramsey number.

It proves fruitful to represent the $N$-person party problem by an $N$-vertex
graph. Each party-goer is identified with a vertex, and a red (blue) edge is 
drawn between a pair of vertices when the corresponding people are acquaintances 
(strangers). Since any two people attending will either know each other or not, 
every pair of vertices is joined by a red or blue edge. The party graph is 
thus the complete graph $K_{N}$ (all vertex pairs joined by an edge) with edges
colored red or blue. Notice that the group of $m$ mutual acquaintances 
(strangers) corresponds to a red $K_{m}$ (blue $K_{n}$) subgraph of $K_{N}$. 
The Ramsey theory result for the party problem becomes a theorem in 
graph theory \cite{Boll}: if the order $N$ of the complete graph $K_{N}$
satisfies $N\geq r(m,n)$, then every red/blue coloring of the edges of $K_{N}$ 
contains either a red $K_{m}$ or a blue $K_{n}$ subgraph.

The classical two-color Ramsey numbers $r(m,n)$ are extremely difficult to
calculate, with only $9$ values currently known \cite{radzi}. It was once hoped
that by considering proper subgraphs $G\subset K_{m}$ and $H\subset K_{n}$, 
generalized Ramsey numbers $r(G,H)$ might prove easier to calculate and inspire
new techiques that would also work for $r(m,n) \equiv r(K_{m},K_{n})$. Although 
these hopes have not been borne out to date, the study of generalized Ramsey 
numbers is now an active, well established part of Ramsey theory. Formally, for 
given graphs $G$ and $H$, the generalized Ramsey number $r(G,H)$ is defined to 
be the smallest positive integer $p$ for which every red/blue edge-coloring of 
the complete graph $K_{p}$ contains either a red $G$ or a blue $H$ subgraph 
\cite{GRS,Boll}. Generalized Ramsey numbers can also be defined for families of 
graphs $\calG$ and $\calH$. Such families typically partition into graph 
isomorphism (GI) classes $\{\calG^{i}\subset\calG\}$ and $\{\calH^{j}\subset
\calH\}$, and associated with each pair of classes is a generalized Ramsey 
number $r(\calG^{i},\calH^{j})$. We write $r(\calG ,\calH )$ for the set of all 
such Ramsey numbers. Early tabulations of generalized Ramsey numbers with 
$G$ and $H$ of order at most $5$ appear in Refs.~\cite{C+H1,C+H2,Clan,Hend}, 
while Ref.~\cite{radzi} presents the current state-of-the-art.

In this paper we present a quantum algorithm for computing generalized Ramsey 
numbers. We reformulate the computation as a combinatorial optimization
problem which is solved using adiabatic quantum optimization; and determine
the Ramsey numbers $r(\calT_{m},\calT_{n})$ for trees of order $m,n=6,7,8$, 
most of which were previously unknown. The quantum algorithm presented here 
generalizes an earlier adiabatic quantum algorithm for classical Ramsey numbers 
$r(m,n)$ \cite{us1} which was used to \textit{experimentally\/} determine a 
number of small Ramsey numbers \cite{us2}. 

The structure of this paper is as follows. In Section~\ref{sec2} we summarize 
the basic concepts from graph theory that will be needed in the remainder of 
the paper. Section~\ref{sec3} then shows how the computation of $r(G,H)$ can be 
transformed into a combinatorial optimization problem whose solution is found 
using adiabatic quantum optimization \cite{aqo}. Calculation of the 
generalized Ramsey numbers $r(\calT_{m},\calT_{n})$ for trees of order 
$m,n=6,7,8$ appears in Section~\ref{sec4}. In the interests of clarity, this 
section focuses on the simplest case with $m,n=6$; the remaining tree Ramsey 
numbers appear in Appendix~\ref{AppB}. The paper closes with a summary of our 
results in Section~\ref{sec5}, and for the reader's convenience, we collect 
previously known results for tree Ramsey numbers in Appendix~\ref{AppA}.

\section{Preliminaries}
\label{sec2}

We begin by reviewing those ideas from graph theory \cite{Boll} that will be 
central to our discussion. In the following all sets will be finite. We denote the 
cardinality of the set $X$ by $|X|$, and the set of all $2$-subsets of $X$ by 
$X^{(2)}$. 

A graph $G$ is specified by a non-empty set of vertices $V_{G}$ and a set of 
edges $E_{G}\subseteq V_{G}^{(2)}$. The order (size) of $G$ is denoted 
$|V_{G}|$ ($|E_{G}|$). A graph $G^{\prime}$ is a subgraph of a graph $G$ iff
$V_{G^{\prime}}\subseteq V_{G}$ and $E_{G^{\prime}}\subseteq E_{G}$. We denote
by $K_{n}$, $P_{n}$, and $K_{1,n-1}$ the complete graph, path, and star of 
order $n$, respectively. Lastly, we denote by $\mathcal{L}_{n}$ the set of
$2^{\binom{n}{2}}$ distinct vertex-labelled graphs with fixed $n$-vertex set, 
and by $\mathcal{U}_{n}$ the set of vertex-unlabelled graphs of order $n$.

Two graphs $G_{1}$ and $G_{2}$ are isomorphic ($G_{1}\cong G_{2}$) iff there 
there exists a bijection $f: V_{G_{1}}\rightarrow V_{G_{2}}$ such that 
$\{ u,v\}\in E_{G_{1}}$ iff $\{ f(u),f(v)\}\in E_{G_{2}}$. The bijection $f$ is 
called an isomorphism of $G_{1}$ and $G_{2}$. We 
write $G_{1}\sqsubseteq G_{2}$ iff there exists a subgraph $G^{\prime}\subseteq 
G_{2}$ such that $G_{1} \cong G^{\prime}$.

A red-blue coloring of the edges of a graph $G$ is a map $c: E_{G}\rightarrow\{
\mathrm{red, blue}\}$. Given a graph $G$ and an edge-coloring $c$ of $G$, the
red subgraph $G^{r}(c)$ has vertex set $V_{G}$ and edge set $\{ e\in E_{G}: 
c(e)=\mathrm{red}\}$. Similarly, the blue subgraph $G^{b}(c)$ has vertex set 
$V_{G}$ and edge set $\{ e\in E_{G}: c(e) =\mathrm{blue}\}$. Finally, we define
the \textit{arrow\/} relation between graphs $F$, $G$, and $H$. We write  
$F\rightarrow (G,H)$ iff, for all edge-colorings $c: E_{F}\rightarrow 
\{\mathrm{red, blue}\}$, either $G\sqsubseteq F^{r}(c)$ or $H\sqsubseteq 
F^{b}(c)$.

With these definitions in place, we are now in a position to define the 
generalized Ramsey numbers.
\begin{definition}
Given graphs $G$ and $H$, the generalized Ramsey number $r(G,H)$  is:
\begin{displaymath}
r(G,H) = \min \{ n\in\mathbb{P}: K_{n}\rightarrow (G,H)\} .
\end{displaymath}
\label{def1}
\end{definition}

A red-blue edge-colored graph $F$ is said to be $(G,H)$-critical iff: (i)~$F$
has order $r(G,H) -1$, and (ii)~$G\not\sqsubseteq F^{r}(c)$ and $H\not
\sqsubseteq F^{b}(c)$.

We collect literature results pertaining to generalized Ramsey numbers for 
certain families of trees in Appendix~\ref{AppA}. These results allow us to
determine which of the tree Ramsey numbers calculated in Section~\ref{sec4} 
and Appendix~\ref{AppB} are new, and provide checks for the rest.

\section{Quantum algorithm for generalized Ramsey numbers}
\label{sec3}

In this Section we present an adiabatic quantum algorithm for computing 
generalized Ramsey numbers. We first show (Section~\ref{sec3_1}) how a 
computation of the generalized Ramsey number $r(G,H)$ can be transformed into a 
combinatorial optimization problem (COP) which  is then solved 
(Section~\ref{sec3_2}) using adiabatic quantum optimization. The resulting 
algorithm generalizes an earlier adiabatic quantum algorithm 
for classical two-color Ramsey numbers \cite{us1} which has been used to 
\textit{experimentally\/} determine a number of small Ramsey numbers \cite{us2}.

\subsection{Generalized Ramsey numbers through combinatorial optimization}
\label{sec3_1}

Red-blue edge-colorings of $K_{N}$ are an essential ingredient in the definition
of $r(G,H)$. Each such coloring can be represented by a Boolean string 
$$e = 
(e_{1,2},\ldots , e_{i,j},\ldots e_{N-1,N})$$ 
of length $\binom{N}{2}$, where $e_{i,j} = 1$ ($0$) if the edge $\{ i,j\}$ 
(with $i<j$) is colored red (blue). For a given coloring $e$ of $K_{N}$, let 
$K^{r}_{N}(e)$ and $K^{b}_{N}(e)$ denote, respectively, its red and blue 
subgraphs. 

Let $e$ be a coloring of $K_{N}$. The following procedure counts the number of 
red subgraphs of $K_{N}^{r}(e)$ that are isomorphic to $G$. To begin, choose 
$|V_{G}|$ vertices from the $N$ vertices of $K_{N}$, and denote this choice 
by $S_{\alpha} = \{ v_{1}, \ldots . v_{|V_{G}|}\}$. Next, let 
$K^{r}_{\alpha}(e)$ be the subgraph of $K_{N}^{r}(e)$ with vertex set 
$S_{\alpha}$ and edge set $E^{r}_{\alpha}(e) = \{ \,\{ i,j\}\, |\, (i,j\in 
S_{\alpha})\bigwedge (i<j)\bigwedge (e_{i,j} = 1)\,\}$. We show below that
the following Boolean function evaluates to $1$ (True) if $G\cong 
K^{r}_{\alpha}(e)$ is True, and to $0$ (False) otherwise:
\begin{equation}
f[G\cong K^{r}_{\alpha}(e)] = \bigvee_{\pi\in\, Sym(V_{G})}\:
                                                \bigwedge_{\{ i,j\}\in\, E_{G}} 
                                                  e_{\pi (i),\pi (j)} .
\label{fGeqn}
\end{equation}
Here $V_{G}$ ($E_{G}$) is the vertex (edge) set of $G$; and $Sym(V_{G})$
is the symmetric group on $V_{G}$. Notice that if $G\cong K_{\alpha}^{r}(e)$
is True, there exists a permutation $\pi$ that transforms $V_{G}\rightarrow 
S_{\alpha}$ and preserves adjacency so that $e_{\pi (i),\pi (j)} = 1$ iff 
$\{ i,j\}\in E_{G}$. Thus the conjunction over $E_{G}$ evaluates to $1$ for 
this permutation, and so the disjunction evaluates to $1$. On the other 
hand, if $G\cong K^{r}_{\alpha}(e)$ is False, no permutation $\pi$ exists 
which preserves adjacency, and so for each permutation, at least one 
$e_{\pi (i),\pi (j)} = 0$ for $\{ i,j\}\in E_{G}$. The conjunction thus 
evaluates to $0$ for all permutations $\pi$, and the disjunction then 
evaluates to $0$. Summing $f[G\cong K^{r}_{\alpha}(e)]$ over all vertex 
choices $S_{\alpha}$ gives the number of red subgraphs of $K^{r}_{\alpha}(e)$ 
that are isomorphic to $G$. Denoting this sum as $\calO_{N} (e; G)$, we have
\begin{equation}
\calO_{N} (e;G) = \sum_{S_{\alpha}}\, 
                                               f[G\cong K^{r}_{\alpha}(e)].
\label{redsum}
\end{equation}
In a similar manner, the number of blue subgraphs of $K^{b}_{N}(e)$ isomorphic
to $H$ is
\begin{equation}
\calO_{N} (e;H) = \sum_{S_{\beta}}\, 
                                               f[H\cong K^{b}_{\alpha}(e)],
\label{bluesum}
\end{equation}
where: (i)~$S_{\beta} = \{ v_{1}, \ldots , v_{|V_{H}|}\}$ is a choice of 
$|V_{H}|$ vertices from the $N$ vertices of $K_{N}$; (ii)~$K^{b}_{\beta}(e)$ 
is the subgraph of $K^{b}_{N}(e)$ with vertex set $S_{\beta}$ and edge set 
$E^{b}_{\beta}(e) = \{ \,\{ i,j\}\, |\, (i,j\in S_{\beta})\bigwedge (i<j)
\bigwedge (e_{i,j} = 0)\,\}$; and
\begin{equation}
f[H\cong K^{b}_{\beta}(e)] = \bigvee_{\pi\in\, Sym(V_{G})}\:
                                                \bigwedge_{\{ i,j\}\in\, E_{G}} 
                                                  \overline{e}_{\pi (i),\pi (j)} ,
\label{fHeqn}
\end{equation}
where $\overline{e}_{\pi (i),\pi (j)} = 1 - e_{\pi (i),\pi (j)}$.

We now define an objective function $\calO_{N}(e; G,H)$ which assigns to each
coloring $e$ (of $K_{N}$) the total number of red and blue subgraphs it contains
that are, respectively, isomorphic to $G$ and $H$:
\begin{equation}
\calO_{N}(e;G,H) = \calO_{N} (e;G) + \calO_{N} (e;H) .
\label{objectivefunc}
\end{equation}
From Ramsey theory we know that if $N<r(G,H)$, then there is a coloring 
$e_{\ast}$ for which $G\not\sqsubseteq K^{r}_{N}(e_{\ast})$ and 
$H\not\sqsubseteq K^{b}_{N}(e_{\ast})$. For this coloring the objective 
function vanishes and so
\begin{equation}
\min_{e}\left[ \calO_{N}(e;G,H)\right] = 0,   \hspace{0.25in} (N<r(G,H)).
\label{belowthresh}
\end{equation}
On the other hand, if $N\geq r(G,H)$, we know that $K_{N}\rightarrow (G,H)$ 
and so
\begin{equation}
\min_{e}\left[\calO_{N}(e;G,H) \right] > 0,  \hspace{0.25in} (N\geq r(G,H)).
\label{abovethresh}
\end{equation}

The above discussion suggests the following COP for $r(G,H)$. Given graphs 
$G$ and $H$, and a positive integer $N$, find a coloring $e_{\ast}$ of $K_{N}$ 
that minimizes the objective function $\calO_{N}(e;G,H)$. As we have just seen, 
if $N< r(G,H)$, the minimum value of the objective function will be $0$, while 
if $N\geq r(G,H)$, the minimum value will be positive. This motivates the 
following classical optimization algorithm for finding $r(G,H)$ which will 
guide our construction of the quantum algorithm in Section~\ref{sec3_2}. 
\begin{enumerate}
\item \textit{Choose $N$ to be a strict lower bound for $r(G,H)$.\/} In 
principle, the probabilistic method \cite{probM} can always be used to produce 
such a lower bound, though in some cases, such lower bounds may already be 
available in the literature (e.~g., see Ref.~\cite{radzi}). 
\item \textit{Solve the COP for the minimum value of $\calO_{N}(e;G,H)$.\/} 
Since $N<r(G,H)$ we know that $\min_{e}\left[\calO_{N}(e;G,H)\right] = 0$. 
\item \textit{Increment $N\rightarrow N+1$ and determine the minimum value of 
$\calO_{N}(e;G,H)$ for this new $N$.\/} If the minimum is zero, continue 
incrementing $N\rightarrow N+1$ and finding the minimum value of the 
new objective function until $\min\left[\calO_{N}(e;G,H)\right]$ first 
becomes positive. When this first occurs, the algorithm returns the current 
value of $N$ as the Ramsey number $r(G,H)$ since this is the smallest $N$ 
for which $K_{N}\rightarrow (G,H)$. 
\end{enumerate}
We next show how this classical optimization algorithm can be promoted to an 
adiabatic quantum optimization for $r(G,H)$. 

\subsection{Adiabatic quantum algorithm for $r(G,H)$}
\label{sec3_2}

Here we show how the classical optimization algorithm for $r(G,H)$ 
(Section~\ref{sec3_1}) can be converted into an quantum algorithm. 

The adiabatic quantum optimization (AQO) algorithm \cite{aqo} exploits the 
adiabatic dynamics of a quantum system to solve COPs. The AQO algorithm uses 
the objective function for the COP to define a problem Hamiltonian $H_{P}$ 
whose ground-state subspace encodes all optimal solutions. The algorithm 
evolves the state of an $L$-qubit register from the ground-state of an initial 
Hamiltonian $H_{i}$ to the ground-state of $H_{P}$ with probability approaching 
$1$ in the adiabatic limit. An appropriate measurement at the end of the 
adiabatic evolution yields a solution of the COP almost certainly. The 
time-dependent Hamiltonian $H(t)$ for AQO is
\begin{equation}
H(t) = A(t/T) H_{i} + B(t/T) H_{P},
\label{aqoHam}
\end{equation}
where $T$ is the algorithm runtime, adiabatic dynamics corresponds to 
$T\rightarrow \infty$, and $A(t/T)$ [$B(t/T)$] is a positive monotonically
decreasing [increasing] function with $A(1) = 0$ [$B(0)=0$].

The point of departure for converting the classical optimization algorithm for 
$r(G,H)$ into an adiabatic quantum algorithm is the set of binary edge-coloring 
strings $e$ introduced in Section~\ref{sec3_1} for graphs of order $N$. Each 
of the $L = \binom{N}{2}$ bits in $e$ is promoted to a qubit so 
that the adiabatic quantum algorithm uses $L$ qubits. The $2^{L}$ strings 
$e$ are used to label the $2^{L}$ computational basis states (CBS) that span 
the $L$-qubit Hilbert space: $e \rightarrow |e\rangle = |e_{0}\cdots 
e_{L-1}\rangle$, with $e_{i} = 0,1$ for $i=0,\ldots , L-1$. The problem 
Hamiltonian $H_{P}$ is defined to be diagonal in the computational basis 
$|e\rangle$, with eigenvalue $\lambda (e) = \calO_{N}(e;G,H)$, where 
$\calO_{N}(e;G,H)$ is the objective function for the classical optimization 
algorithm for $r(G,H)$:
\begin{equation}
H_{P} |e\rangle = \calO_{N}(e;G,H) |e\rangle .
\label{probHam}
\end{equation}
Notice that the smallest eigenvalue (viz.~ground-state energy) of $H_{P}$
will be zero iff there exists a coloring $e_{\ast}$ with no red subgraph 
isomorphic to $G$ or blue subgraph isomorphic to $H$. The initial Hamiltonian 
$H_{i}$ is chosen to be
\begin{equation}
H_{i} = -\sum_{k=0}^{L-1} \sigma^{k}_{x} ,
\label{initHam}
\end{equation}
where $\sigma^{k}_{x}$ acts like a NOT operator on the $k^{th}$ qubit,
\begin{displaymath}
\sigma_{x}^{k}|e_{0}\cdots e_{k}\cdots e_{L-1}\rangle = 
     |e_{0}\cdots (e_{k}\oplus 1)\cdots e_{L-1}\rangle ,
\end{displaymath}
where $\oplus$ indicates binary addition. The ground-state of 
$H_{i}$ is easily shown to be the uniform superposition of $L$-qubit CBS.

As with the classical optimization algorithm for $r(G,H)$, the adiabatic quantum
algorithm begins by setting the graph order $N$ equal to a strict lower bound 
for $r(G,H)$, obtained using the probabilistic method, or a lower bound from 
the literature. The AQO algorithm is run on $L_{N}= \binom{N}{2}$ qubits, 
and at the end of the adiabatic evolution, the qubits are measured in the 
computational basis. The result is a binary string $e_{\ast}$ of length $L_{N}$.
In the \textit{adiabatic limit\/} ($T\rightarrow\infty$), the string $e_{\ast}$ 
will be an optimal string, almost certainly, with $\calO_{N}(e;G,H) = 0$ since 
$N<r(G,H)$. The value of $N$ is now incremented $N\rightarrow N+1$, the AQO 
algorithm is re-run on $L_{N+1}$ qubits, and the qubits measured in the 
computational basis at the end of adiabatic evolution. This process is 
repeated until the objective function value for the measured string is first 
positive. When this first occurs, in the adiabatic limit, the current $N$ 
value will be equal to $r(G,H)$, almost certainly. Note that any real 
application of AQO will only be approximately adiabatic. Thus the probability 
that the measured string $e_{\ast}$ will be an optimal string is $1-\epsilon$. 
In this case, the algorithm must be run $k\sim\mathcal{O}(\ln [1-\delta ]/
\ln\epsilon)$ times so that, with probability $\delta > 1-\epsilon$, at least 
one of the measurement outcomes will be an optimal string. We can make 
$\delta$ arbitrarily close to $1$ by choosing $k$ sufficiently large. This then
gives an adiabatic quantum algorithm for computing generalized Ramsey 
numbers.

\section{Numerical results for $r(\calT_{m},\calT_{n})$}
\label{sec4}

In this Section we numerically determine the generalized Ramsey numbers 
$r(\calT_{m},\calT_{n})$ associated with trees of order $m,n=6,7,8$. These
Ramsey numbers are of interest as many are unknown, and only determined to
within loose lower and upper bounds \cite{radzi}. Ideally, these Ramsey numbers 
would be found by simulating the quantum dynamics of the AQO algorithm 
presented in Section~\ref{sec3_2}. However, the exponential growth of 
Hilbert space dimension with number of qubits makes simulation of quantum 
systems with more than $20$ qubits impracticable \cite{aqo,fg1,fg2}. From 
Ref.~\cite{radzi}, the Ramsey numbers for $6$-vertex trees satisfy $7\leq 
r(\calT_{6},\calT_{6})\leq 25$. Thus simulating the AQO algorithm at the 
lower bound is already impractical as this requires $\binom{7}{2} = 21$ 
qubits. The situation is even worse for the other tree Ramsey numbers 
listed above. However, the classical optimization algorithm of 
Section~\ref{sec3_1} does allow us to determine these tree Ramsey numbers, 
and as we shall see below, most of the tree Ramsey numbers found are new.

In Section~\ref{sec4_1} we discuss the methodology and complexity of our
numerical computation. In the interests of clarity we limit our presentation 
of numerical results in Section~\ref{sec4_2} to $r(\calT_{6},\calT_{6})$; the 
remaining tree Ramsey numbers are presented in Appendix~\ref{AppB}.
Specifically, the Ramsey numbers $r(\calT_{7},\calT_{n})$ with $n=6,7$ appear
in Section~\ref{AppB_1}; and $r(\calT_{8},\calT_{n})$ with $6\leq n\leq 8$ 
appear in Section~\ref{AppB_2}. Section~\ref{sec4_2} and Appendix~\ref{AppB_1}
also present, for each input pair of GI classes, the number of non-isomorphic 
critical graphs, and at the Ramsey threshold $N = r(\calT^{i}_{m},
\calT_{n}^{j})$, the number of non-isomorphic optimal graphs and their 
associated minimum objective function values.

\subsection{Sources of complexity}
\label{sec4_1}

The difficulty of calculating Ramsey numbers was noted in the Introduction,
and the above optimization algorithm does not evade this difficulty. Here we 
describe three sources of exponential complexity which the COP contains, 
and discuss how they impact the numerical work presented in 
Section~\ref{sec4_2} and Appendix~\ref{AppB}.

For a given coloring $e$ of $K_{N}$, the algorithm examines all choices of 
$m$-sets $S_{\alpha}$ and $n$-sets $S_{\beta}$. There are $\binom{N}{m}$
and $\binom{N}{n}$ such choices which, respectively, scale exponentially with 
$m$ and $n$. As $m,n\leq 8$ in the numerical work presented in this paper, 
this source of intractability proved managable.

The second source of intractability arises from the need to consider all 
possible two-colorings of $K_{N}$. There are $2^{\binom{N}{2}}$ such colorings
which is super-exponential in $N$. Note, however, that (two-)colorings of $K_{N}$ 
that are isomorphic to a given coloring $e$ contain a red $G$ or a blue $H$
iff $e$ does. Thus, when calculating $r(G,H)$, we only need to consider 
vertex-unlabelled colorings of $K_{N}$. Since there are far fewer 
unlabelled colorings of $K_{N}$ than labelled colorings 
(see Table~\ref{table4_1}),
\begin{table}
\caption{\label{table4_1}The number of unlabelled ($u_{N}$) and labelled 
($l_{N}$) colorings of $K_{N}$ \cite{OEIS}.}
\begin{center}
\begin{tabular}{|l|l|l|} \hline
$N$ & $u_N$  & $l_N = 2^{\binom{N}{2}}$\mbox{\rule{0cm}{1.5em}} \\ \hline
$1$ & 1 & 1\\
$2$ & 2 & 2\\
$3$ & 4 & 8\\
$4$ & 11 & 64\\
$5$ & 34 & 1024\\
$6$ & 156 & 32768\\
$7$ & 1044 & 2097152\\
$8$ & 12346 & 268435456\\
$9$ & 274668 & 68719476736\\
$10$ & 12005168 & 35184372088832\\
$11$ & 1018997864 & 36028797018963968\\
$12$ & 165091172592 & 73786976294838206464\\
$13$ & 50502031367952 & 302231454903657293676544\\
$14$ & 29054155657235488 & 2475880078570760549798248448\\ \hline
\end{tabular}
\end{center}
\end{table}
it was possible to exhaustively examine all unlabelled colorings of $K_{N}$ for 
$N\leq 11$. For a given $N$, the graph isomorphism algorithm NAUTY \cite{nauty} 
was used to generate the unlabelled colorings of $K_{N}$. To go to larger $N$ 
(viz.~$N\geq 12$), it was necessary to give up on exhaustive examination of 
colorings to find the objective function minimum, and instead work with the 
heuristic algorithm Tabu search \cite{tabu}. If, for a given $N$, Tabu search 
returned a coloring $e_{\ast}$ with $\calO_{N}(e_{\ast};G,H) = 0$, then we 
know that $e_{\ast}$ does not contain a red $G$ or a blue $H$, and so $r(G,H) 
> N$. However, if the smallest objective value returned by Tabu search is 
positive, we cannot rule out that Tabu search missed a coloring with vanishing 
objective. In this case, absent further information, the most that can be 
concluded is that $r(G,H) \geq N$. We return to this point in 
Appendix~\ref{AppB_2}. 

The final source of exponential complexity arises when computing the look-up 
tables for $f[G\cong K^{r}_{\alpha}(e)]$ and $f[H\cong K^{b}_{\beta}(e)]$.
As discussed above, each choice $S_{\alpha}$ ($S_{\beta}$) of $m$ ($n$) 
vertices gives rise to a subgraph $K^{r}_{\alpha}(e)$ ($K^{b}_{\beta}(e)$) 
which must be examined to see if it is isomorphic to $G$ ($H$). A separate 
look-up table was used to store the values of $f[G\cong K^{r}_{\alpha}(e)]$ 
and $f[H\cong K^{b}_{\beta}(e)]$. Naively, each table requires an entry for 
each of the $2^{\binom{m}{2}}$ and $2^{\binom{n}{2}}$ possible 
$K^{r}_{\alpha}(e)$ and $K^{b}_{\beta}(e)$, respectively. When $m$ 
and/or $n$ equals $8$, this becomes unmanagable. The solution is again to 
store only unlabelled graphs in the look-up tables for $f[G\cong K^{r}_{\alpha}
(e)]$ and $f[H\cong K^{b}_{\beta}(e)]$. NAUTY was used to find all unlabelled 
graphs $\mathcal{G}$ of order $8$, and for each $\mathcal{G}$, 
Eq.~(\ref{fGeqn}) and/or Eq.~(\ref{fHeqn}) was used to evaluate  
$f[G\cong \mathcal{G}]$ and/or $f[H\cong \mathcal{G}]$, depending upon whether
$m$ and/or $n$ was equal to $8$. Then, for a given coloring $e$ of $K_{N}$, 
to determine whether $G\cong K^{r}_{\alpha}(e)$ or $H\cong K^{b}_{\beta}(e)$, 
NAUTY was used to find the unlabelled graph isomorphic to $K^{r}_{\alpha}(e)$ 
($K^{b}_{\beta}(e)$), and the look-up table value for the unlabelled graph 
used to determine whether $G\cong K^{r}_{\alpha}(e)$ 
($H\cong K^{b}_{\beta}(e)$).

By combining all of the above mitigation procedures, we were able to handle
complete graphs $K_{N}$ with $N\leq 14$, and graphs $G$ and $H$ corresponding 
to trees with $6\leq |V_{G}|,\; |V_{H}| \leq 8$. 

\subsection{Tree Ramsey numbers $r(\calT_{6},\calT_{6})$}
\label{sec4_2}

\begin{figure}
\begin{center}
\mbox{\subfigure[$\mathcal{T}^{1}_{6}$]{
\includegraphics[trim=3.5cm 12cm 0 6cm,clip, width=.33\textwidth]{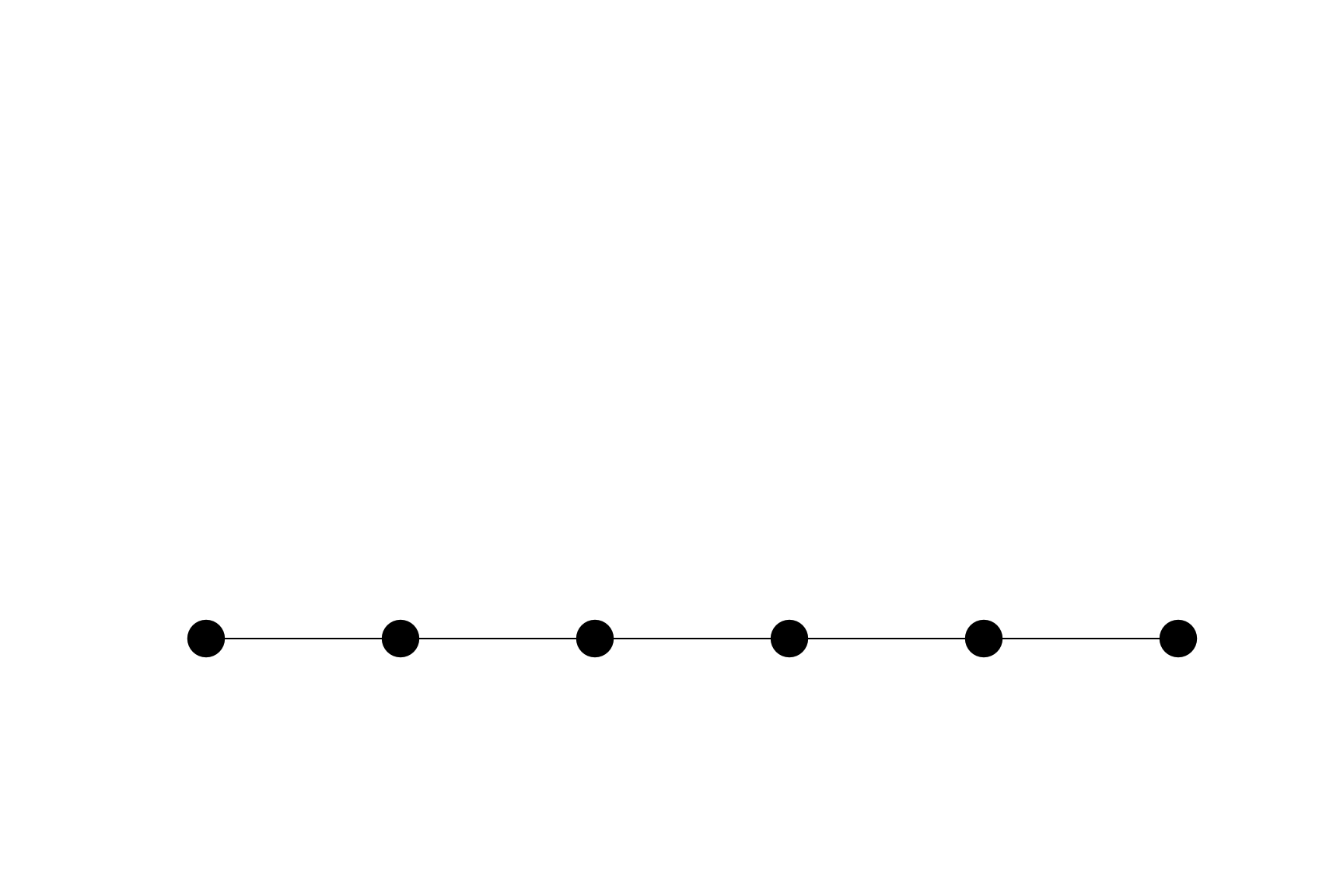}
}
\quad
\subfigure[$\mathcal{T}^{2}_{6}$]{
\includegraphics[trim=3.5cm 12cm 0 6cm,clip, width=.33\textwidth]{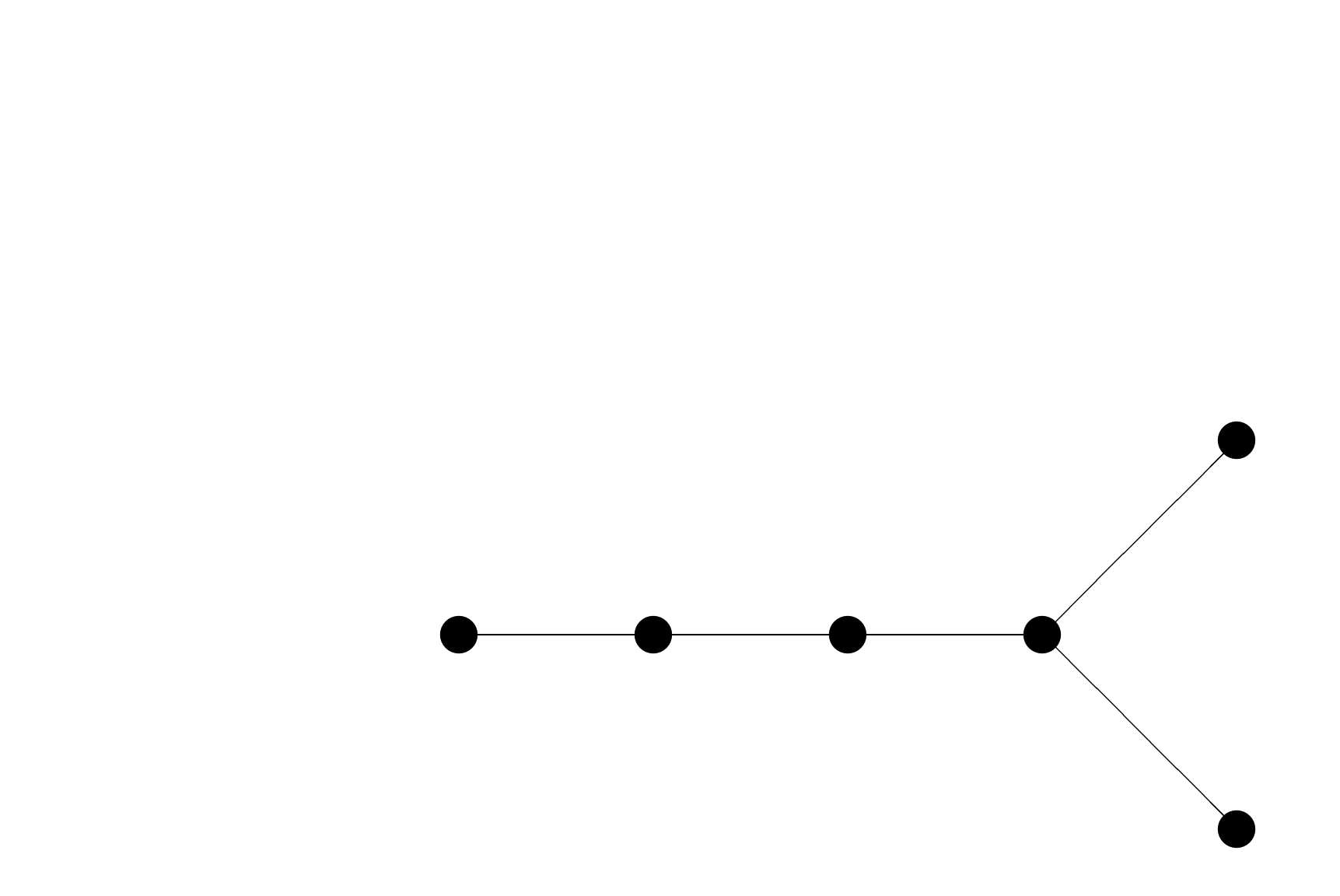}
}
\quad
\subfigure[$\mathcal{T}^{3}_{6}$]{
\includegraphics[trim=3.5cm 12cm 0 6cm,clip, width=.33\textwidth]{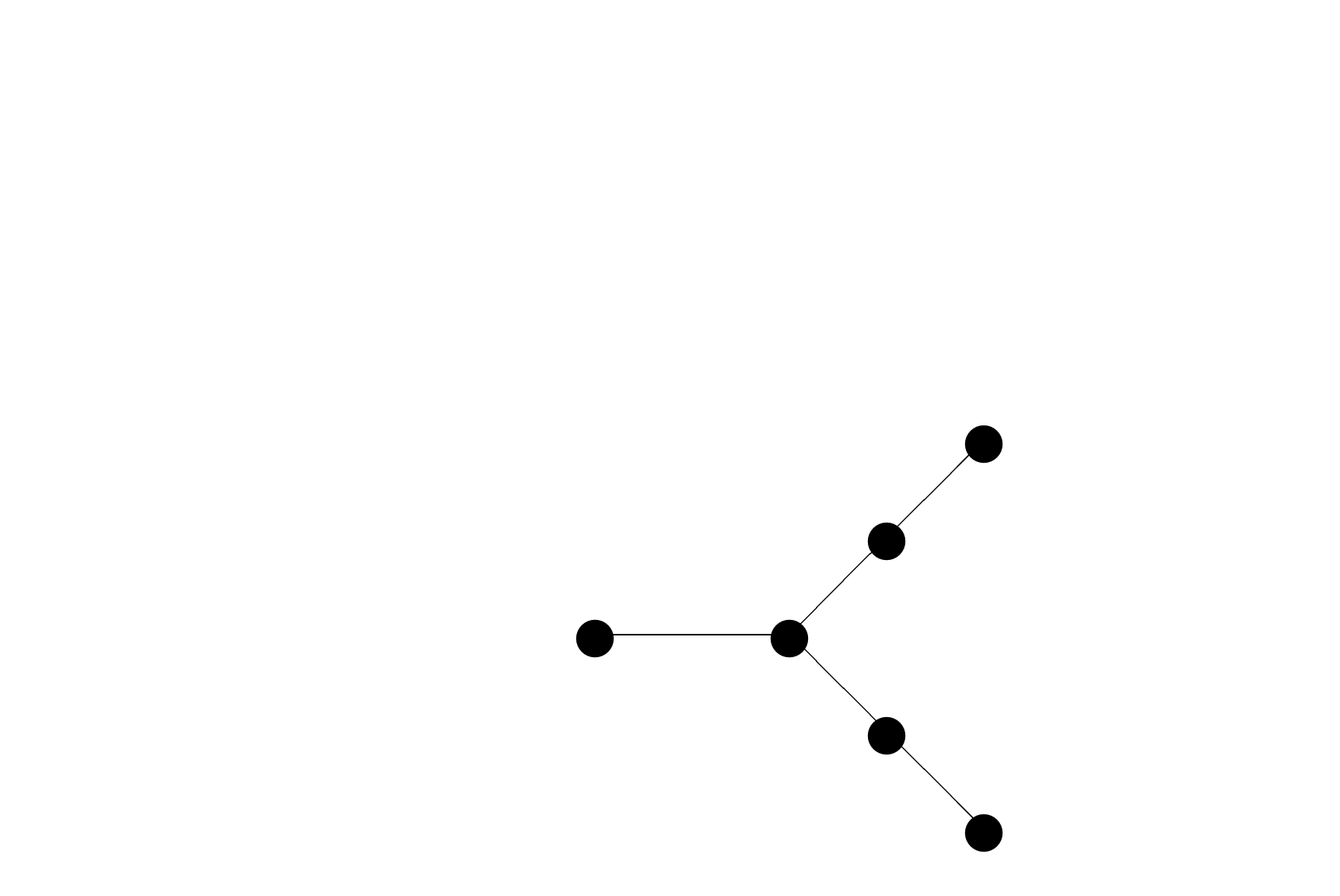}
}
}\\
\mbox{\subfigure[$\mathcal{T}^{4}_{6}$]{
\includegraphics[trim=3.5cm 12cm 0 6cm,clip, width=.33\textwidth]{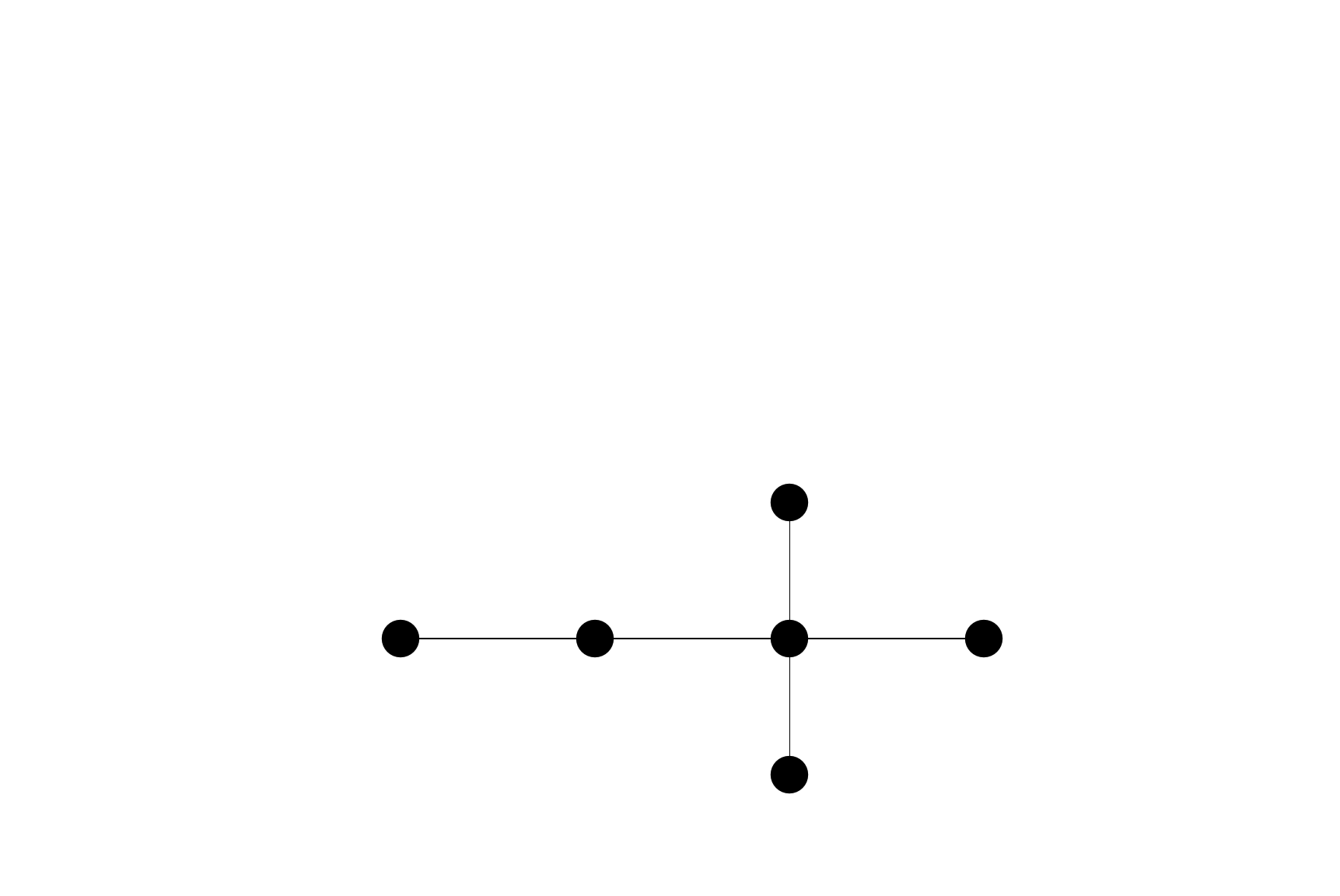}
}
\quad
\subfigure[$\mathcal{T}^{5}_{6}$]{
\includegraphics[trim=3.5cm 12cm 0 6cm,clip, width=.33\textwidth]{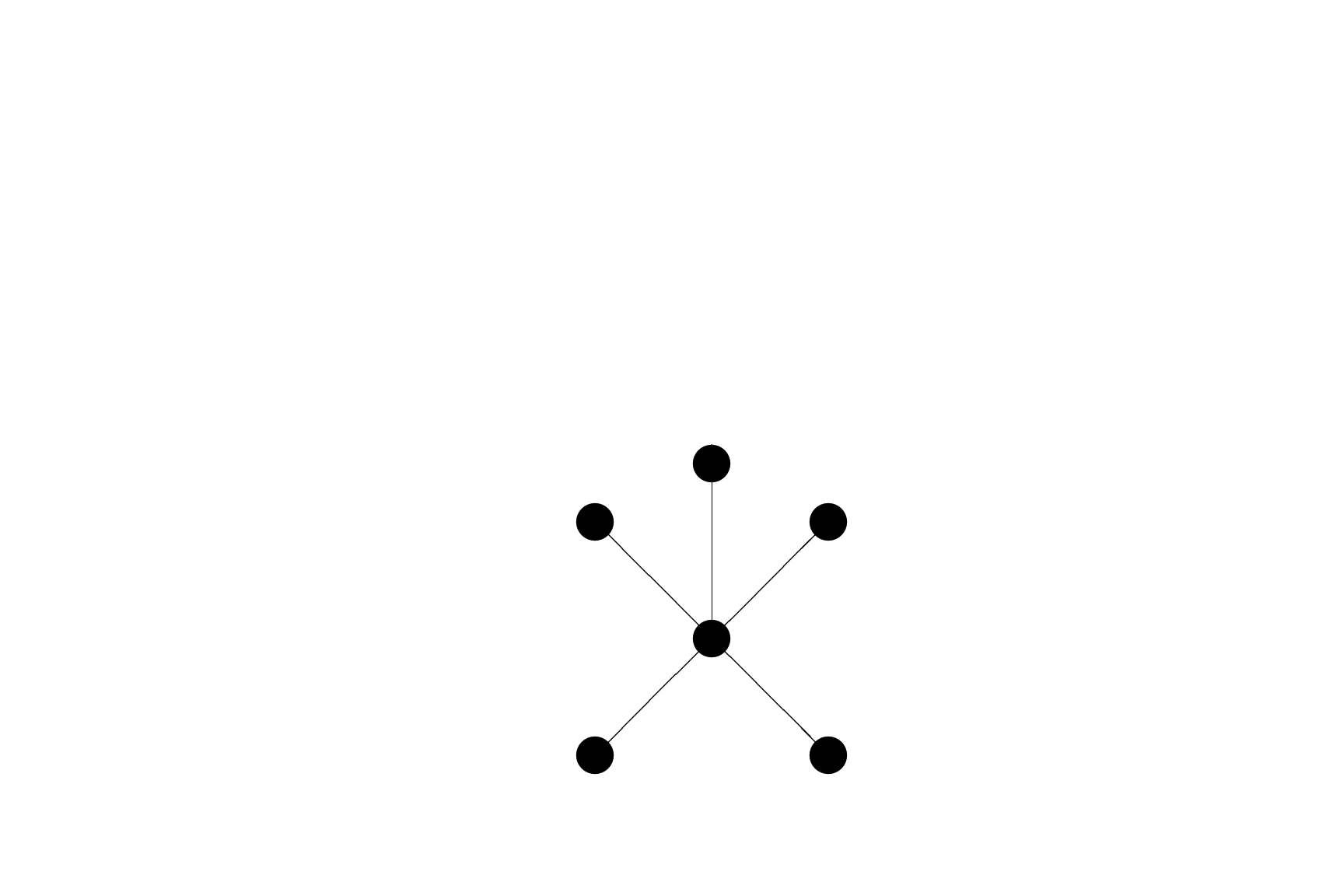}
}
\quad
\subfigure[$\mathcal{T}^{6}_{6}$]{
\includegraphics[trim=3.5cm 12cm 0 6cm,clip, width=.33\textwidth]{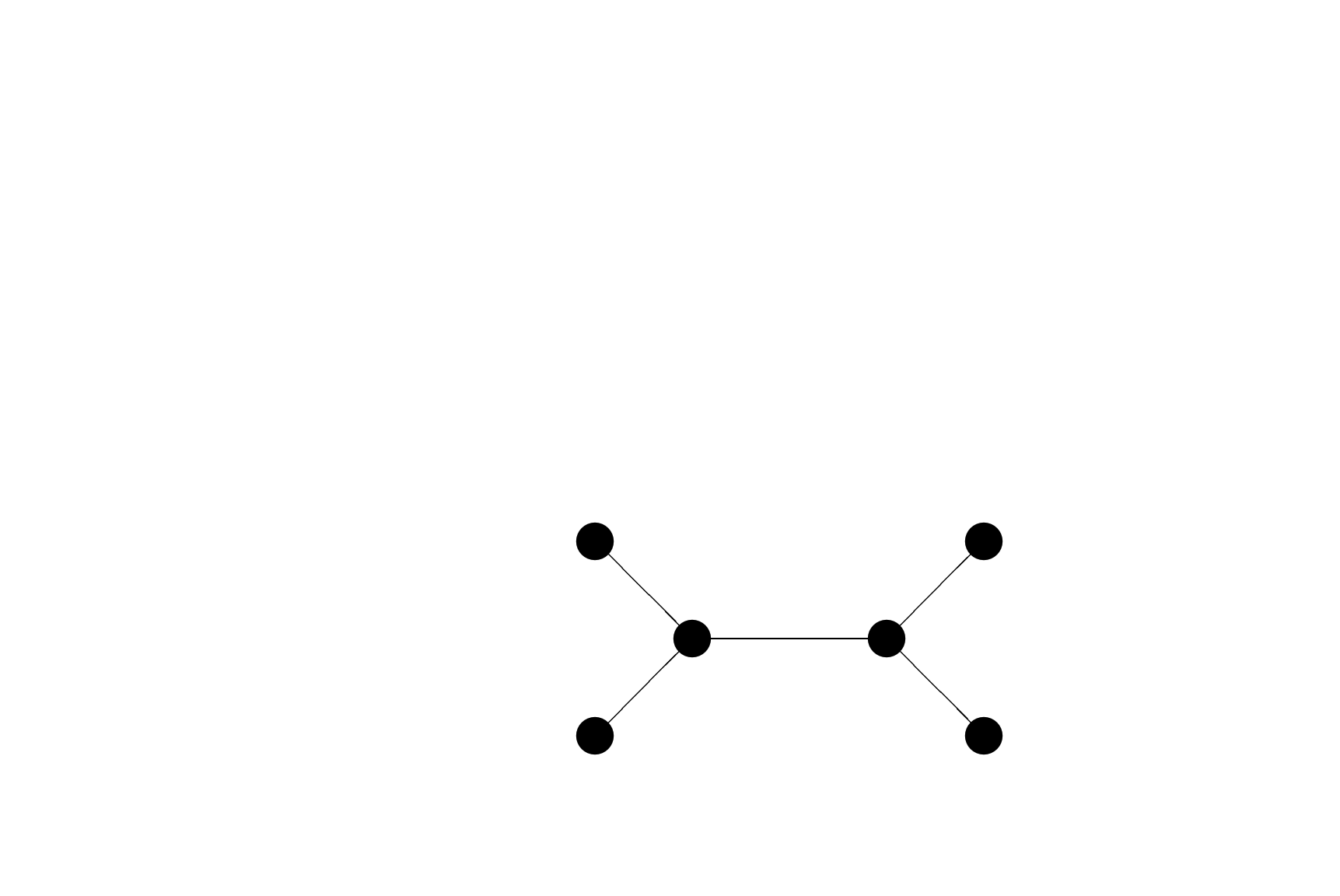}
}
}
\caption{\label{fig4_1} Graph isomorphism classes $\calT^{j}_{6}$ ($j=1,\ldots 
6$) for trees of order $6$.}
\end{center}
\end{figure}

Trees of order $6$ partition into six GI classes \cite{harary} which we denote
by $\{ \calT^{j}_{6} : j = 1,\ldots 6\}$, and show as unlabelled graphs in 
Figure~\ref{fig4_1}.
All GI classes except $\calT_{6}^{3}$ correspond to known unlabelled graphs
(Section~\ref{sec2} and Appendix~\ref{AppA}):
\begin{equation}
\begin{array}{lll}
\calT_{6}^{1} = P_{6} = S^{(4)}_{1,1}; 
  & \hspace{0.5in} \calT_{6}^{4} = S^{(2)}_{1,3}; \hspace{0.5in}&
                \calT_{6}^{6} = S^{(2)}_{2,2}.  \\
\calT_{6}^{2} = S^{(3)}_{1,2} ; & \hspace{0.5in} \calT_{6}^{5} = K_{1,5}; & 
\end{array}
\end{equation}
Using the numerical procedure described in Section~\ref{sec4_1} we determined 
the tree Ramsey numbers $r(\calT_{6}^{i},\calT_{6}^{j})$ for $i,j = 1, \ldots , 
6$ which we displayed in Table~\ref{table4_2}
\begin{table}
\caption{\label{table4_2}Numerical results for tree Ramsey numbers 
$r(\calT_{6}^{i},\calT_{6}^{j})$ with $1\leq i,j \leq 6$. Table rows (columns) 
are labelled by $i$ ($j$). Only the upper triangular table entries are shown as 
the lower triangular entries follow from $r(\calT_{6}^{j},\calT_{6}^{i}) =
r(\calT_{6}^{i},\calT_{6}^{j})$. A superscript ``x'' on a table entry indicates 
that Theorem~A.x of Appendix~\ref{AppA} applies, and so these tree Ramsey 
numbers were known prior to this work. The reader can verify that our 
numerical results are in agreement with the theorems of Appendix~\ref{AppA}. 
The remaining $24$ tree Ramsey numbers (to the best of our knowledge) are new.}
\begin{center}
\begin{tabular}{|c|cccccc|} \hline
$\mathbf{r(\calT_{6}^{i},\calT_{6}^{j})}$ & \rule{0cm}{1.5em} & & & & & \\
$i\; \backslash \; j$ & 1 & 2 & 3 & 4 & 5 & 6 \\ \hline
1\rule{0cm}{1.5em}  &  $8^{\,2,6}$  & 8 & 8 & 8 & $9^{\,4}$ & 8     \\
2  &  & 7 & 8 & 7 & $9^{\,4}$ & 8   \\
3  &  &  & 8 & 8 & $9^{\,4}$ & 8    \\
4  &  &  &  &  $7^{\,7}$ & $9^{\,4}$ & 8    \\
5  &  &  &  &  &  $10^{\, 3}$ & 9    \\
6  & \rule[-1em]{0cm}{1em} &  &  &  &  &  $8^{\, 7}$  \\ \hline
\end{tabular}
\end{center}
\end{table}
Only the upper triangular table entries are shown as the lower triangular 
entries follow from $r(\calT_{6}^{j},\calT_{6}^{i}) =r(\calT_{6}^{i},
\calT_{6}^{j})$. A superscript ``x'' on a table entry indicates that 
Theorem~A.x in Appendix~\ref{AppA} applies, and so these tree Ramsey numbers 
were known prior to this work. The reader can verify that our numerical results 
are in agreement with the theorems of Appendix~\ref{AppA}. The remaining $24$ 
tree Ramsey numbers (to the best of our knowledge) are new.

As explained in Section~\ref{sec4_1}, for $m,n=6$, and for graphs of order
$7\leq N\leq 10$, our numerical procedure can exhaustively search over 
all non-isomorphic graphs, and so can find the number of non-isomorphic 
optimal graphs and their associated objective function value. For graphs with 
order $N = r(\calT_{6}^{i},\calT_{6}^{j}) -1$, the optimal graphs are 
($\calT_{6}^{i},\calT_{6}^{j}$)-critical graphs (Section~\ref{sec2}). 
Table~\ref{table4_3}
\begin{table}
\caption{\label{table4_3}Numerical results for the number of non-isomorphic 
critical graphs $N_{c}(\calT_{6}^{i},\calT_{6}^{j})$ with $1\leq i,j \leq 6$. 
Table rows (columns) are labelled by $i$ ($j$), and for table entry ($i,j$), the 
graph order is $r(\calT_{6}^{i},\calT_{6}^{j}) - 1$. Only the upper triangular 
table entries are shown as the lower triangular entries follow from symmetry 
under interchange of colors: $N_{c}(\calT_{6}^{j},\calT_{6}^{i}) = 
N_{c}(\calT_{6}^{i},\calT_{6}^{j})$.}
\begin{center}
\begin{tabular}{|c|cccccc|} \hline
$\mathbf{N_{c}(\calT_{6}^{i},\calT_{6}^{j})}$ & \rule{0cm}{1.5em} & & & & & \\
$i\; \backslash \; j$ & 1 & 2 & 3 & 4 & 5 & 6 \\ \hline
1  &  4 &     2   &    4    &   2      &  1           &    3 \\
2  &       &     8  &     2    &   9    &  1           &     1 \\
3  &       &             &     4    &   2      &  1           &    3 \\
4  &       &             &             &   14  &  6     &     1 \\
5  &      &              &             &             & 16  &     1 \\
6  &    &   &  &   &               &     2\\\hline
\end{tabular}
\end{center}
\end{table}
lists the number of non-isomorphic critical graphs for each pairing of GI 
classes $(\calT_{6}^{i},\calT_{6}^{j})$. These graphs are all found to have 
vanishing objective function value, which is expected, since critical graphs 
have order $N < r(\calT_{6}^{i},\calT_{6}^{j})$.

At the Ramsey threshold, $N = r(\calT_{6}^{i},\calT_{6}^{j})$, optimal
graphs first acquire a non-vanishing objective function value (see 
Section~\ref{sec3_1}). In Tables~\ref{table4_4} 
\begin{table}
\caption{\label{table4_4}Numerical results for the number of non-isomorphic 
optimal graphs $N_{opt}(\calT_{6}^{i},\calT_{6}^{j})$ with $1\leq i,j \leq 6$. 
Table rows (columns) are labelled by $i$ ($j$), and for table entry ($i,j$), the 
graph order is $r(\calT_{6}^{i},\calT_{6}^{j})$. Only the upper triangular table 
entries are shown as the lower triangular entries follow from symmetry under 
interchange of colors: $N_{opt}(\calT_{6}^{j},\calT_{6}^{i}) = 
N_{opt}(\calT_{6}^{i},\calT_{6}^{j})$.}
\begin{center}
\begin{tabular}{|c|cccccc|} \hline
$\mathbf{N_{opt}(\calT_{6}^{i},\calT_{6}^{j})}$ & \rule{0cm}{1.5em} & & & & & \\
$i\; \backslash \; j$ & 1 & 2 & 3 & 4 & 5 & 6 \\ \hline
1 &  4   &       2    &     4   &       2   &      1   &       3 \\
2 &       &     8    &      2   &      5   &      1   &       1 \\
3 &        &             &     4   &       2   &      1   &       3 \\
4 &       &             &           &       2   &      1   &       1 \\
5 &       &             &           &            &   8   &      1 \\
6 &       &             &           &            &            &       2 \\ \hline
\end{tabular}
\end{center}
\end{table}
and \ref{table4_5}
\begin{table}
\caption{\label{table4_5}Numerical results for the minimum objective function 
value $\calO_{N}(e_{\ast};\calT_{6}^{i},\calT_{6}^{j})$ at the Ramsey threshold
with $1\leq i,j \leq 6$. Table rows (columns) are labelled 
by $i$ ($j$), and for table entry ($i,j$), the graph order is $r(\calT_{6}^{i},
\calT_{6}^{j})$. Only the upper triangular table entries are shown as the 
lower triangular entries follow from symmetry of the objective function under 
interchange of colors.}
\begin{center}
\begin{tabular}{|c|cccccc|} \hline
$\mathbf{\calO_{N}(e_{\ast};\calT_{6}^{i},\calT_{6}^{j})}$ & 
\rule{0cm}{1.5em} & & & & & \\
$i\; \backslash \; j$ & 1 & 2 & 3 & 4 & 5 & 6 \\ \hline
1 & 1  &   1  &   1  &   1  &   4  &   1 \\
2 &       &   1  &   1  &   1  &   4  &   1 \\
3 &       &      &   1  &   1  &   4  &   1 \\
4 &        &      &      &   1  &   4  &   1 \\
5 &        &      &      &      &   5  &   4 \\
6 &        &      &      &      &      &   1\\\hline
\end{tabular}
\end{center}
\end{table}
we list, respectively, the number of non-isomorphic optimal graphs 
for each pairing of GI classes $(\calT_{6}^{i},\calT_{6}^{j})$ and the 
corresponding minimum objective function value.

As noted earlier, in the interests of clarity, we present the remainder of our 
tree Ramsey number results in Appendix~\ref{AppB}.

\section{Summary}
\label{sec5}

In this paper we presented an adiabatic quantum algorithm for computing
generalized Ramsey numbers. We showed how such a computation can be reformulated
as a combinatorial optimization problem whose solution is found using adiabatic
quantum optimization. We determined all generalized Ramsey numbers for trees of 
order $6$-$8$, resulting in $1600$ tree Ramsey numbers, of which (to the best of 
our knowledge) $1479$ are new. All results are consistent with a conjectured 
upper bound on tree Ramsey numbers \cite{fss}. 

\begin{acknowledgements}
F. G. thanks T. Howell III for continued support.
\end{acknowledgements}

\begin{coi}
{\footnotesize\textbf{Conflicts of Interest}~The authors declare 
that they have no conflict of interest.}
\end{coi}

\appendix

\section{Tree Ramsey numbers - literature survey}
\label{AppA}

In this Appendix we quote five theorems from the literature pertaining to 
generalized Ramsey numbers for certain families of trees. These theorems:
(i)~allow us to identify which of our tree Ramsey numbers are new; and 
(ii)~provide checks on the remainder of our results. Section~\ref{sec2} 
provides a brief review of the graph theory concepts needed in this paper.

\begin{theorem}[Gerecs\'{e}r and Gy\'{a}rf\'{a}s \cite{ger&gya}]
For paths $P_{m}$ and $P_{n}$ with $2\leq m\leq n$,
\begin{displaymath}
r(P_{m},P_{n}) = n + \bigg\lfloor \frac{m}{2}\bigg\rfloor - 1.
\end{displaymath}
\label{Fact1}
\end{theorem}
\begin{theorem}[Harary \cite{hara}]
For stars $K_{1,m-1}$ and $K_{1,n-1}$ with $m,n \geq 2$ and diameter $2$,
\begin{displaymath}
r(K_{1,m-1}, K_{1,n-1}) = \left\{ \begin{array}{ll}
                                                           m+n-3, & \mathrm{odd}\; m,n\\
                                                          m+n -2, & \mathrm{otherwise.}
                                                      \end{array}  \right.  
\end{displaymath}
\label{Fact2}
\end{theorem}
\begin{theorem}[Cockayne \cite{cocka}]
If $T_{m}$ is a tree of order $m$ containing a vertex of degree one adjacent to
a vertex of degree two, then $r(T_{m},K_{1,n-1}) = m+n-3$ ($n\geq 2$), provided 
one of the following holds:\\
\begin{description}
\item[(1)]  $n - 1 \equiv 0,2 \;\bmod (m-1)$ ;
\item[(2)] $n - 1 \not\equiv 1 \;\bmod (m-1)$ and $n-1 \geq (m-3)^{2}$;
\item[(3)] $n-1 \not\equiv 1 \;\bmod (m-1)$ and $n-1 \equiv 1 \;\bmod (m-2)$;
\item[(4)] $n-1 \equiv m-2 \;\bmod (m-1)$ and $n-1 > m-2$.\\
\end{description}
\label{Fact4}
\end{theorem}

The following definition proves convenient.
\begin{definition}
Suppose $a,b\geq 1$ and $k\geq 2$. Then $S^{(k)}_{a,b}$ is the graph obtained 
from the disjoint stars $K_{1,a}$ and $K_{1,b}$ by joining their centers with a 
path of order $k$ (viz.\ length $k-1$).
\label{def2}
\end{definition}

Thus the order of $S^{(k)}_{a,b}$ is $a+b+k$ and the path $P_{n} = 
S^{(n-2)}_{1,1}$ for $n\geq 4$.

\begin{theorem}[Burr and Erd\"{o}s \cite{Bur&Erd}]
Consider the graph $S^{(4)}_{a,b}$ with $a\geq b\geq 1$ and diameter $5$. Then
\begin{displaymath}
r(S^{(4)}_{a,b},S^{(4)}_{a,b}) = \max\{ 2a+3,a+2b+5\} .
\end{displaymath}
\label{Fact5}
\end{theorem}
\begin{theorem}[Grossman, Harary, and Klawe \cite{Gros&Hara&Klaw}]
Consider the graph $S^{(2)}_{a,b}$ with $a\geq b\geq 1$ and diameter $3$. Then
\begin{displaymath}
r(S^{(2)}_{a,b},S^{(2)}_{a,b}) \geq 
      \left\{  \begin{array}{ll}
                      \max\{ 2a+1,a+2b+2\}, & a\;\mathrm{odd, and}\; b = 1,2\\
                     \max\{ 2a+2,a+2b+2\}, & \mathrm{otherwise.}
                  \end{array}  \right. 
\end{displaymath}
Furthermore,
\begin{displaymath}
r(S^{(2)}_{a,b},S^{(2)}_{a,b}) =
      \left\{  \begin{array}{ll}
                      \max\{ 2a+1,a+2b+2\}, & a\;\mathrm{odd, and}\; b = 1,2\\
                     \max\{ 2a+2,a+2b+2\}, &  a\;\mathrm{even, or}\; b\geq 3\;
                        \mathrm{provided}\; a\leq\sqrt{2}b\;\mathrm{or}\; 
                           a\geq 3b.
                  \end{array}  \right. 
\end{displaymath}
\label{Fact6}
\end{theorem}

\section{Remaining tree Ramsey numbers}
\label{AppB}

In this Appendix we present our remaining tree Ramsey number results.
Appendix~\ref{AppB_1} contains our results for $r(\calT_{7},\calT_{n})$ with
$n=6,7$; and Appendix~\ref{AppB_2} contains $r(\calT_{8},\calT_{n})$ for 
$n = 6,7,8$. Appendix~\ref{AppB_1} also present, for each input pair of GI 
classes, the number of non-isomorphic critical graphs, and at the Ramsey 
threshold $N = r(\calT^{i}_{m},\calT_{n}^{j})$, the number of non-isomorphic 
optimal graphs and their associated minimum objective function values.

\subsection{Tree Ramsey numbers $r(\calT_{7},\calT_{n})$ for $n=6,7$}
\label{AppB_1}

\begin{figure}
\begin{center}
\mbox{
\subfigure[$\mathcal{T}^{1}_{7}$]{
\includegraphics[trim=3.5cm 18cm 0 6cm,clip, width=.33\textwidth]{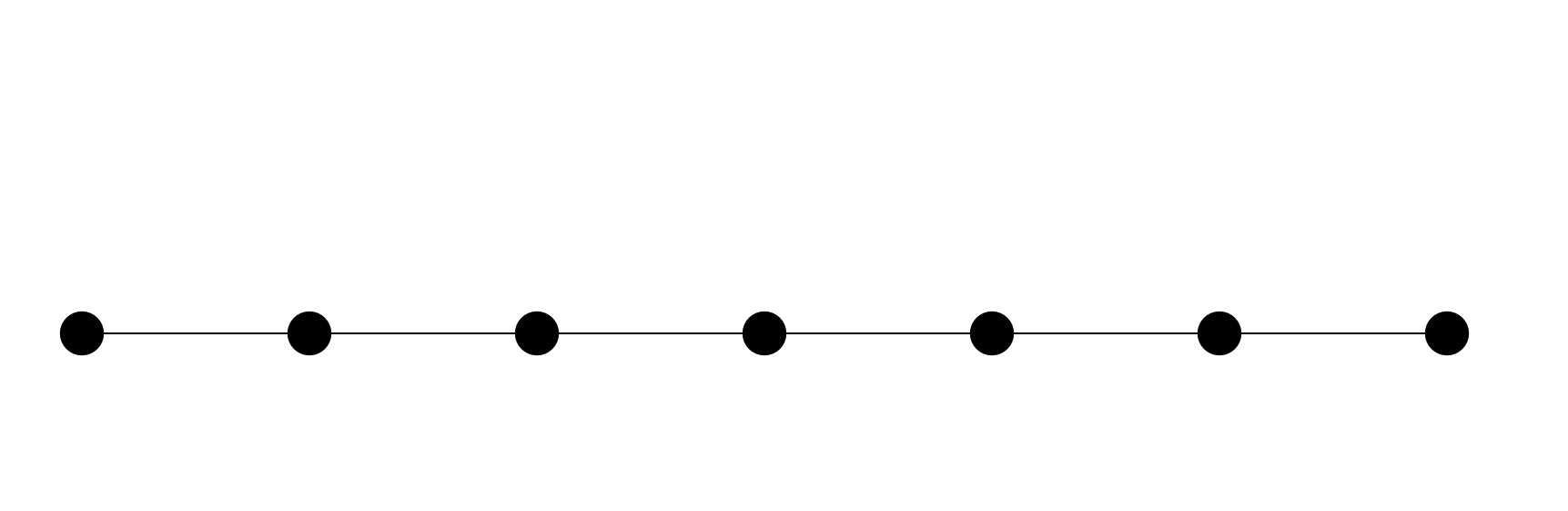}
}
\quad
\subfigure[$\mathcal{T}^{2}_{7}$]{
\includegraphics[trim=3.5cm 18cm 0 6cm,clip, width=.33\textwidth]{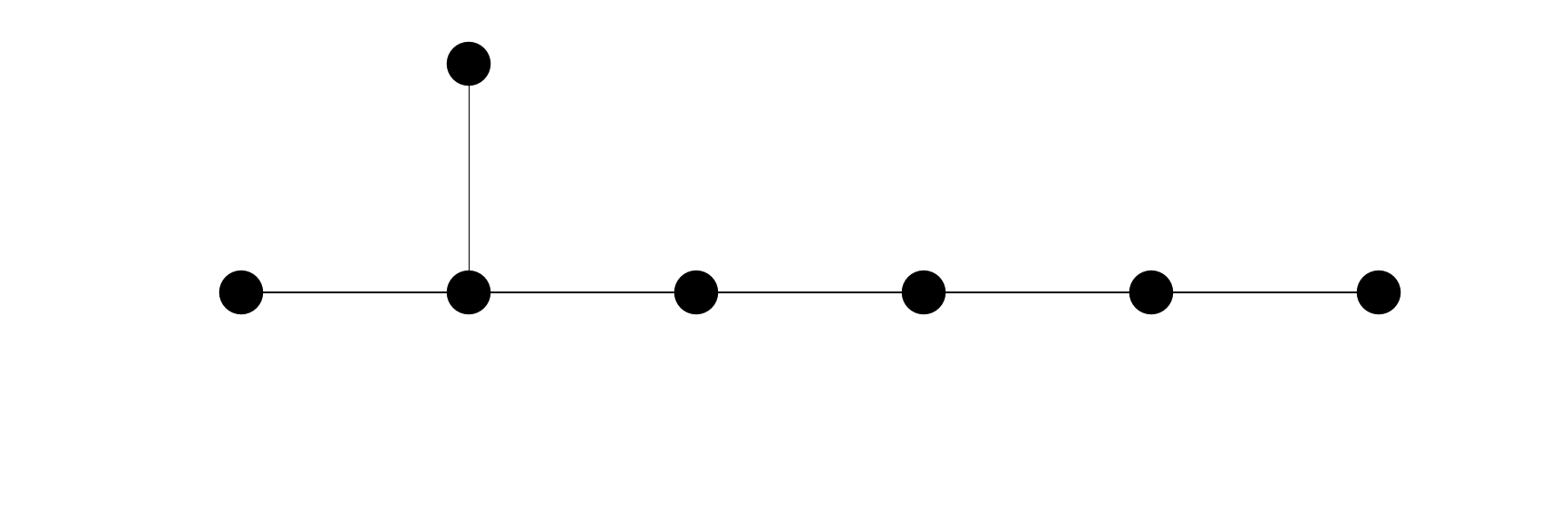}
}
\quad
\subfigure[$\mathcal{T}^{3}_{7}$]{
\includegraphics[trim=3.5cm 18cm 0 6cm,clip, width=.33\textwidth]{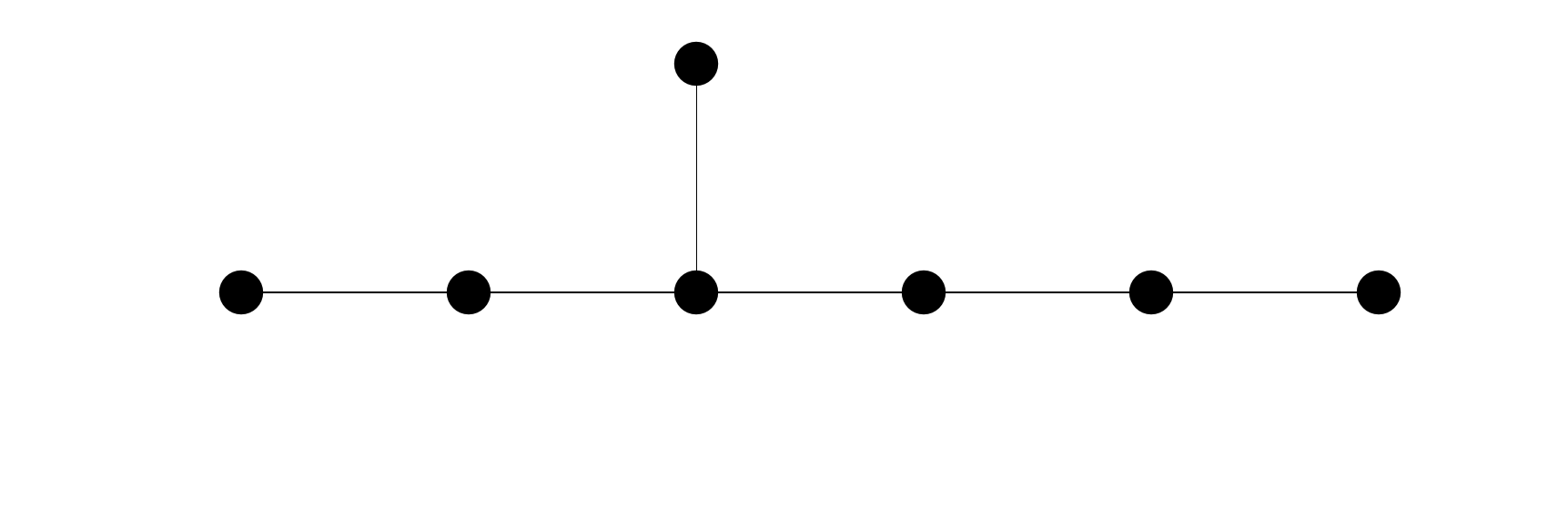}
}
}\\
\mbox{\subfigure[$\mathcal{T}^{4}_{7}$]{
\includegraphics[trim=3.5cm 18cm 0 6cm,clip, width=.33\textwidth]{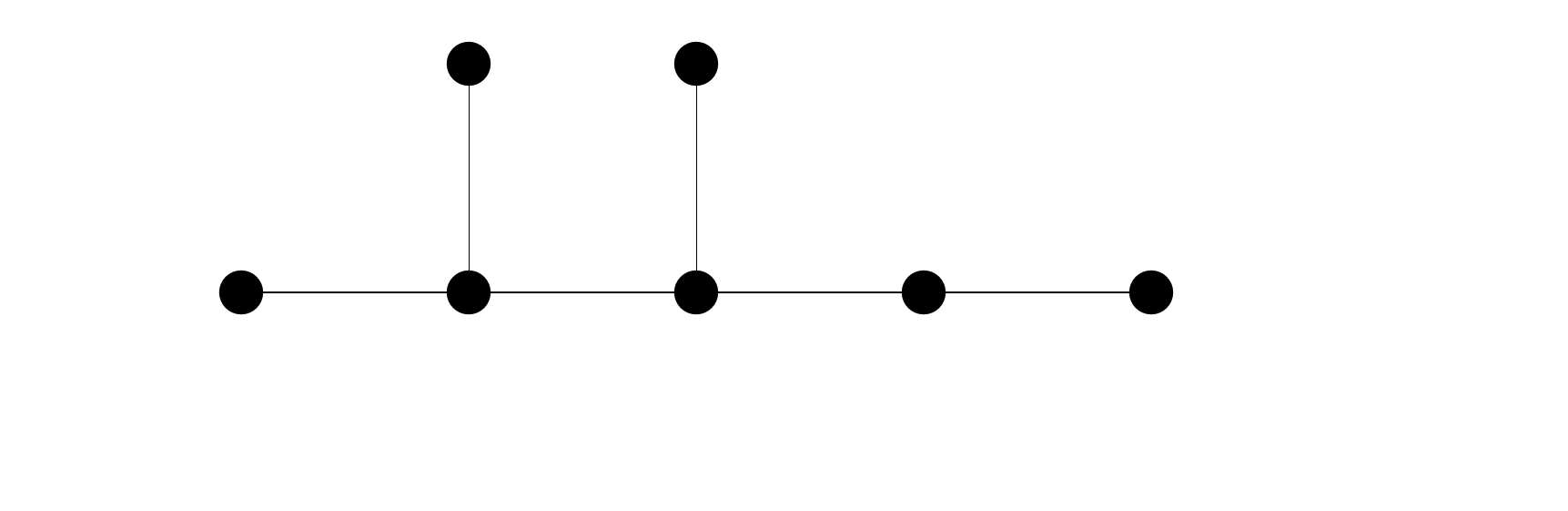}
}
\quad
\subfigure[$\mathcal{T}^{5}_{7}$]{
\includegraphics[trim=3.5cm 18cm 0 6cm,clip, width=.33\textwidth]{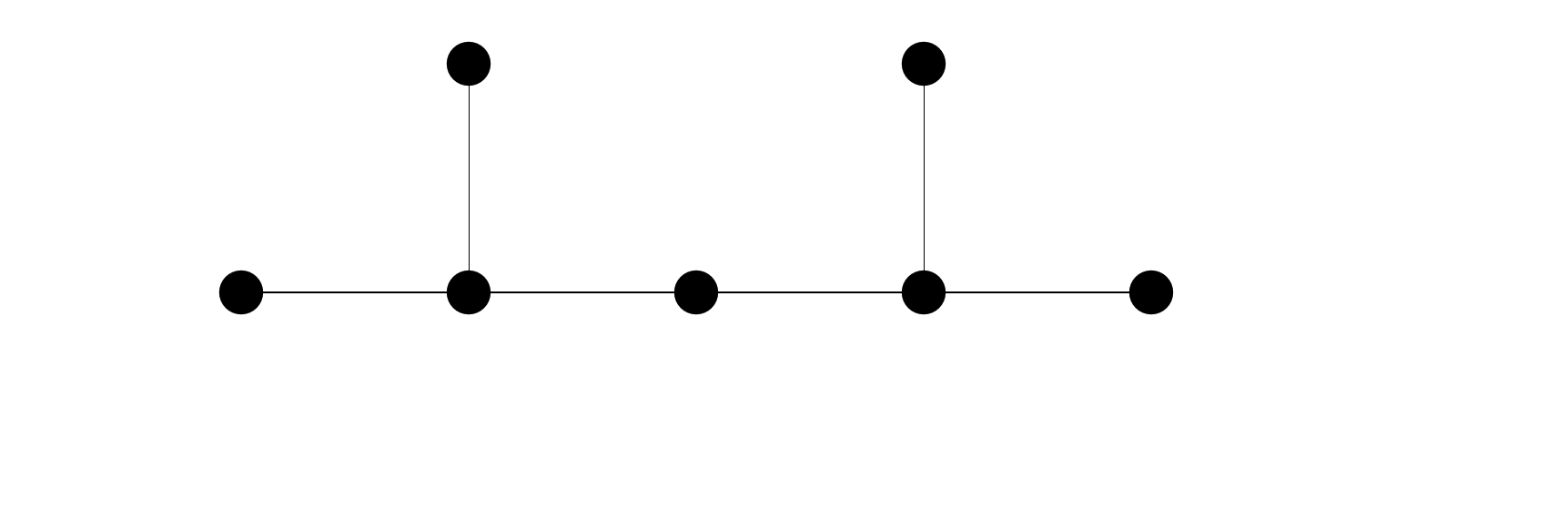}
}
\quad
\subfigure[$\mathcal{T}^{6}_{7}$]{
\includegraphics[trim=3.5cm 17cm 0 6cm,clip, width=.33\textwidth]{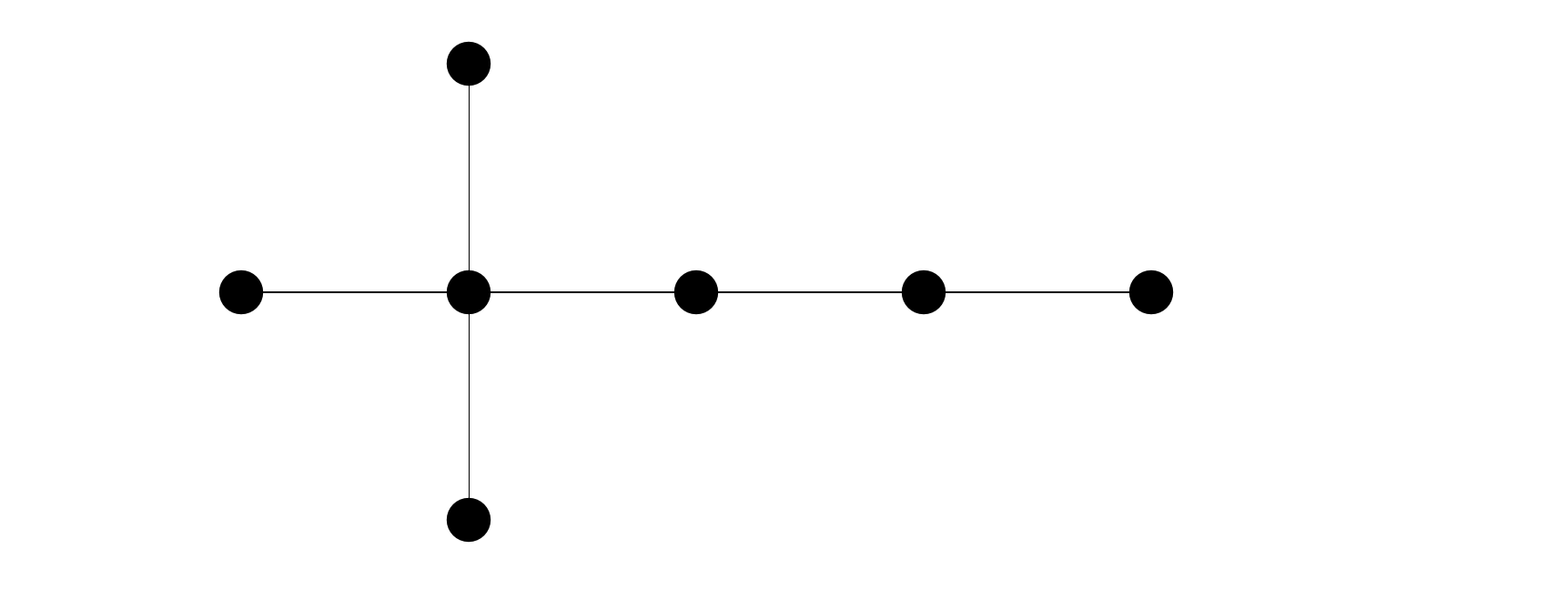}
}
}\\
\mbox{\subfigure[$\mathcal{T}^{7}_{7}$]{
\includegraphics[trim=3.5cm 17cm 0 6cm,clip, width=.33\textwidth]{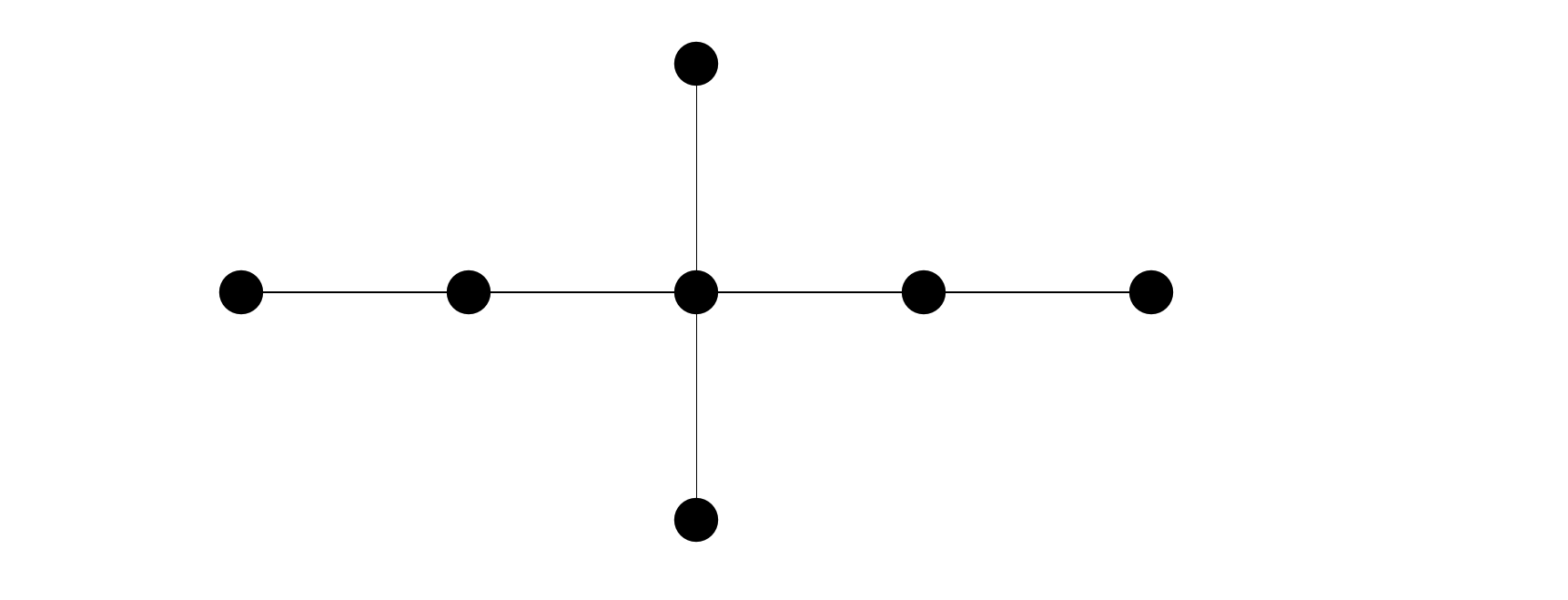}
}
\quad
\subfigure[$\mathcal{T}^{8}_{7}$]{
\includegraphics[trim=3.5cm 17cm 0 6cm,clip, width=.33\textwidth]{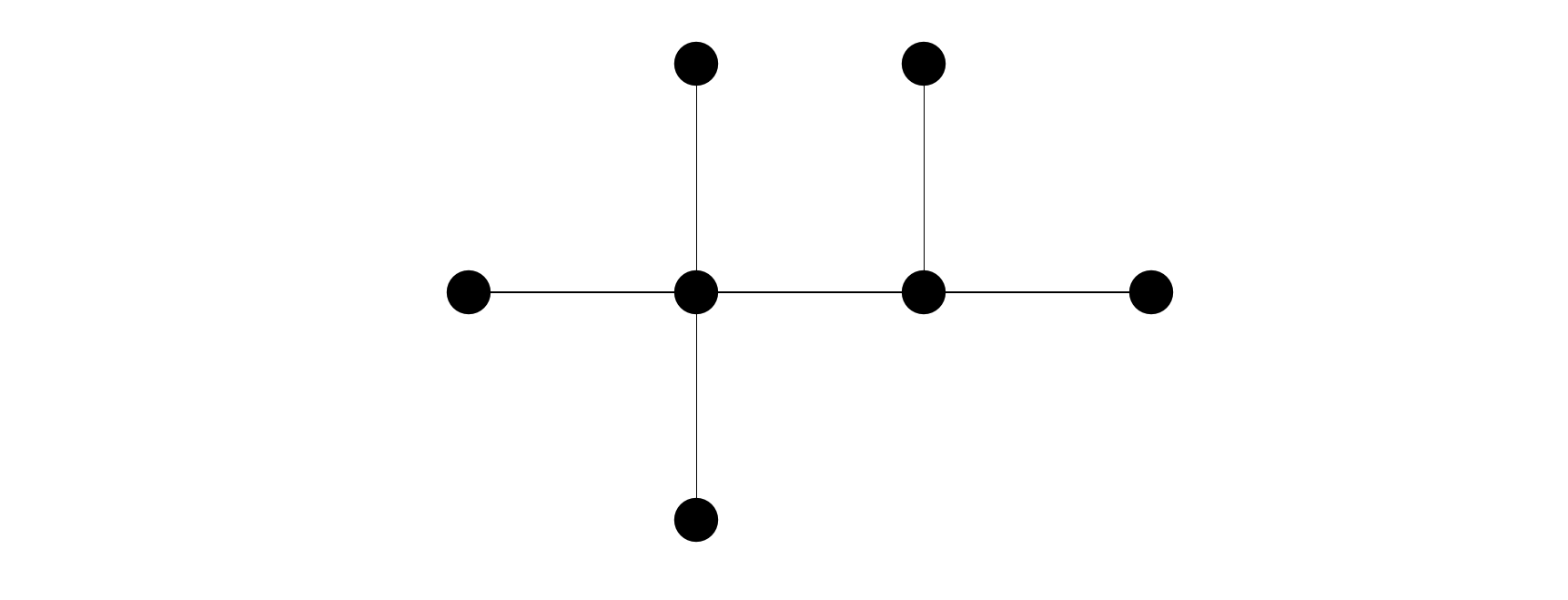}
}
\quad
\subfigure[$\mathcal{T}^{9}_{7}$]{
\includegraphics[trim=3.5cm 18cm 0 6cm,clip, width=.33\textwidth]{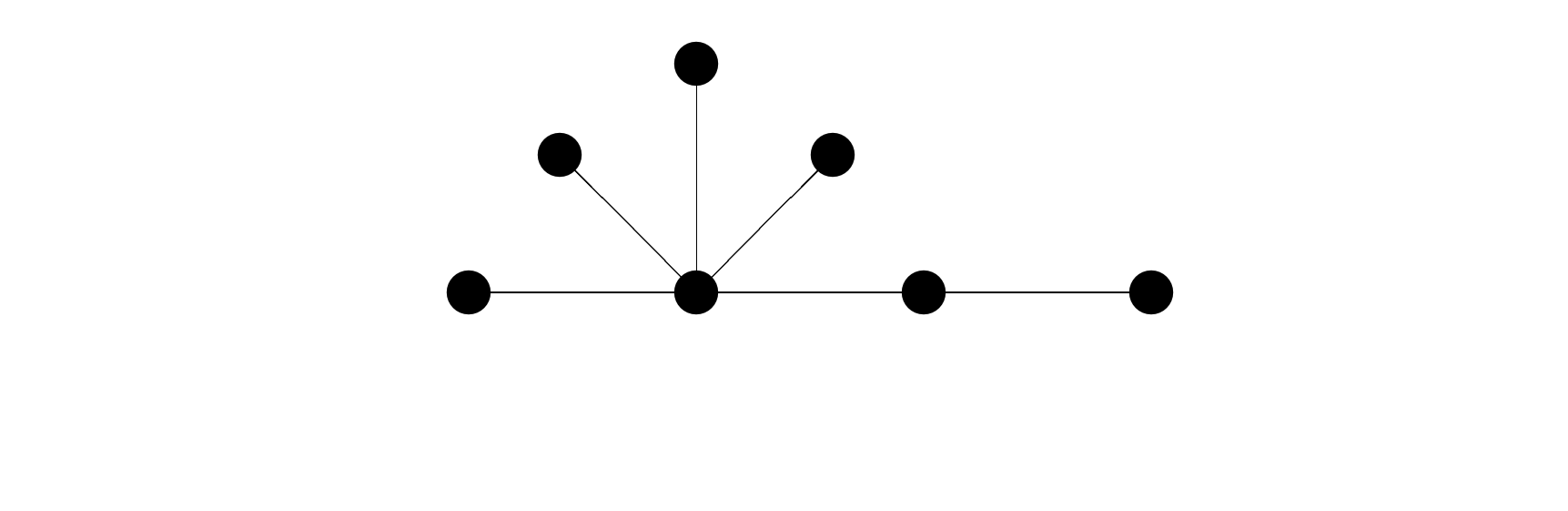}
}
}\\
\mbox{\subfigure[$\mathcal{T}^{10}_{7}$]{
\includegraphics[trim=3.5cm 17cm 0 6cm,clip, width=.33\textwidth]{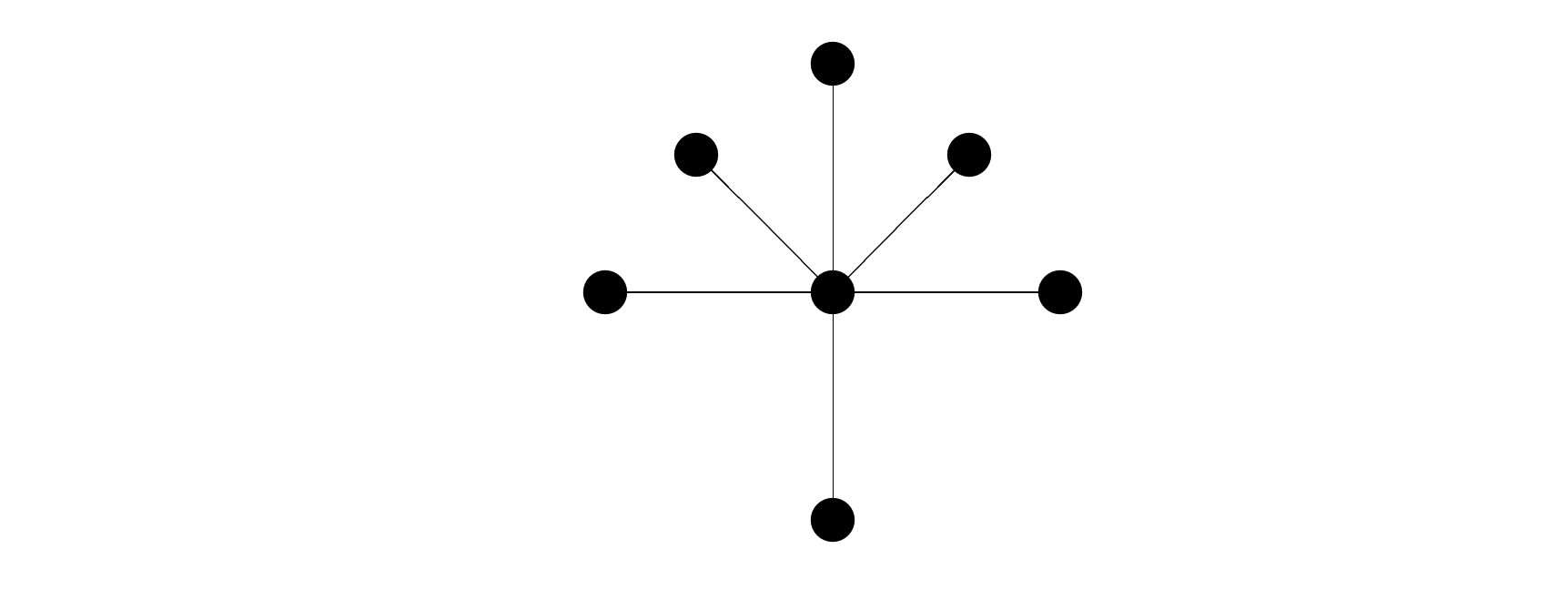}
}
\quad
\subfigure[$\mathcal{T}^{11}_{7}$]{
\includegraphics[trim=3.5cm 18cm 0 3cm,clip, width=.33\textwidth]{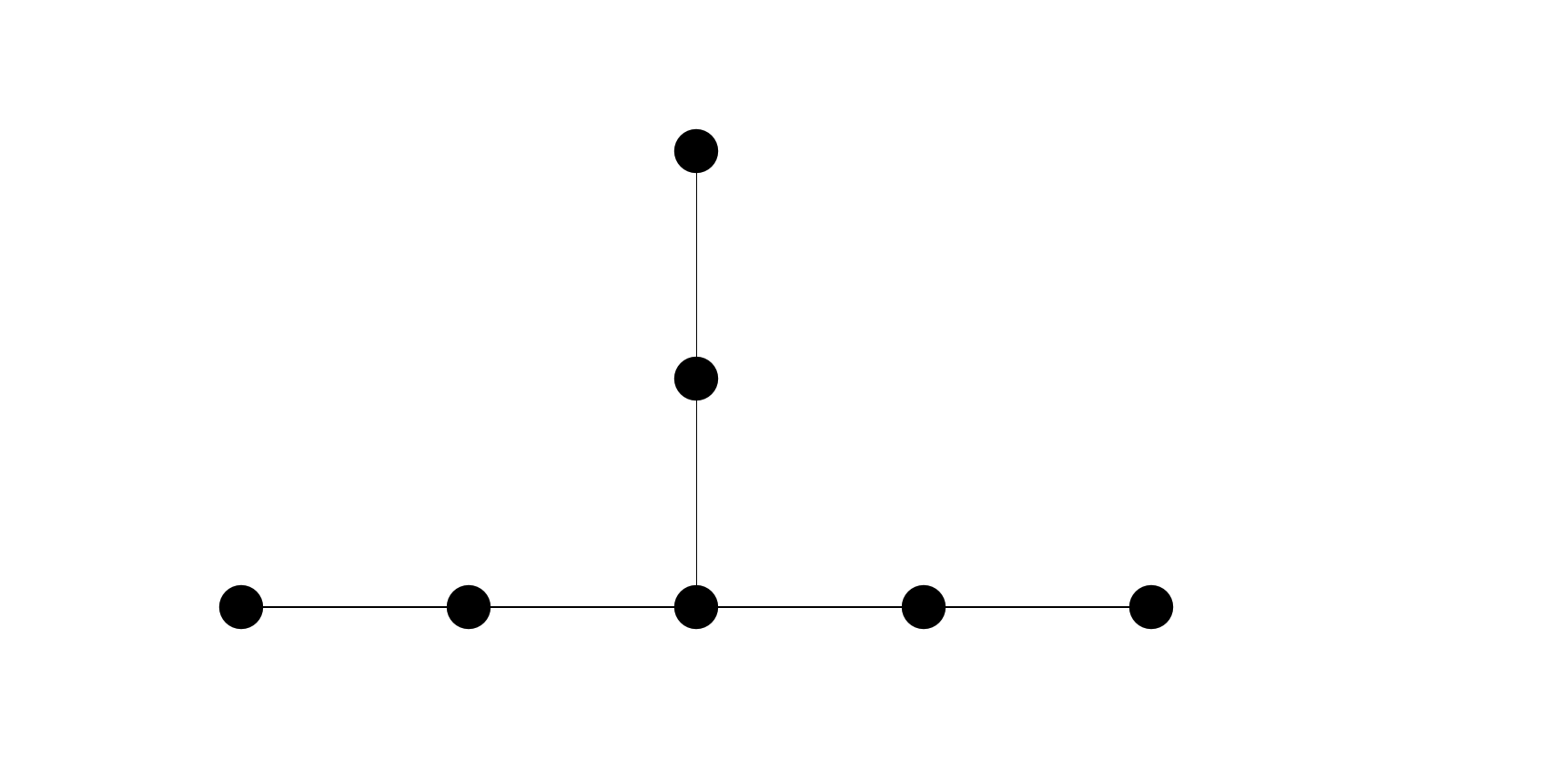}
}
}
\caption{\label{figB_1} Graph isomorphism classes $\calT^{j}_{7}$ ($j=1,\ldots 
11$) for trees of order $7$.}
\end{center}
\end{figure}

Trees of order $7$ partition into eleven GI classes \cite{harary} which we 
denote by $\{ \calT^{j}_{7} : j = 1,\ldots 11\}$, and show as unlabelled 
graphs in Figure~\ref{figB_1}.
Seven of these GI classes correspond to known (unlabelled) graphs 
(Section~\ref{sec2} and Appendix~\ref{AppA}):
\begin{equation}
\begin{array}{lll}
\calT_{7}^{1} = P_{7} = S^{(5)}_{1,1}; 
  & \hspace{0.5in}  \calT_{7}^{6} =   S^{(3)}_{3,1}; \hspace{0.5in} &
               \calT_{7}^{10} = K_{1,6}.   \\
\calT_{7}^{2} = S^{(4)}_{2,1} ; & 
            \hspace{0.5in}\calT_{7}^{8} = S^{(2)}_{3,2}; \hspace{0.5in} & \\
\calT_{7}^{5} = S^{(3)}_{2,2}; & 
    \hspace{0.5in} \calT_{7}^{9} = S^{(2)}_{4,1}; \hspace{0.5in}  &
\end{array}
\end{equation}

\subsubsection{$r(\calT_{7}^{i},\calT_{6}^{j})$}
\label{AppB_1_1}

Using our numerical procedure (Section~\ref{sec4_1}) we determined the tree 
Ramsey numbers $r(\calT_{7}^{i},\calT_{6}^{j})$  for $1\leq i\leq 11$ and
$1\leq j\leq 6$ which are displayed in Table~\ref{tableB_1}.
\begin{table}
\caption{\label{tableB_1}Numerical results for tree Ramsey numbers 
$r(\calT_{7}^{i},\calT_{6}^{j})$ with $1\leq i \leq 11$ and $1\leq j \leq 6$. 
Table rows (columns) are labelled by $i$ ($j$). A superscript ``x'' on a table 
entry indicates that Theorem~A.x of Appendix~\ref{AppA} applies, and so 
these tree Ramsey numbers were known prior to this work. The reader can 
verify that our numerical results are in agreement with the theorems of 
Appendix~\ref{AppA}. The remaining $64$ tree Ramsey numbers (to the best of our
knowledge) are new.}
\begin{center}
\begin{tabular}{|c|cccccc|} \hline
$\mathbf{r(\calT_{7}^{i},\calT_{6}^{j})}$ & \rule{0cm}{1.5em} & & & & & \\
$i\; \backslash \; j$ & 1 & 2 & 3 & 4 & 5 & 6 \\ \hline
1\rule{0cm}{1.5em}  & $9^{\,2}$  & 8 & 9 & 8 & 9 & 9 \\
2  &  9 & $8$ & 9 & 8 & 9 & 9 \\
3  & 9 & 8 & 9 & 8 & 9 & 9\\
4  & 9 & 8 & 9 & 8 & 9 & 9\\
5  & 9 & 9 & 9 & 9 & 9 & 9\\
6  & 9 & 9 & 9 & 9 & 9 & 9\\
7  & 9 & 8 & 9 & 8 & 9 & 9\\
8  & 9 & 8 & 9 & 8 & 9 & 9 \\
9  & 9 & 9 & 9 & 9 & 10 & 9\\
10 & 11 & 11 & 11 & 11 & $11^{\,3}$ & 11\\
11 & 9 & 8 & 9 & 8 & 9 & 9 \\\hline
\end{tabular}
\end{center}
\end{table}
A superscript ``x'' on a table entry indicates that Theorem~A.x of 
Appendix~\ref{AppA} applies, and so these tree Ramsey numbers were known prior 
to this work. The reader can verify that our numerical results are in agreement 
with the theorems of Appendix~\ref{AppA}. The remaining $64$ tree Ramsey 
numbers (to the best of our knowledge) are new.

As explained in Section~\ref{sec4_1}, for trees of order $6$ and $7$, and for
graphs of order $8\leq N\leq 11$, our numerical procedure can exhaustively 
search over all non-isomorphic graphs, and so can find the number of 
non-isomorphic optimal graphs and their associated objective function value. 
For graphs with order $N = r(\calT_{6}^{i},\calT_{6}^{j}) -1$, the optimal 
graphs are ($\calT_{7}^{i},\calT_{6}^{j}$)-critical graphs (Section~\ref{sec2}).
Table~\ref{tableB_2}
\begin{table}
\caption{\label{tableB_2}Numerical results for the number of non-isomorphic 
($\calT_{7}^{i},\calT_{6}^{j}$)-critical graphs $N_{c}(\calT_{7}^{i},
\calT_{6}^{j})$ with $1\leq i\leq 11$ and $1\leq j \leq 6$. Table rows (columns)
are labelled by $i$ ($j$), and for table entry ($i,j$), the graph order is 
$r(\calT_{7}^{i},\calT_{6}^{j}) - 1$.}
\begin{center}
\begin{tabular}{|c|cccccc|} \hline
$\mathbf{N_{c}(\calT_{7}^{i},\calT_{6}^{j})}$ & \rule{0cm}{1.5em} & & & & & \\
$i\; \backslash \; j$ & 1 & 2 & 3 & 4 & 5 & 6 \\ \hline
1  & 2 & 8 & 2 & 5 & 1 & 1 \\
2 & 2 & 6 & 2 & 3 & 1 & 1 \\
3 & 2 & 6 & 2 & 3 & 1 & 1 \\
4 & 2 & 6 & 2 & 3 & 1 & 1 \\
5 & 3 & 1 & 3 & 1 & 6 & 2 \\
6 & 3 & 1 & 3 & 1 & 10 & 2 \\
7 & 2 & 6 & 2 & 3 & 7 & 1 \\
8 & 2 & 5 & 2 & 5 & 8 & 1 \\
9 & 3 & 1 & 3 & 6 & 16 & 2 \\
10 & 1 & 1 & 1 & 1 & 60 & 1 \\
11 & 2 & 9 & 2 & 7 & 3 & 1 \\\hline
\end{tabular}
\end{center}
\end{table}
lists the number of non-isomorphic critical graphs for each pairing of GI 
classes $(\calT_{7}^{i},\calT_{6}^{j})$. These graphs are all found to have 
vanishing objective function value, which is expected, since critical graphs 
have order $N < r(\calT_{7}^{i},\calT_{6}^{j})$.

At the Ramsey threshold, $N = r(\calT_{7}^{i},\calT_{6}^{j})$, optimal
graphs first acquire a non-vanishing objective function value (see 
Section~\ref{sec3_1}). In Tables~\ref{tableB_3} 
\begin{table}
\caption{\label{tableB_3}Numerical results for the number of non-isomorphic 
optimal graphs $N_{opt}(\calT_{7}^{i},\calT_{6}^{j})$ with $1\leq i \leq 11$ 
and $1\leq j \leq 6$. Table rows (columns) are labelled by $i$ ($j$), and for 
table entry ($i,j$), the graph order is $r(\calT_{7}^{i},\calT_{6}^{j})$.}
\begin{center}
\begin{tabular}{|c|cccccc|} \hline
$\mathbf{N_{opt}(\calT_{7}^{i},\calT_{6}^{j})}$ & \rule{0cm}{1.5em} & & & & & \\
$i\; \backslash \; j$ & 1 & 2 & 3 & 4 & 5 & 6 \\ \hline
1 & 2 & 8 & 2 & 5 & 1 & 1 \\
2 & 2 & 6 & 2 & 3 & 1 & 1 \\
3 & 2 & 6 & 2 & 3 & 1 & 1 \\
4 & 2 & 6 & 2 & 3 & 1 & 1 \\
5 & 2 & 1 & 2 & 1 & 1 & 1 \\
6 & 2 & 1 & 2 & 1 & 1 & 1 \\
7 & 2 & 6 & 2 & 3 & 1 & 1 \\
8 & 2 & 5 & 2 & 2 & 1 & 1 \\
9 & 2 & 1 & 2 & 1 & 1 & 1 \\
10 & 1 & 1 & 1 & 1 & 1 & 1 \\
11 & 2 & 8 & 2 & 5 & 1 & 1 \\ \hline
\end{tabular}
\end{center}
\end{table}
and \ref{tableB_4}
\begin{table}
\caption{\label{tableB_4}Numerical results for the minimum objective function 
value $\calO_{N}(e_{\ast};\calT_{7}^{i},\calT_{6}^{j})$ at the Ramsey threshold
with $1\leq i \leq 11$ and $1\leq j \leq 6$. Table rows (columns) are labelled 
by $i$ ($j$), and for table entry ($i,j$), the graph order is $r(\calT_{7}^{i},
\calT_{6}^{j})$.}
\begin{center}
\begin{tabular}{|c|cccccc|} \hline
$\mathbf{\calO_{N}(e_{\ast};\calT_{7}^{i},\calT_{6}^{j})}$ & 
\rule{0cm}{1.5em} & & & & & \\
$i\; \backslash \; j$ & 1 & 2 & 3 & 4 & 5 & 6 \\ \hline
1 & 1 & 1 & 1 & 1 & 4 & 1 \\
2 & 1 & 1 & 1 & 1 & 4 & 1 \\
3 & 1 & 1 & 1 & 1 & 4 & 1 \\
4 & 1 & 1 & 1 & 1 & 4 & 1 \\
5 & 1 & 6 & 1 & 6 & 4 & 1 \\
6 & 1 & 6 & 1 & 6 & 4 & 1 \\
7 & 1 & 1 & 1 & 1 & 4 & 1 \\
8 & 1 & 1 & 1 & 1 & 4 & 1 \\
9 & 1 & 6 & 1 & 6 & 5 & 1 \\
10 & 6 & 6 & 6 & 6 & 6 & 6 \\
11 & 1 & 1 & 1 & 1 & 4 & 1 \\ \hline
\end{tabular}
\end{center}
\end{table}
we list, respectively, the number of non-isomorphic optimal graphs 
for each pairing of GI classes $(\calT_{7}^{i},\calT_{6}^{j})$ and the 
corresponding minimum objective function value.

\subsubsection{$r(\calT_{7}^{i},\calT_{7}^{j})$}
\label{AppB_1_2}

Using the numerical procedure described in Section~\ref{sec4_1}, we determined 
the tree Ramsey numbers $r(\calT_{7}^{i},\calT_{7}^{j})$  for $1\leq i,j\leq 11$ 
which are displayed in Table~\ref{tableB_5}.
\begin{table}
\caption{\label{tableB_5}Numerical results for tree Ramsey numbers 
$r(\calT_{7}^{i},\calT_{7}^{j})$ with $1\leq i,j \leq 11$. Table rows (columns) 
are labelled by $i$ ($j$). Only the upper triangular table entries are shown as 
the lower triangular entries follow from $r(\calT_{7}^{j},\calT_{7}^{i}) =
r(\calT_{7}^{i},\calT_{7}^{j})$. A superscript ``x'' on a table entry indicates 
that Theorem~A.x of Appendix~\ref{AppA} applies, and so these tree Ramsey 
numbers were known prior to this work. The reader can verify that our 
numerical results are in agreement with the theorems of Appendix~\ref{AppA}. 
The remaining $102$ tree Ramsey numbers (to the best of our knowledge) are 
new.}
\begin{center}
\begin{tabular}{|c|ccccccccccc|} \hline
$\mathbf{r(\calT_{7}^{i},\calT_{7}^{j})}$ & \rule{0cm}{1.5em} & & & & & & & & 
                                                                       & & \\
$i\; \backslash \; j$ & 1 & 2 & 3 & 4 & 5 & 6 & 7 & 8 & 9 & 10 & 11 \\ \hline
1\rule{0cm}{1.5em} & $9^{\,2}$  & 9 & 9 & 9 & 9 & 9 & 9 & 9 & 9 & $11^{\,4}$ 
                                                                            & 9\\ 
2 & & $9^{\,6}$ & 9 & 9 & 9 & 9 & 9 & 9 & 9 & $11^{\,4}$ & 9\\
3 & & & 9 & 9 & 9 & 9 & 9 & 9 & 9 & $11^{\,4}$ & 9  \\
4 & & & & 9 & 9 & 9 & 9 & 9 & 9 & $11^{\,4}$ & 9 \\ 
5 & & & & & 9 & 9 & 9 & 9 & 9 & $11$ & 9 \\
6 & & & & & & 9 & 9 & 9 & 9 & $11^{\,4}$ & 9  \\
7 &  & & & & & & 9 & 9 & 9 & $11^{\,4}$ & 9 \\
8 & & & & & & & & $9^{\,7}$ & 9 & $11$ & 9 \\
9 & & & & & & & & & $10^{\,7}$ & $11^{\,4}$ & 9 \\  
10 & & & & & & & & & & $11^{\,3}$ & 11 \\
11 &  \rule[-1em]{0cm}{1em} & & & & & & & & &  & 9 \\  \hline
\end{tabular}
\end{center}
\end{table}
Only the upper triangular table entries are shown as the lower triangular 
entries follow from $r(\calT_{7}^{j},\calT_{7}^{i}) =r(\calT_{7}^{i},
\calT_{7}^{j})$. A superscript ``x'' on a table entry indicates that Theorem~A.x
in Appendix~\ref{AppA} applies, and so these tree Ramsey numbers were known
prior to this work. The reader can verify that our numerical results are in 
agreement with the theorems of Appendix~\ref{AppA}. The remaining $102$ 
tree Ramsey numbers (to the best of our knowledge) are new.

As explained in Section~\ref{sec4_1}, for $m,n=7$, and for graphs of order
$9\leq N\leq 11$, our numerical procedure can exhaustively search over 
all non-isomorphic graphs, and so can find the number of non-isomorphic 
optimal graphs and their associated objective function value. For graphs with 
order $N = r(\calT_{7}^{i},\calT_{7}^{j}) -1$, the optimal graphs are 
($\calT_{7}^{i},\calT_{7}^{j}$)-critical graphs (Section~\ref{sec2}). 
Table~\ref{tableB_6}
\begin{table}
\caption{\label{tableB_6}Numerical results for the number of non-isomorphic 
critical graphs $N_{c}(\calT_{7}^{i},\calT_{7}^{j})$ with $1\leq i,j \leq 11$. 
Table rows (columns) are labelled by $i$ ($j$), and for table entry ($i,j$), the 
graph order is $r(\calT_{7}^{i},\calT_{7}^{j}) - 1$. Only the upper triangular 
table entries are shown as the lower triangular entries follow from symmetry 
under interchange of colors: $N_{c}(\calT_{7}^{j},\calT_{7}^{i}) = 
N_{c}(\calT_{7}^{i},\calT_{7}^{j})$.}
\begin{center}
\begin{tabular}{|c|ccccccccccc|} \hline
$\mathbf{N_{c}(\calT_{7}^{i},\calT_{7}^{j})}$ & \rule{0cm}{1.5em} &  &  & 
     &  & &  & & & &  \\
$i\; \backslash \; j$ & 1 & 2 & 3 & 4 & 5 & 6 & 7 & 8 & 9 & 10 & 11 \\ \hline
1 &  8  & 6 & 6 & 6 & 5 & 5 & 6 & 5 & 5 & 1 & 8\\
2 & & 4 & 4 & 4 & 3 & 3 & 4 & 3 & 3 & 1 & 6\\
3 & & & 4 & 4 & 3 & 3 & 4 & 3 & 3 & 1 & 6\\ 
4 & & & & 4 & 3 & 3 & 4 & 3 & 3 & 1 & 6\\
5 & & & & & 2 & 2 & 4 & 2 & 6 & 1 & 6\\
6 & & & & & & 2 & 4 & 2 & 10 & 1 & 6\\  
7 &  & & & & & & 4 & 3 & 9 & 1 & 6\\
8 & & & & & & & & 2 & 9 & 1 & 5 \\
9 & & & & & & & & & 16 & 60 & 7 \\ 
10 & & & & & & & & & & 9638 & 1 \\
11 & &&&&&&&&&  & 8 \\ \hline
\end{tabular}
\end{center}
\end{table}
lists the number of non-isomorphic critical graphs for each pairing of GI 
classes $(\calT_{7}^{i},\calT_{7}^{j})$. These graphs are all found to have 
vanishing objective function value, which is expected, since critical graphs 
have order $N < r(\calT_{7}^{i},\calT_{7}^{j})$.

At the Ramsey threshold, $N = r(\calT_{7}^{i},\calT_{7}^{j})$, optimal
graphs first acquire a non-vanishing objective function value (see 
Section~\ref{sec3_1}). In Tables~\ref{tableB_7} 
\begin{table}
\caption{\label{tableB_7}Numerical results for the number of non-isomorphic 
optimal graphs $N_{opt}(\calT_{7}^{i},\calT_{7}^{j})$ with $1\leq i,j \leq 11$. 
Table rows (columns) are labelled by $i$ ($j$), and for table entry ($i,j$), the 
graph order is $r(\calT_{7}^{i},\calT_{7}^{j})$. Only the upper triangular table 
entries are shown as the lower triangular entries follow from symmetry under 
interchange of colors: $N_{opt}(\calT_{7}^{j},\calT_{7}^{i}) = 
N_{opt}(\calT_{7}^{i},\calT_{7}^{j})$.}
\begin{center}
\begin{tabular}{|c|ccccccccccc|} \hline
$\mathbf{N_{opt}(\calT_{7}^{i},\calT_{7}^{j})}$ & \rule{0cm}{1.5em} &  &  & 
   &  & & & & & &  \\
$i\; \backslash \; j$ & 1 & 2 & 3 & 4 & 5 & 6 & 7 & 8 & 9 & 10 & 11\\ \hline
1 & 8   &     6   &     6   &    6   &   4    &  4   &  6   &  5   &  4  &   1 &  8 \\
2 &  &   4   &  4   &  4   &  2    &  2   &  4   &  3   &  2  &  1 &  6 \\
3 &  &  &  4   & 4   &  2    & 2   &  4   & 3   & 2  &  1 & 6 \\
4 &  &  &   &     4   &  2    &  2   &  4   &  3   &  2  &  1 &   6 \\
5 &  &  &   &  &   2    &  2   &  2   &  1   &  6  &  1 &   4 \\
6 &  &  &   &  &  &  2   &  2   &  1   &   6  &  1 &  4 \\
7 &  &  &   &  &  &  &   4   &  3   &  2  &  1 &  6 \\
8 &  &  &   &  &  &  &  &  2   &  1  &  1 &  5 \\
9 &  &  &   &  &  &  &  &  &  22  &  1 &  4 \\
10 &  & &  &  &  &  &  &  &  & 7502 &  1 \\
11 &  & &  &  &  &  &  &  &  & &      8 \\ \hline
\end{tabular}
\end{center}
\end{table}
and \ref{tableB_8}
\begin{table}
\caption{\label{tableB_8}Numerical results for the minimum objective function 
value $\calO_{N}(e_{\ast};\calT_{7}^{i},\calT_{7}^{j})$ at the Ramsey threshold
with $1\leq i,j \leq 11$. Table rows (columns) are labelled 
by $i$ ($j$), and for table entry ($i,j$), the graph order is $r(\calT_{7}^{i},
\calT_{7}^{j})$. Only the upper triangular table entries are shown as the 
lower triangular entries follow from symmetry of the objective function under 
interchange of colors.}
\begin{center}
\begin{tabular}{|c|ccccccccccc|} \hline
$\mathbf{\calO_{N}(e_{\ast};\calT_{7}^{i},\calT_{7}^{j})}$ & 
\rule{0cm}{1.5em} & & & & & & & & & & \\
$i\; \backslash \; j$ & 1 & 2 & 3 & 4 & 5 & 6 & 7 & 8 & 9 & 10 & 11 \\ \hline
 1 &    1   &  1  &   1  &   1  &   1  &   1  &   1  &   1  &   1  &   5  &   1 \\
 2 &       &  1  &   1  &   1  &   1  &   1  &   1  &   1  &   1  &   5  &   1 \\
 3 &       &     &   1  &   1  &   1  &   1  &   1  &   1  &   1  &   5  &   1 \\
4 &         &     &      &   1  &   1  &   1  &   1  &   1  &   1  &   5  &   1 \\
5 &         &     &      &      &   6  &   6  &   1  &   1  &   6  &   5  &   1 \\
6 &         &     &      &      &      &   6  &   1  &   1  &   6  &   5  &   1 \\
7 &         &     &      &      &      &      &   1  &   1  &   1  &   5  &   1 \\
8 &         &     &      &      &      &      &      &   1  &   1  &   5  &   1 \\
9 &         &     &      &      &      &      &      &      &  20  &   5  &   1 \\
10 &         &     &      &      &      &      &      &      &      &   1  &   5 \\
11 &         &     &      &      &      &      &      &      &      &      &   1\\ \hline
\end{tabular}
\end{center}
\end{table}
we list, respectively, the number of non-isomorphic optimal graphs 
for each pairing of GI classes $(\calT_{7}^{i},\calT_{7}^{j})$ and the 
corresponding minimum objective function value.

\subsection{$r(\calT_{8},\calT_{n})$ for $n = 6,7,8$}
\label{AppB_2}

\begin{figure}
\begin{center}
\mbox{
\subfigure[$\mathcal{T}^{1}_{8}$]{
\includegraphics[trim=3.5cm 17cm 0 6cm,clip, width=.33\textwidth]{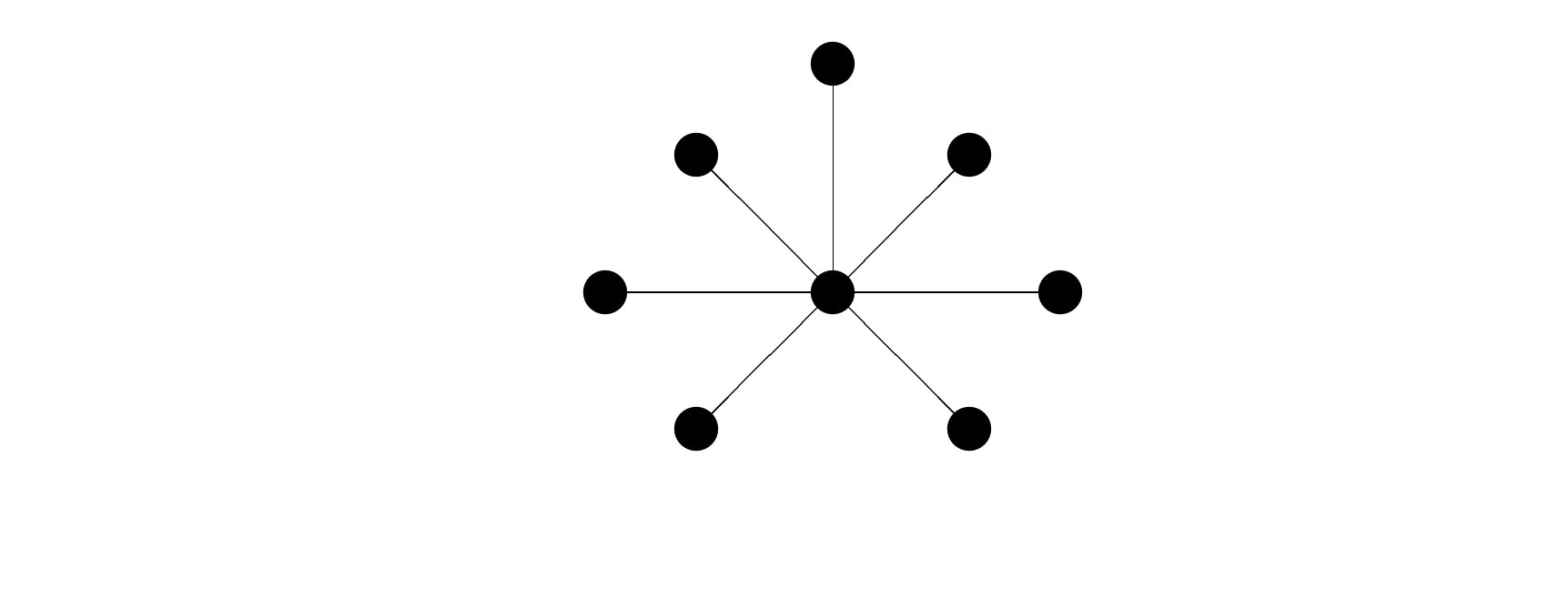}
}
\quad
\subfigure[$\mathcal{T}^{2}_{8}$]{
\includegraphics[trim=3.5cm 17cm 0 6cm,clip, width=.33\textwidth]{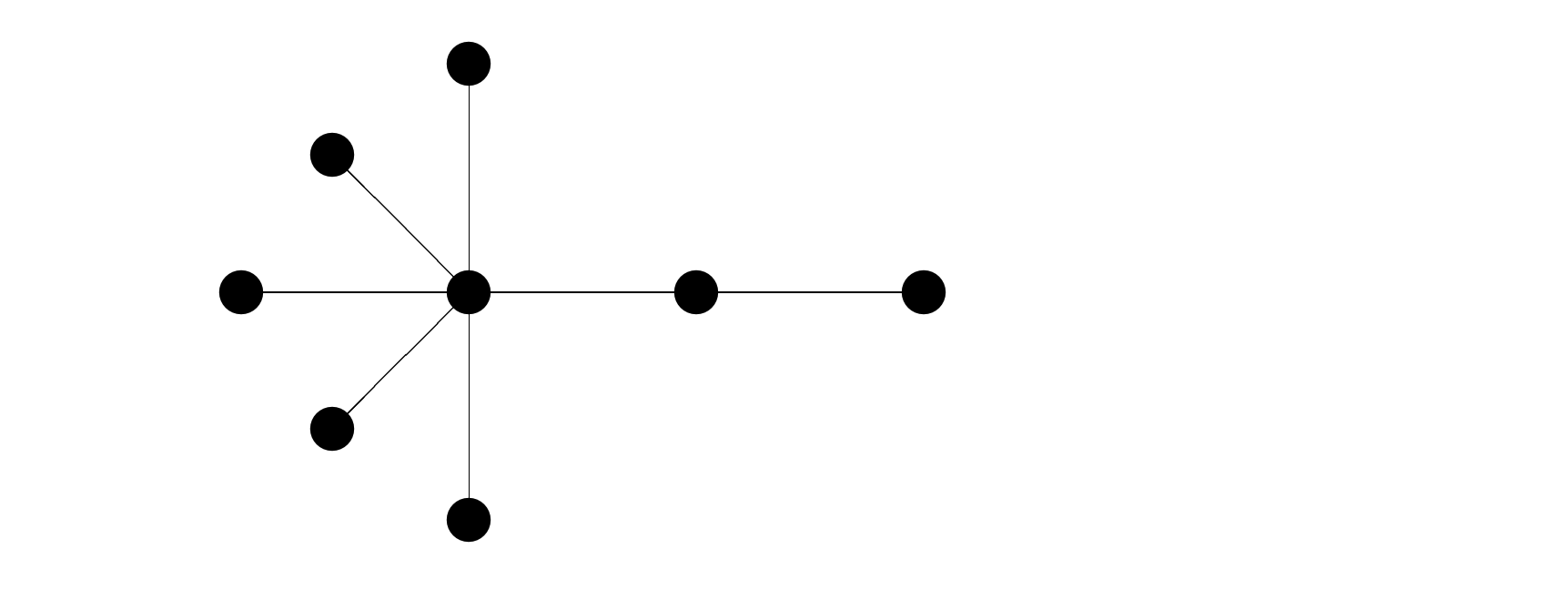}
}
\quad
\subfigure[$\mathcal{T}^{3}_{8}$]{
\includegraphics[trim=3.5cm 17cm 0 6cm,clip, width=.33\textwidth]{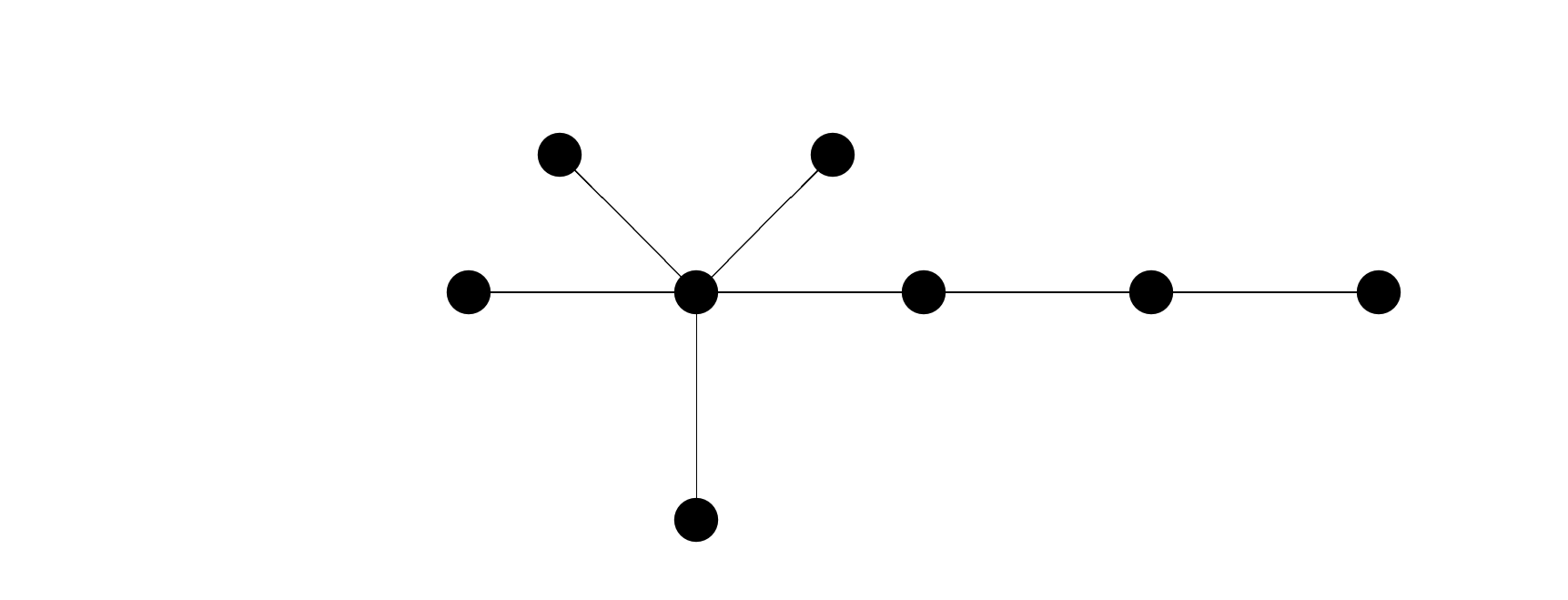}
}
}\\
\mbox{\subfigure[$\mathcal{T}^{4}_{8}$]{
\includegraphics[trim=3.5cm 17cm 0 7cm,clip, width=.33\textwidth]{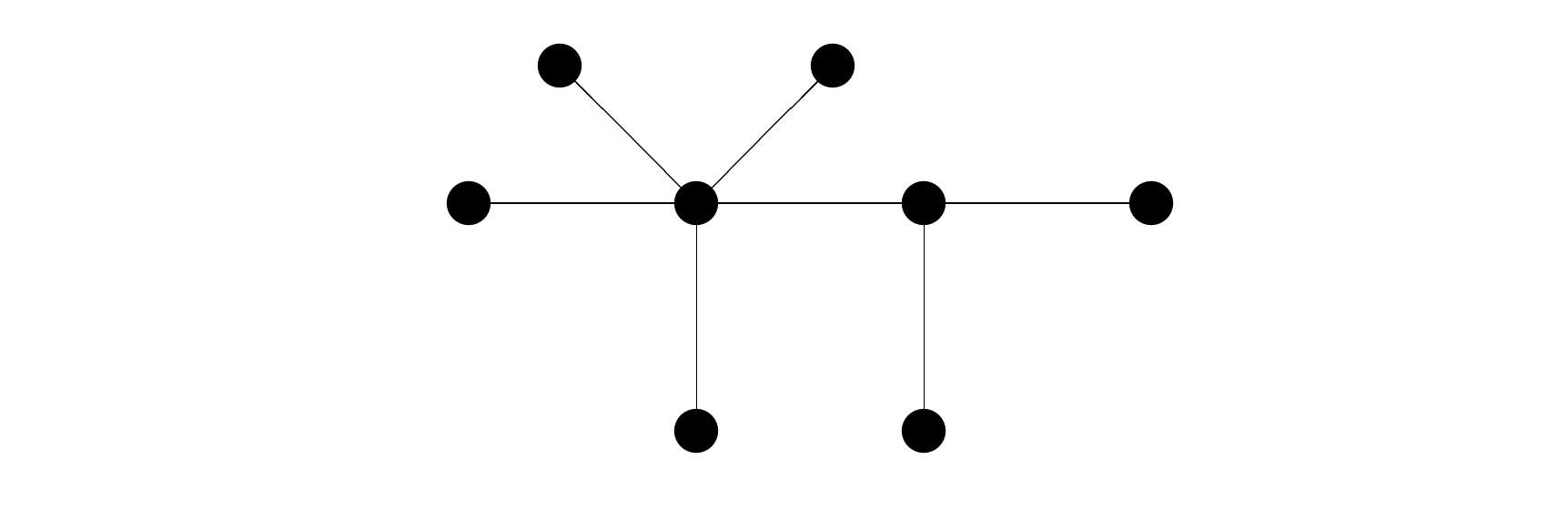}
}
\quad
\subfigure[$\mathcal{T}^{5}_{8}$]{
\includegraphics[trim=3.5cm 17cm 0 6cm,clip, width=.33\textwidth]{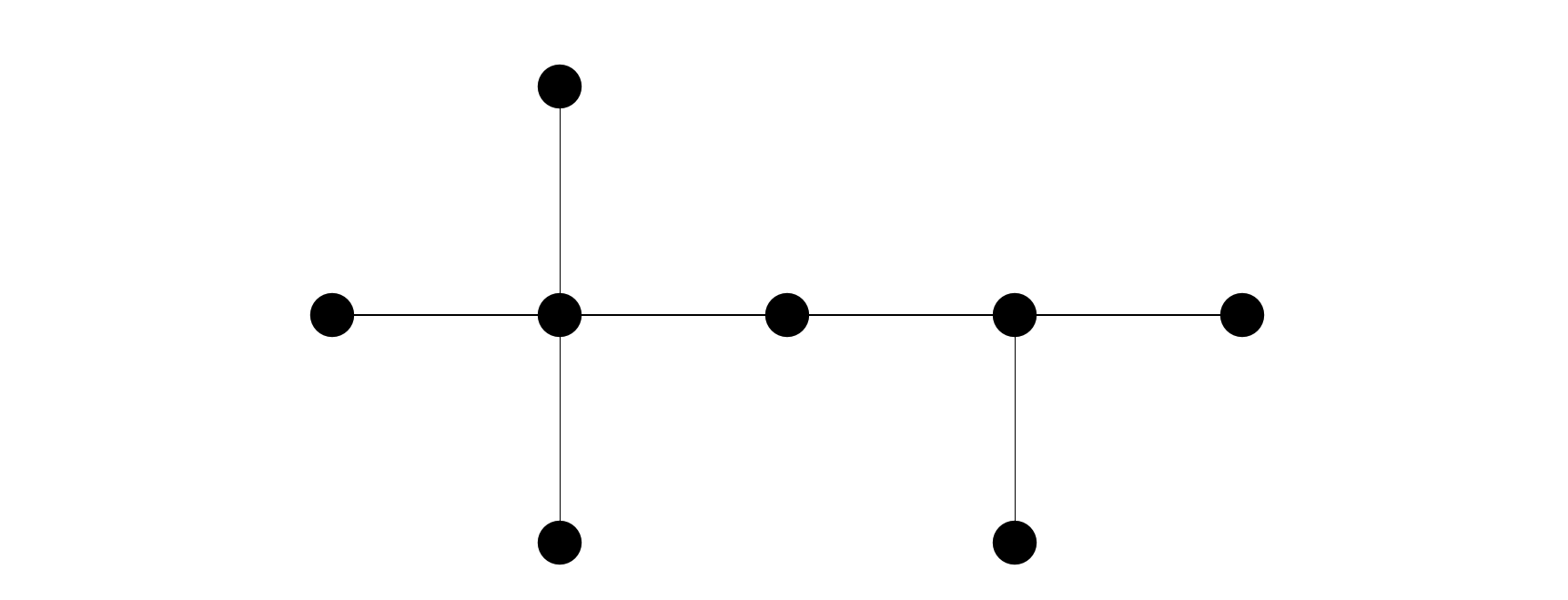}
}
\quad
\subfigure[$\mathcal{T}^{6}_{8}$]{
\includegraphics[trim=3.5cm 17cm 0 6cm,clip, width=.33\textwidth]{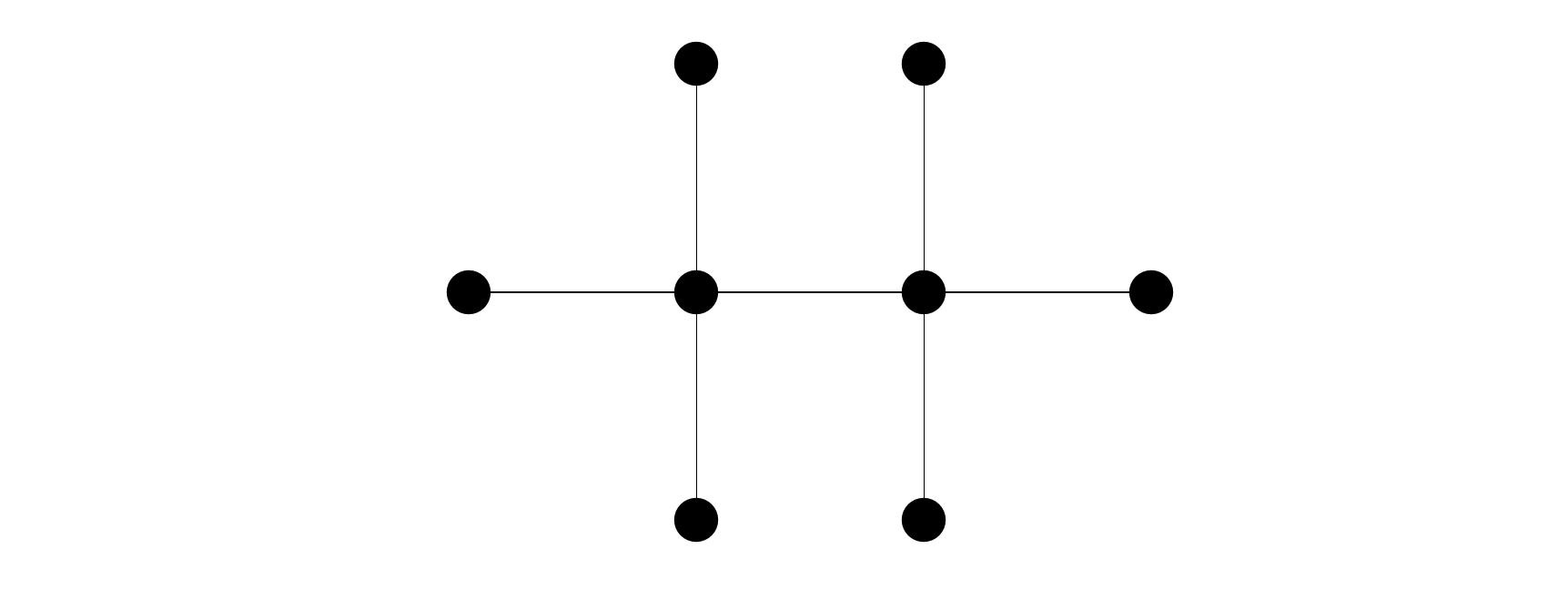}
}
}\\
\mbox{\subfigure[$\mathcal{T}^{7}_{8}$]{
\includegraphics[trim=3.5cm 15cm 0 7cm,clip, width=.33\textwidth]{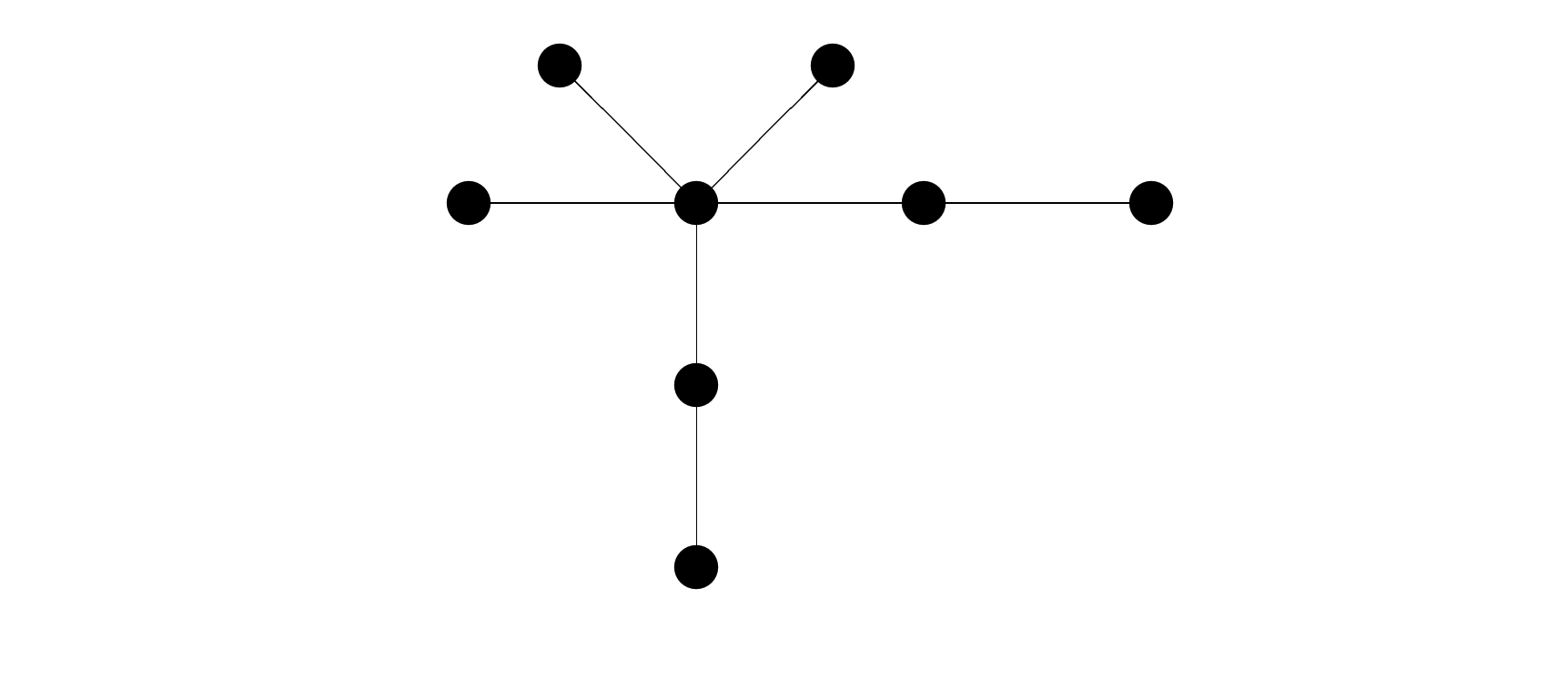}
}
\quad
\subfigure[$\mathcal{T}^{8}_{8}$]{
\includegraphics[trim=3.5cm 15cm 0 6cm,clip, width=.33\textwidth]{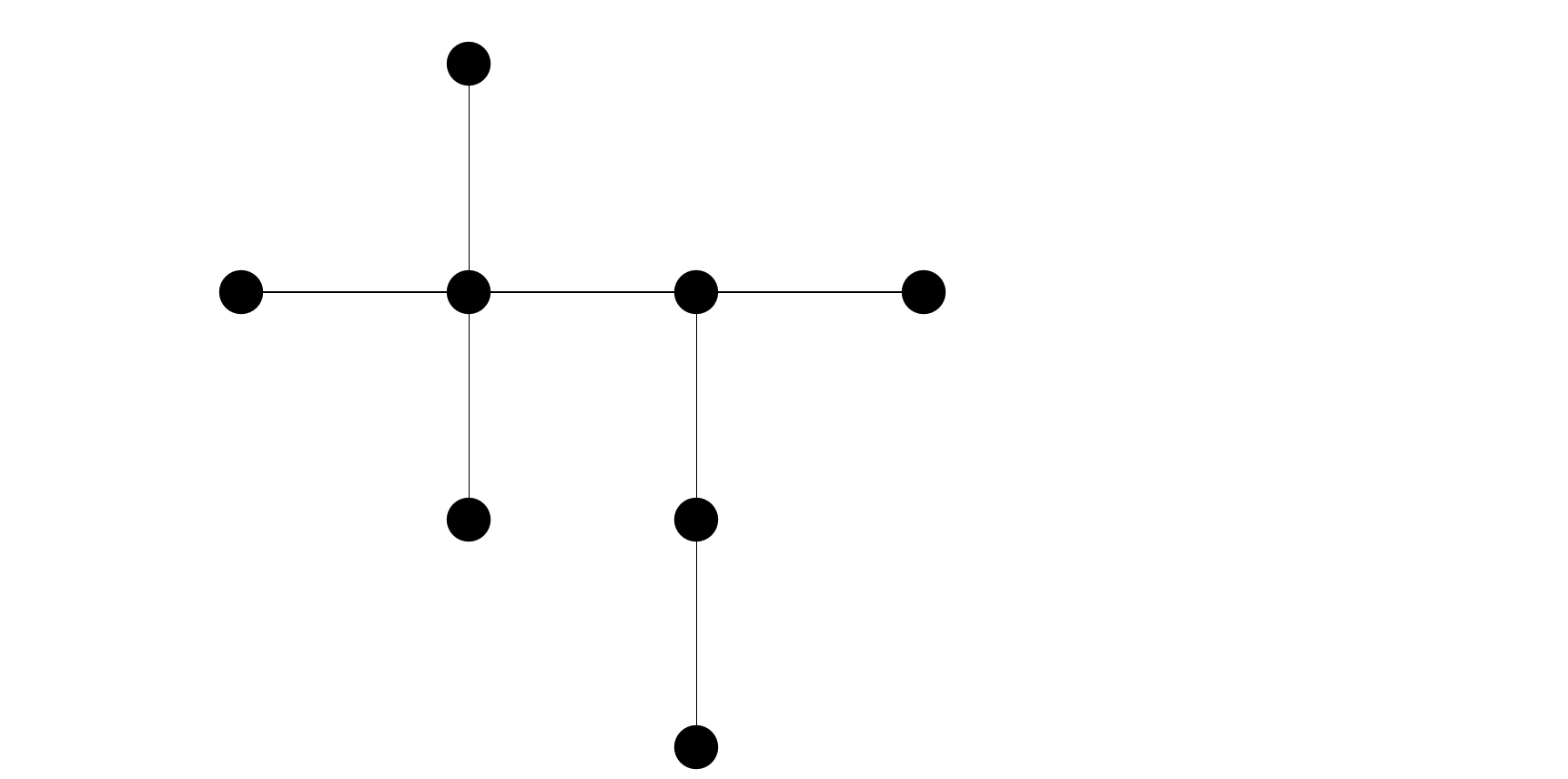}
}
\quad
\subfigure[$\mathcal{T}^{9}_{8}$]{
\includegraphics[trim=3.5cm 15cm 0 6cm,clip, width=.33\textwidth]{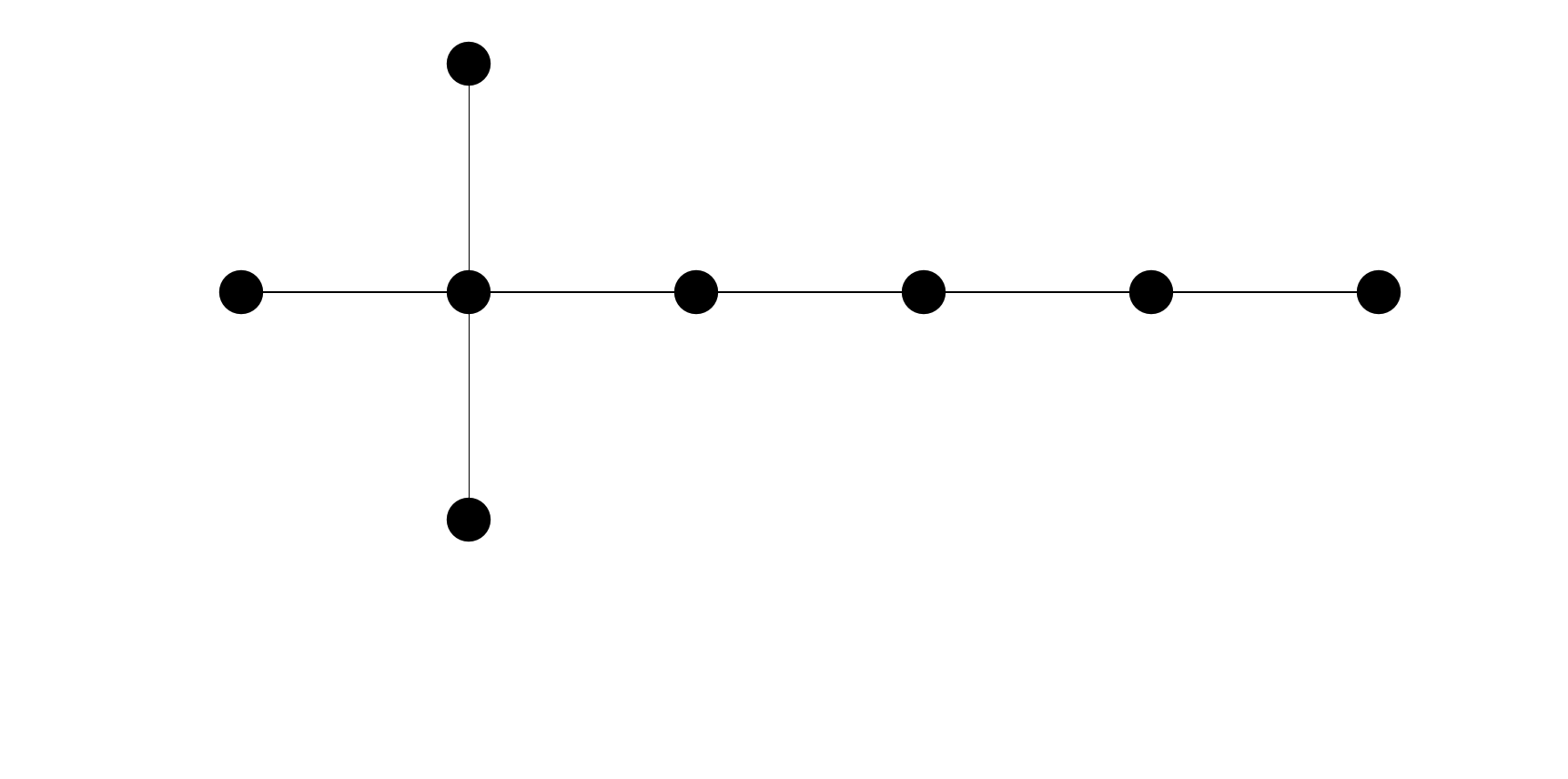}
}
}\\
\mbox{\subfigure[$\mathcal{T}^{10}_{8}$]{
\includegraphics[trim=3.5cm 14.5cm 0 6.5cm,clip, width=.33\textwidth]{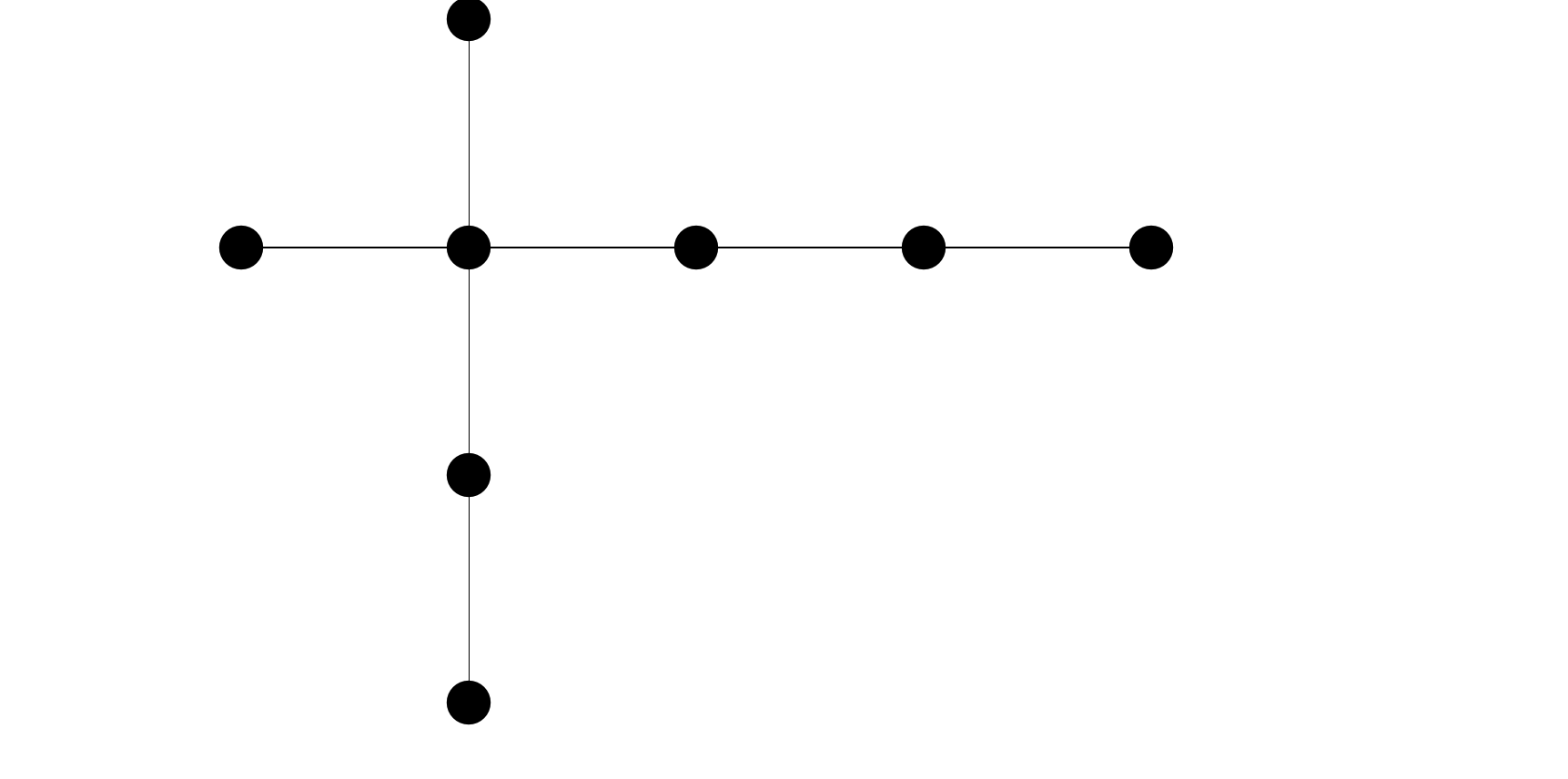}
}
\quad
\subfigure[$\mathcal{T}^{11}_{8}$]{
\includegraphics[trim=3.5cm 12cm 0 7cm,clip, width=.33\textwidth]{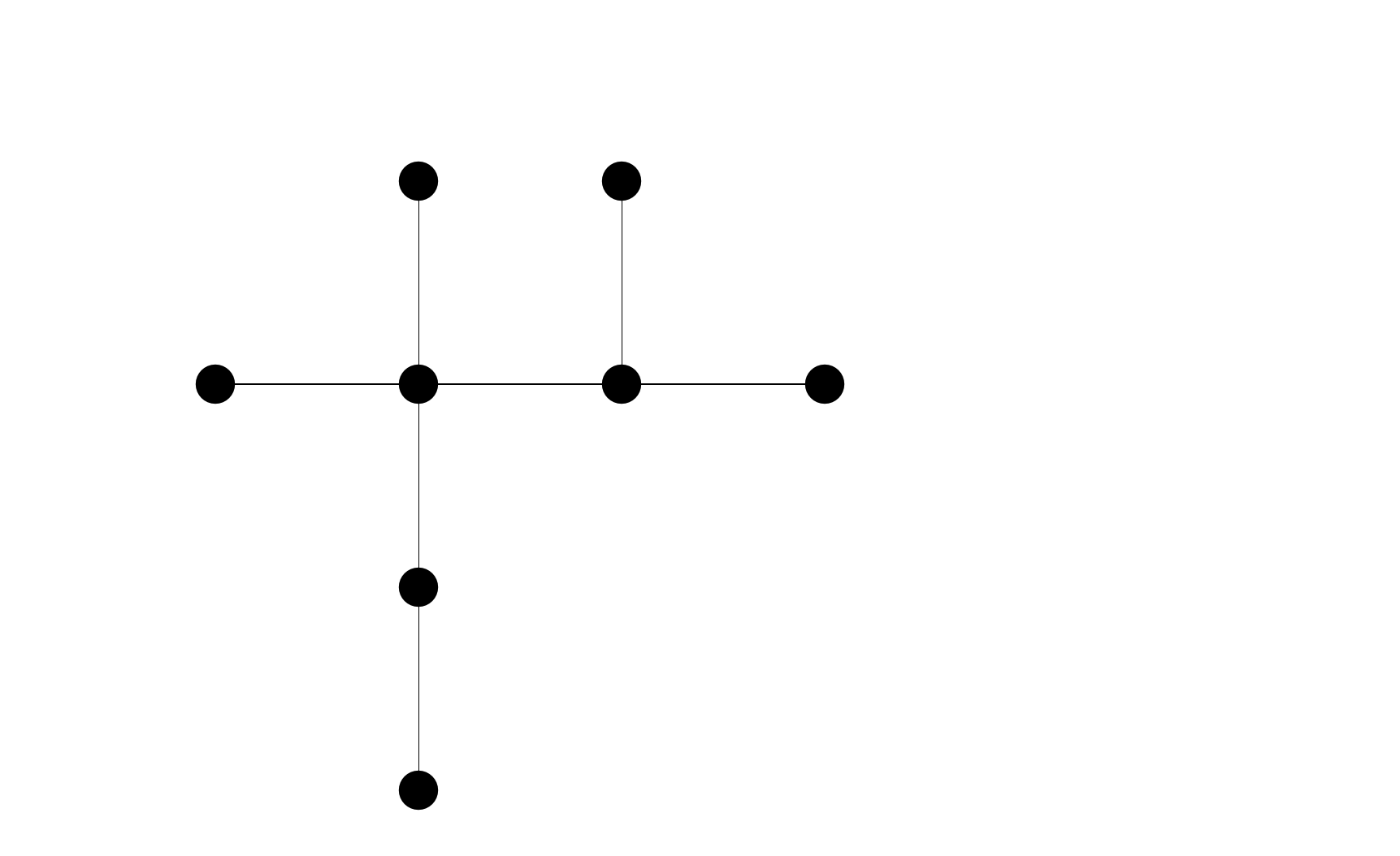}
}
\quad
\subfigure[$\mathcal{T}^{12}_{8}$]{
\includegraphics[trim=3.5cm 14.5cm 0 6cm,clip, width=.33\textwidth]{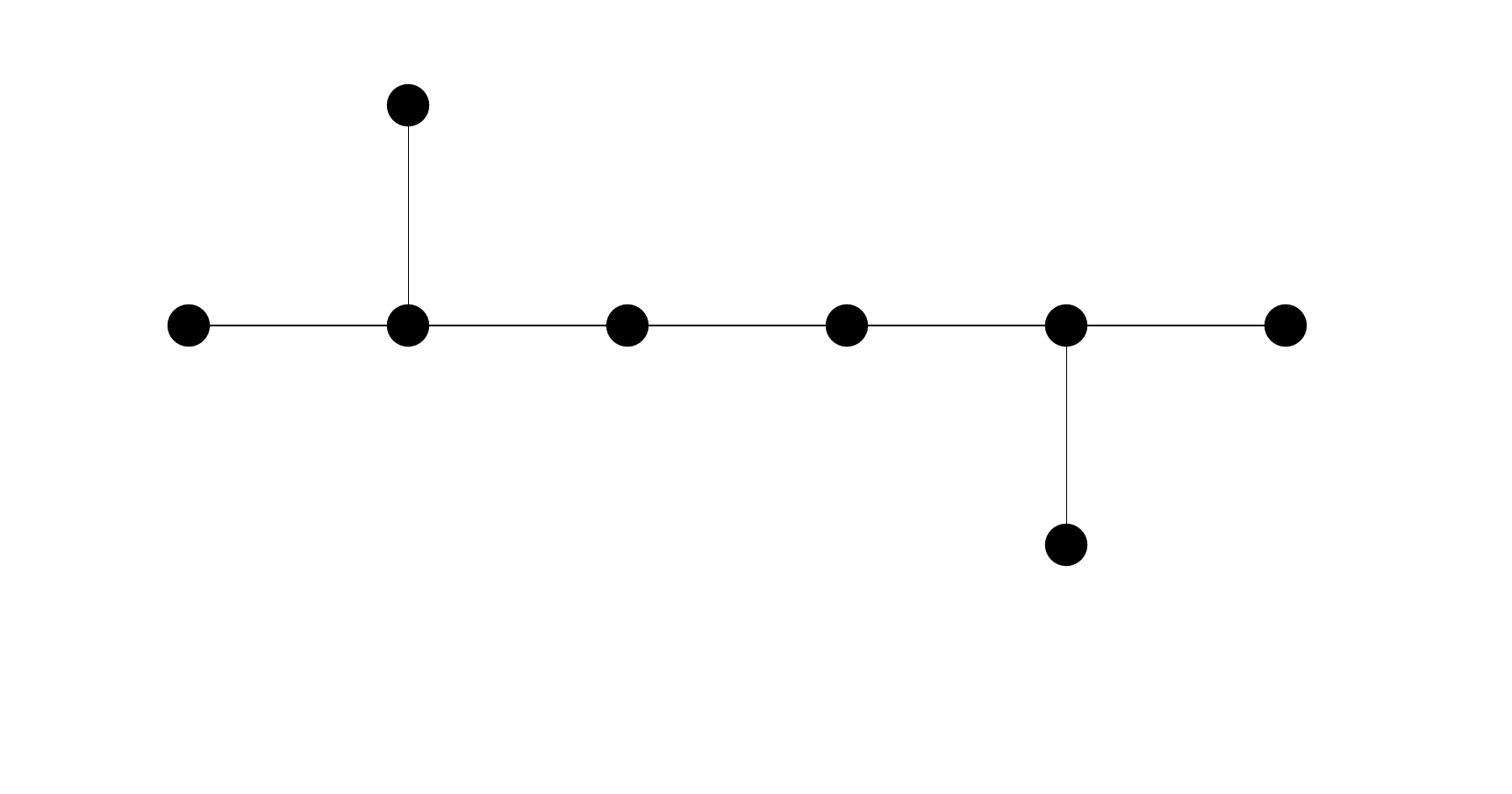}
}
}\\
\mbox{\subfigure[$\mathcal{T}^{13}_{8}$]{
\includegraphics[trim=3.5cm 17cm 0 6.5cm,clip, width=.33\textwidth]{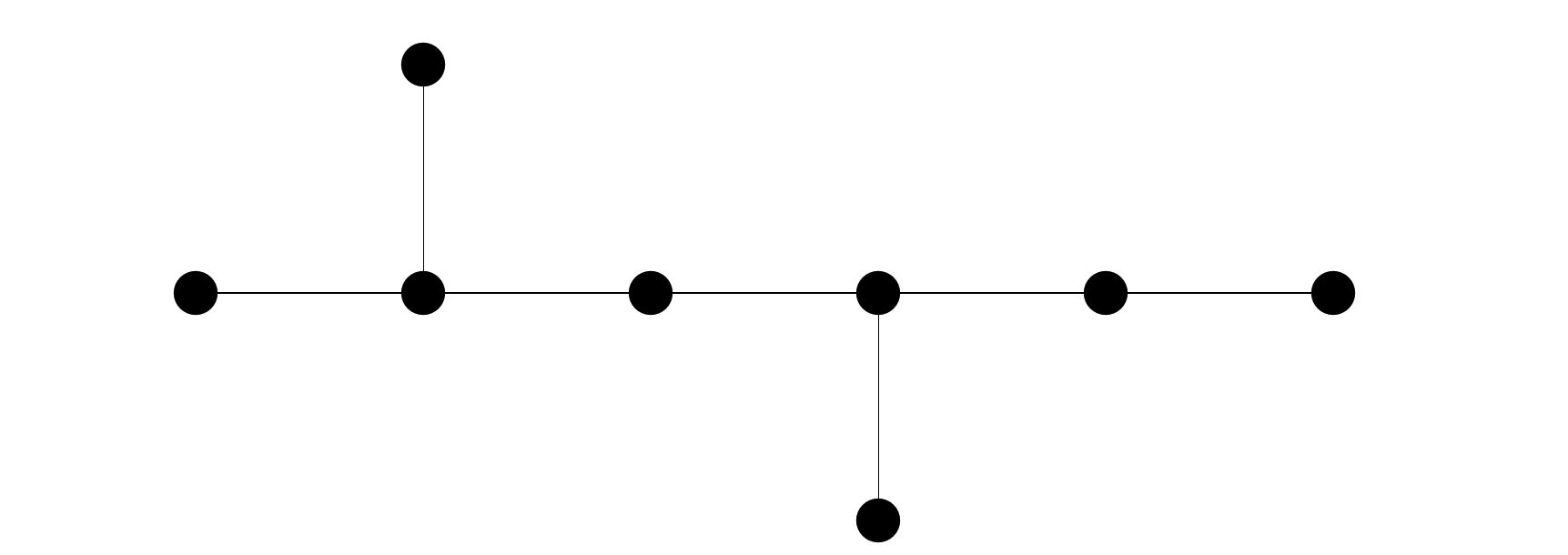}
}
\quad
\subfigure[$\mathcal{T}^{14}_{8}$]{
\includegraphics[trim=3.5cm 17.1cm 0 6cm,clip, width=.33\textwidth]{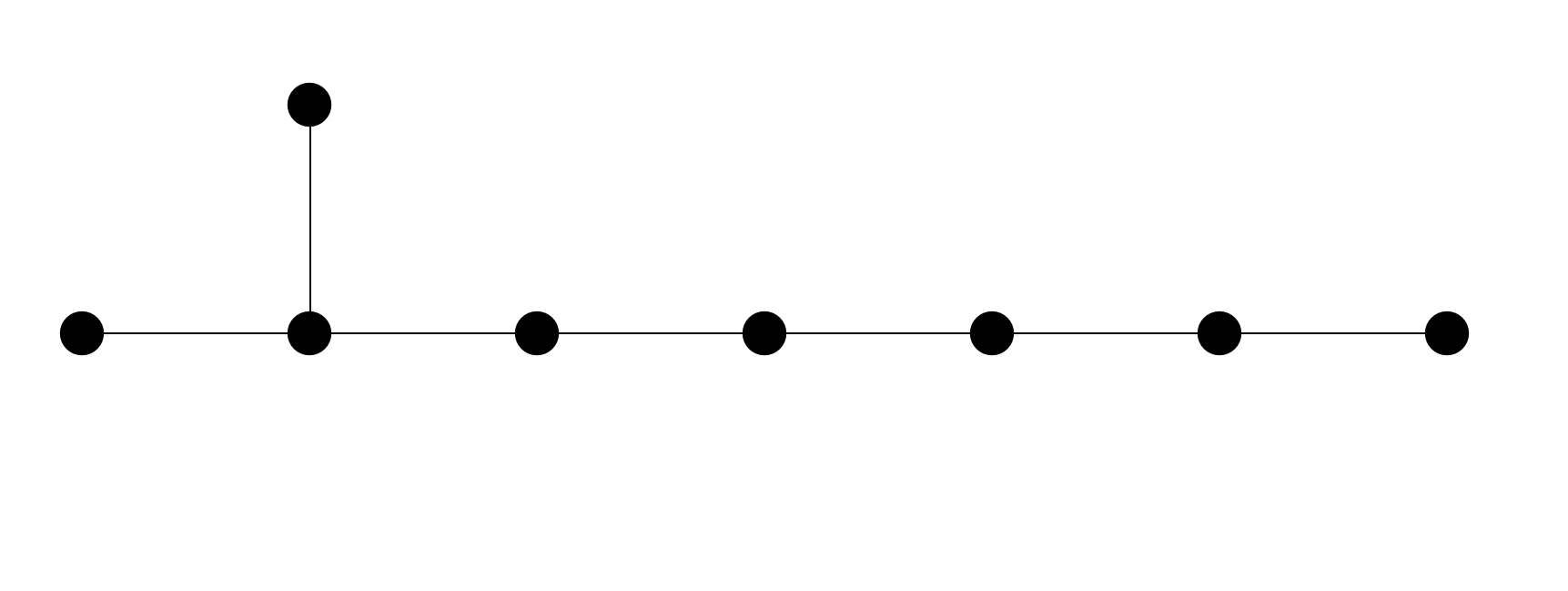}
}
\quad
\subfigure[$\mathcal{T}^{15}_{8}$]{
\includegraphics[trim=3.5cm 19.5cm 0 6cm,clip, width=.33\textwidth]{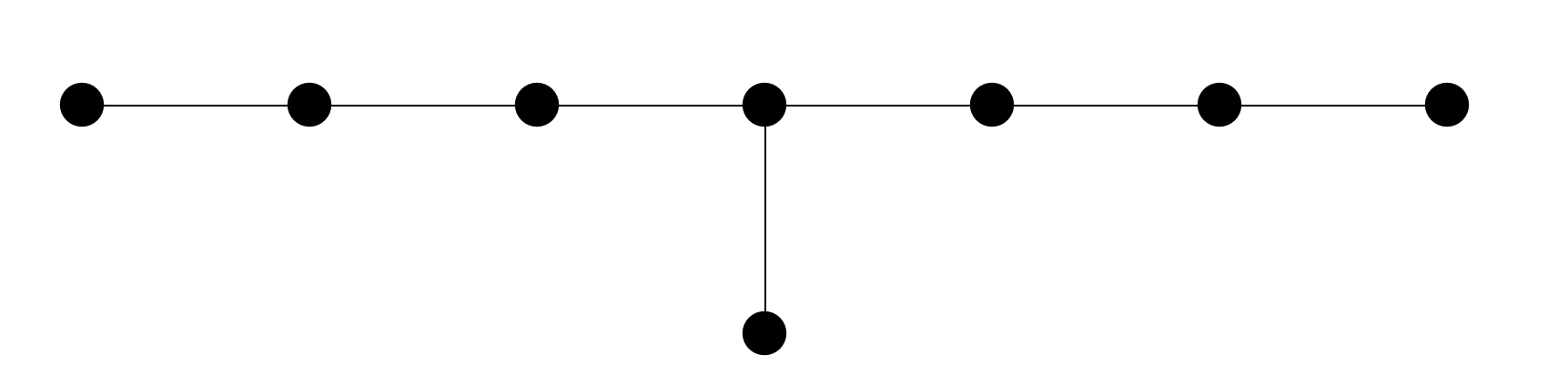}
}
}\\
\mbox{\subfigure[$\mathcal{T}^{16}_{8}$]{
\includegraphics[trim=3.5cm 17cm 0 6cm,clip, width=.33\textwidth]{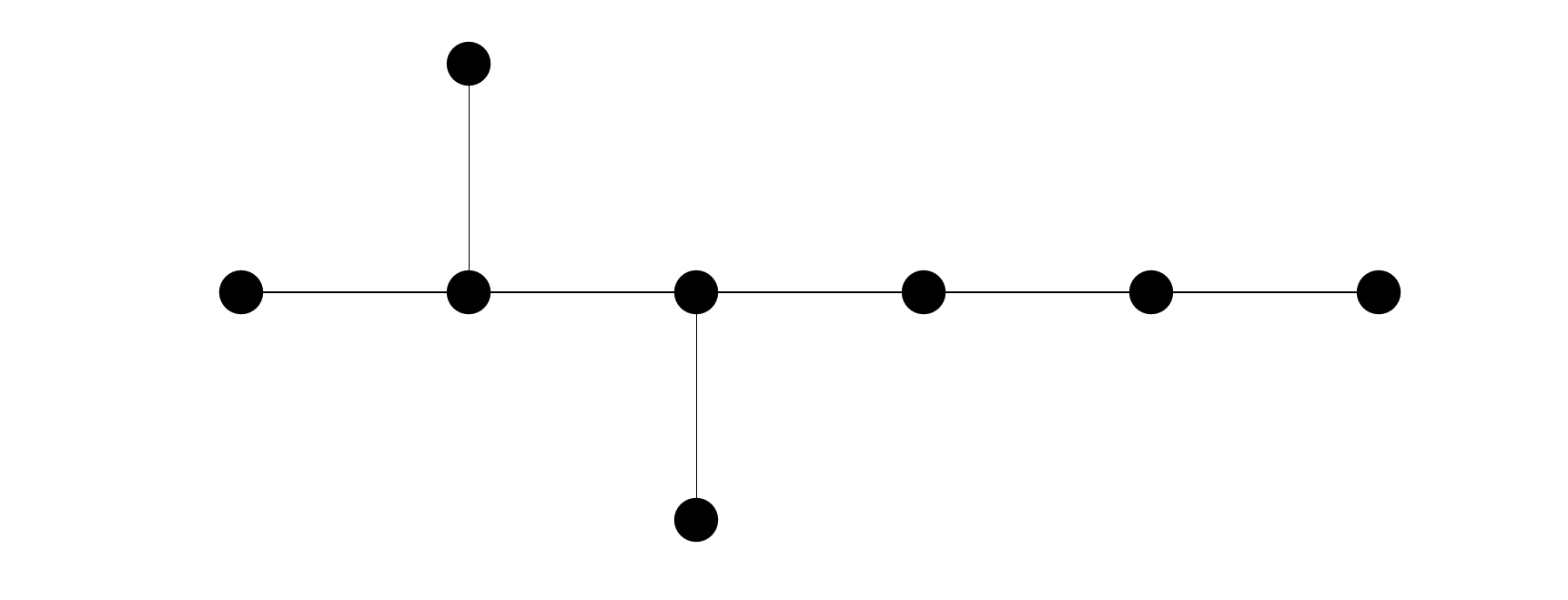}
}
\quad
\subfigure[$\mathcal{T}^{17}_{8}$]{
\includegraphics[trim=3.5cm 17.1cm 0 6cm,clip, width=.33\textwidth]{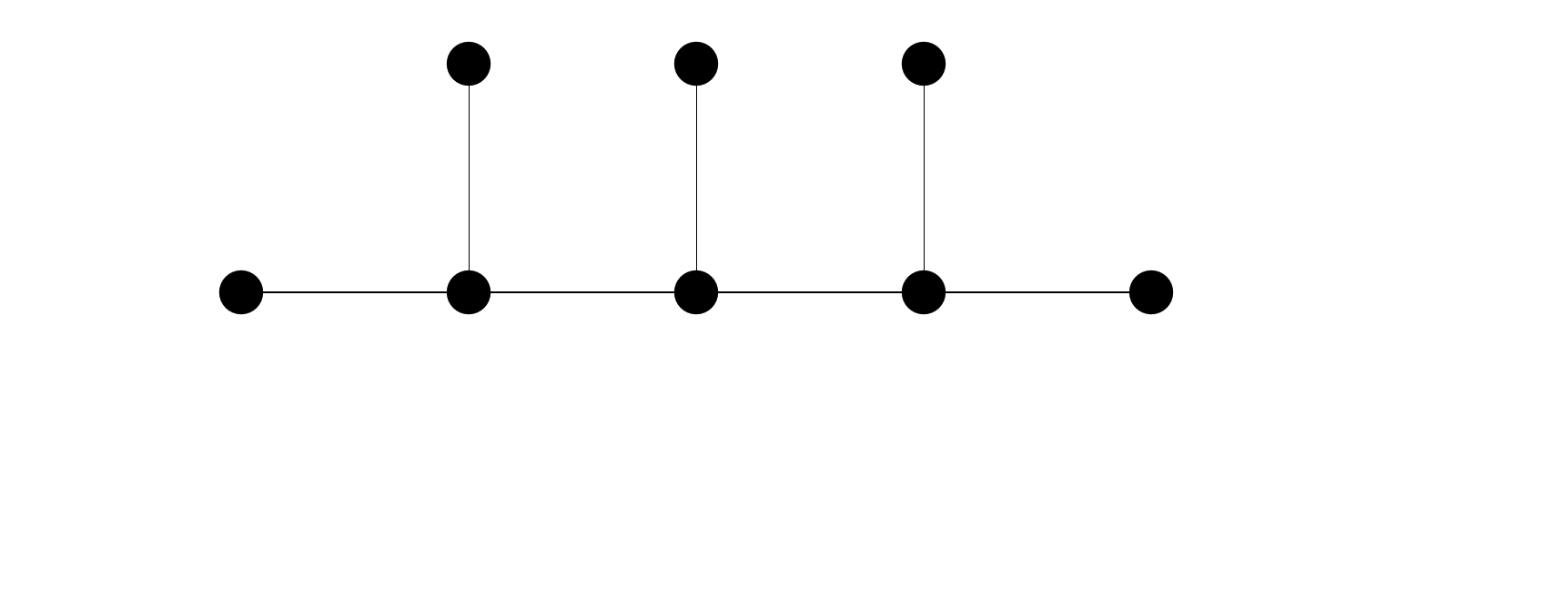}
}
\quad
\subfigure[$\mathcal{T}^{18}_{8}$]{
\includegraphics[trim=3.5cm 14cm 0 6cm,clip, width=.33\textwidth]{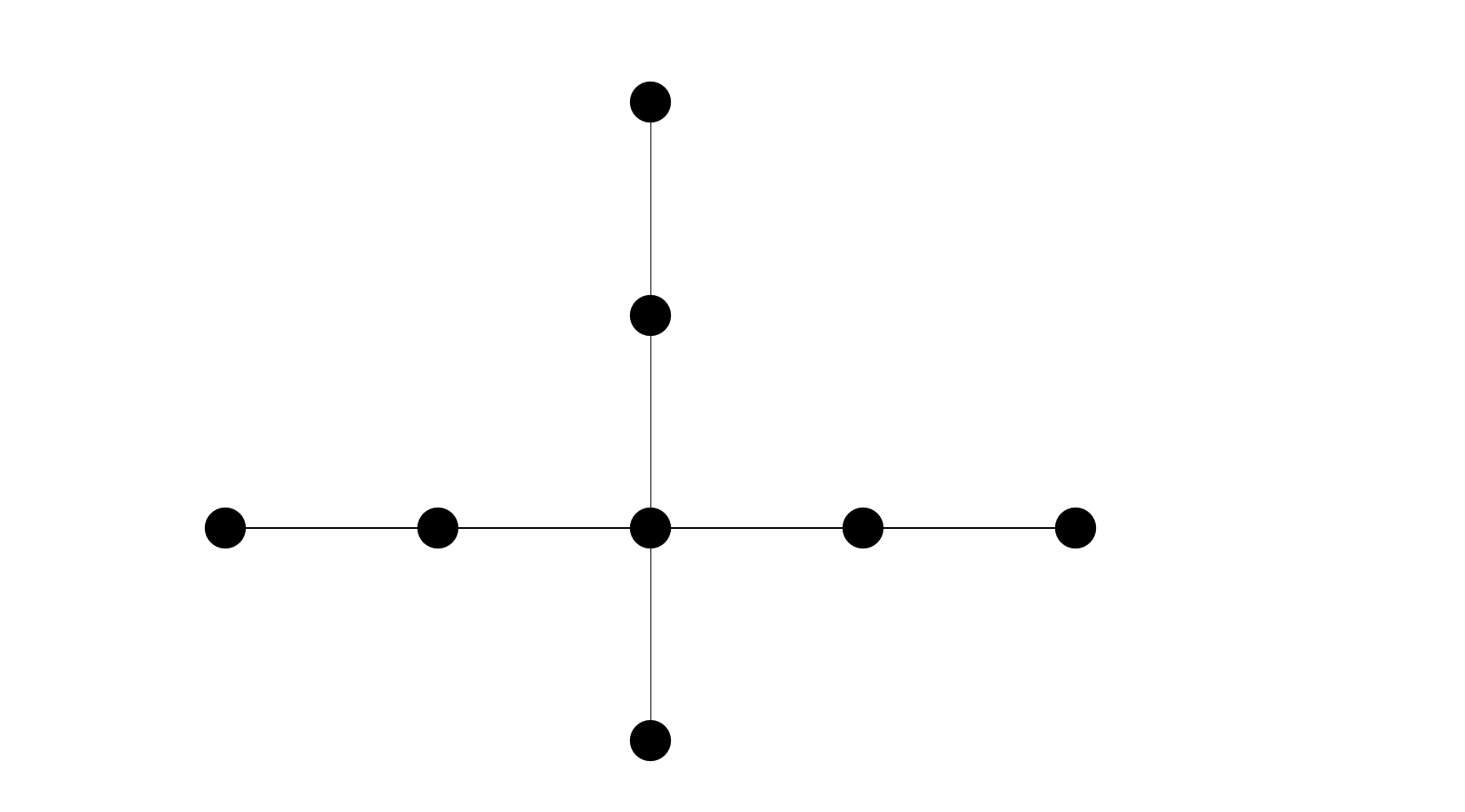}
}
}\\
\mbox{\subfigure[$\mathcal{T}^{19}_{8}$]{
\includegraphics[trim=3.5cm 19cm 0 6cm,clip, width=.33\textwidth]{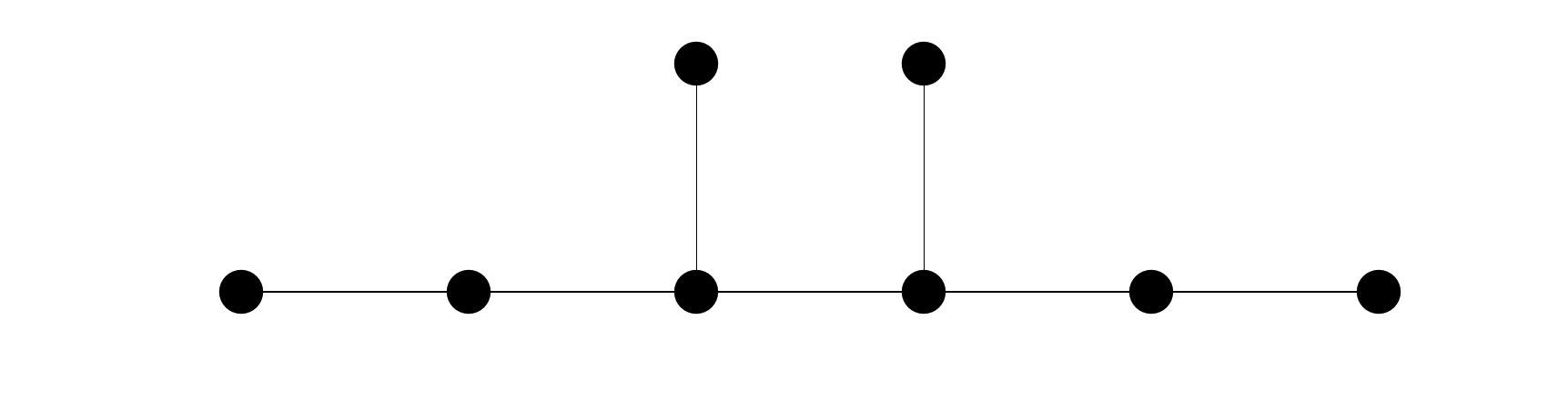}
}
\quad
\subfigure[$\mathcal{T}^{20}_{8}$]{
\includegraphics[trim=3.5cm 18.6cm 0 6cm,clip, width=.33\textwidth]{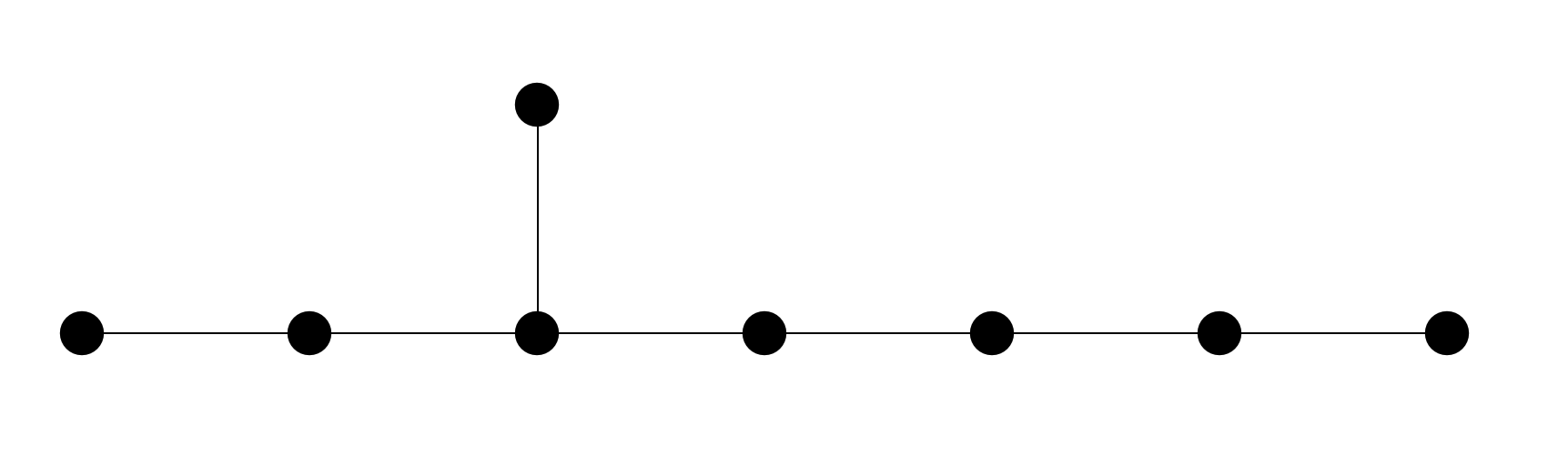}
}
\quad
\subfigure[$\mathcal{T}^{21}_{8}$]{
\includegraphics[trim=3.5cm 16.5cm 0 6cm,clip, width=.33\textwidth]{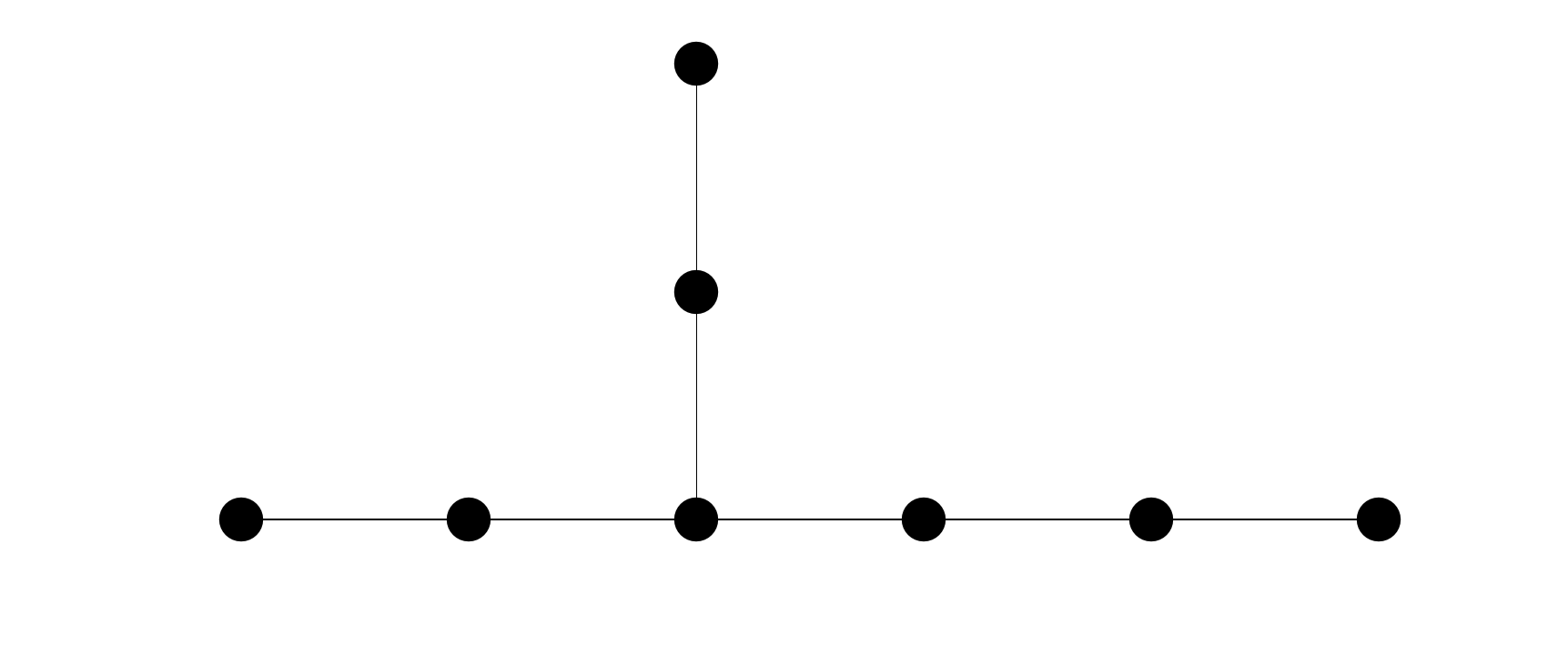}
}
}\\
\mbox{\subfigure[$\mathcal{T}^{22}_{8}$]{
\includegraphics[trim=3.5cm 14cm 0 6cm,clip, width=.33\textwidth]{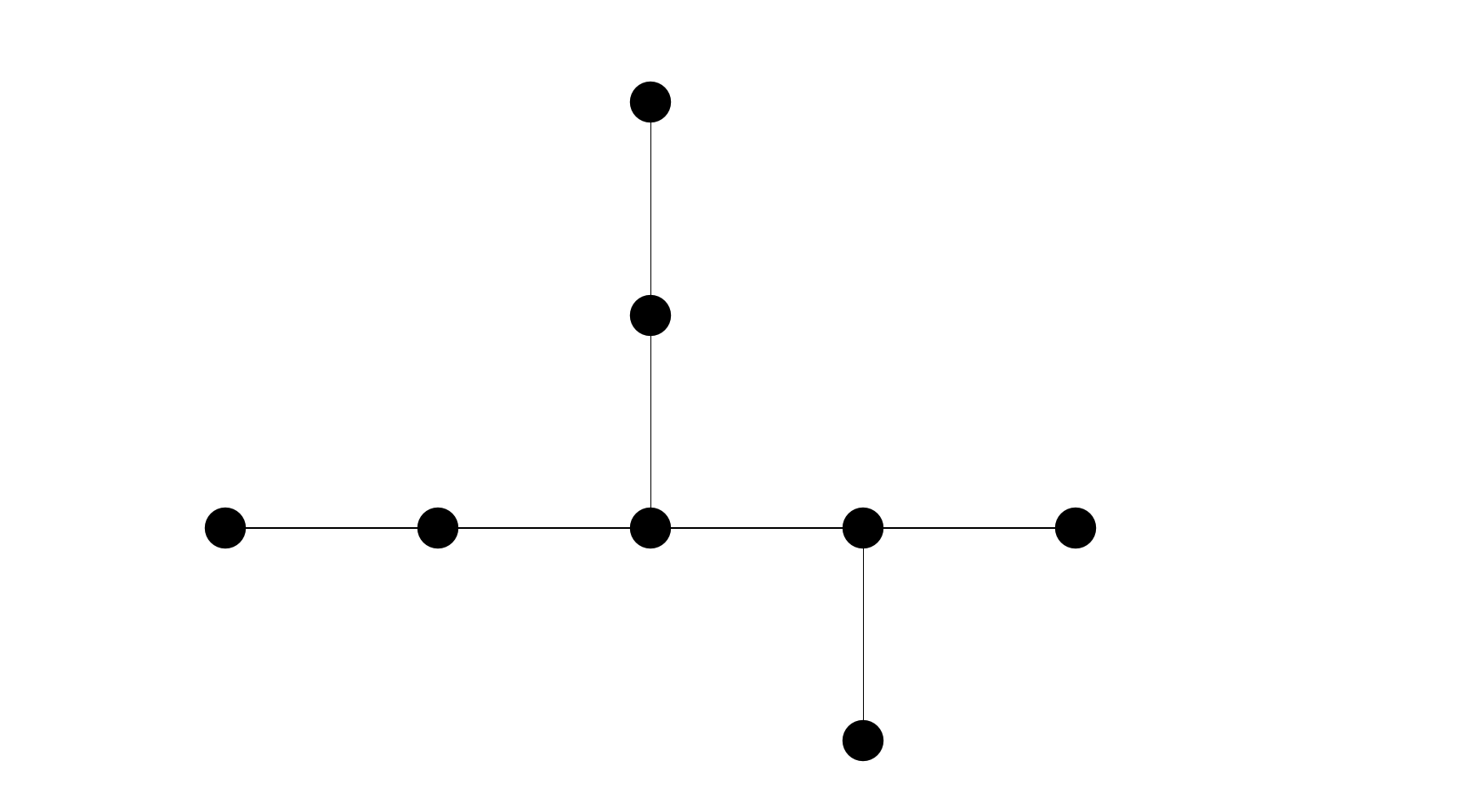}
}
\quad
\subfigure[$\mathcal{T}^{23}_{8}$]{
\includegraphics[trim=3.5cm 17cm 0 6cm,clip, width=.33\textwidth]{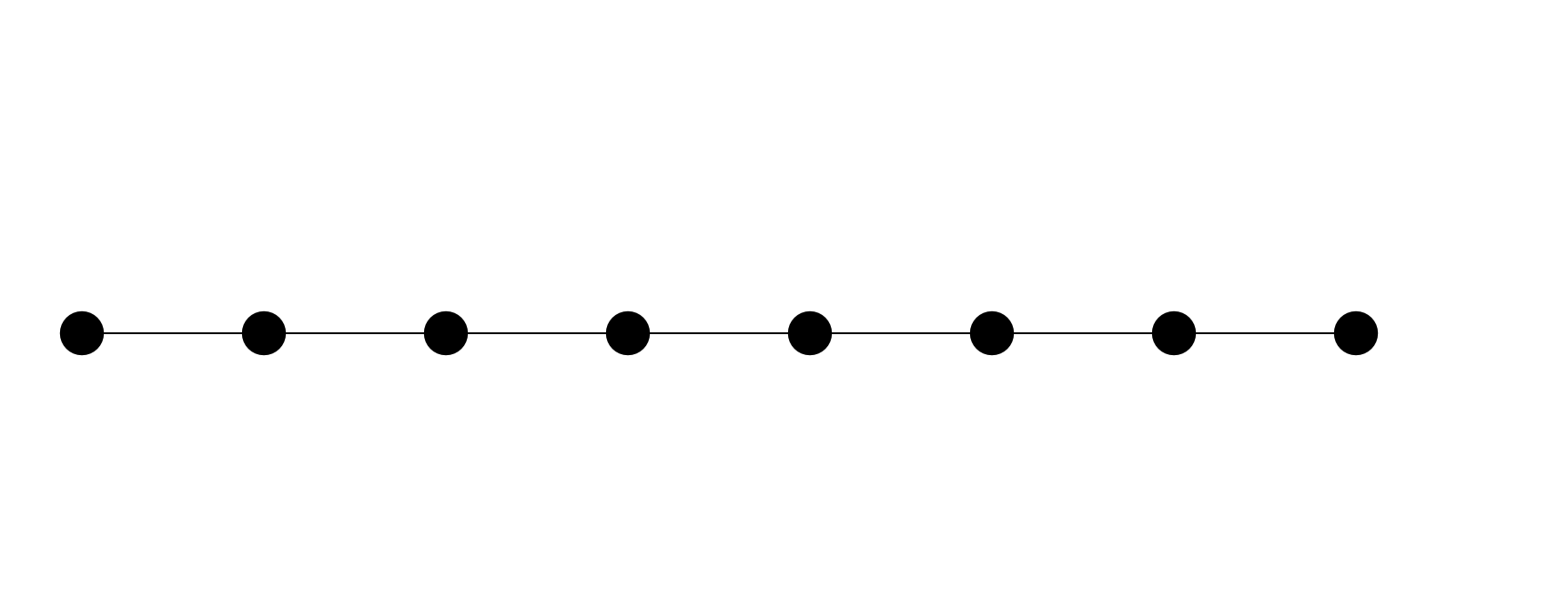}
}
}
\caption{\label{figB_2} Graph isomorphism classes $\calT^{j}_{8}$ ($j=1,\ldots 
23$) for trees of order $8$.}
\end{center}
\end{figure}

Trees of order $8$ partition into twenty-three GI classes \cite{harary}
which we denote by $\{ \calT^{j}_{8} : j = 1,\ldots 23\}$, and show as 
unlabelled graphs in Figure~\ref{figB_2}.
Ten of these GI classes correspond to known (unlabelled) graphs 
(Section~\ref{sec2} and Appendix~\ref{AppA}):
\begin{equation}
\begin{array}{llll}
 \calT_{8}^{1} = K_{1,7}; &  
     \hspace{0.2in} \calT_{8}^{4} = S^{(2)}_{4,2}; \hspace{0.2in}  &
          \hspace{0.2in} \calT_{8}^{9} = S^{(4)}_{3,1}; \hspace{0.2in}  & 
             \calT_{8}^{23} = P_{8} = S^{(6)}_{1,1}.               \\

 \calT_{8}^{2} = S^{(2)}_{5,1}; &   
    \hspace{0.2in} \calT_{8}^{5} = S^{(3)}_{3,2}; \hspace{0.2in} &
        \hspace{0.2in} \calT_{8}^{12} = S^{(4)}_{2,2}; \hspace{0.2in} &
                  \\

\calT_{8}^{3} =   S^{(3)}_{4,1}; & 
   \hspace{0.2in} \calT_{8}^{6} = S^{(2)}_{3,3}; \hspace{0.2in} &
              \hspace{0.2in} \calT_{8}^{14} = S^{(5)}_{2,1}; \hspace{0.2in} & 
                      \\
\end{array}
\end{equation}

\subsubsection{$r(\calT_{8}^{i},\calT_{6}^{j})$}
\label{AppB_2_1}

The numerical procedure described in Section~\ref{sec4_1} was used to 
determine the tree Ramsey numbers $r(\calT_{8}^{i},\calT_{6}^{j})$ for 
$1\leq i\leq 23$ and $1\leq j\leq 6$, and the results are displayed in 
Table~\ref{tableB_9}.
\begin{table}
\caption{\label{tableB_9}Numerical results for tree Ramsey numbers 
$r(\calT_{8}^{i},\calT_{6}^{j})$ with $1\leq i \leq 23$ and $1\leq j \leq 6$. 
Table rows (columns) are labelled by $i$ ($j$). A superscript ``x'' on a table 
entry indicates that Theorem~A.x of Appendix~\ref{AppA} applies, and so 
these tree Ramsey numbers were known prior to this work. The reader can 
verify that our numerical results are in agreement with the theorems of 
Appendix~\ref{AppA}. The remaining $132$ tree Ramsey numbers (to the best of
our knowledge) are new.}
\begin{center}
\begin{tabular}{|c|cccccc|} \hline
$\mathbf{r(\calT_{8}^{i},\calT_{6}^{j})}$ & \rule{0cm}{1.5em} & & & & & \\
$i\; \backslash \; j$ & 1 & 2 & 3 & 4 & 5 & 6 \\ \hline
1\rule{0cm}{1.5em} & $11^{4}$ & $11^{4}$ & $11^{4}$ & $11^{4}$ & $12^{3}$ 
                    & 11 \\
2 & 11 & 11 & 11 & 11 & 11 & 11 \\
3 & 11 & 11 & 11 & 11 & 11 & 11 \\
4 & 10 & 9 & 10 & 9 & 10 & 10 \\
5 & 11 & 11 & 11 & 11 & 11 & 11 \\
6 & 10 & 9 & 10 & 9 & 10 & 10 \\
7 & 10 & 9 & 10 & 9 & 10 & 10 \\
8 & 10 & 9 & 10 & 9 & 9 & 10 \\
9 & 10 & 9 & 10 & 9 & 9 & 10 \\
10 & 10 & 9 & 10 & 9 & 9 & 10 \\
11 & 10 & 9 & 10 & 9 & 9 & 10 \\
12 & 10 & 9 & 10 & 9 & 9 & 10 \\
13 & 10 & 9 & 10 & 9 & 9 & 10 \\
14 & 10 & 9 & 10 & 9 & 9 & 10 \\
15 & 10 & 9 & 10 & 9 & 9 & 10 \\
16 & 10 & 9 & 10 & 9 & 9 & 10 \\
17 & 10 & 9 & 10 & 9 & 9 & 10 \\
18 & 10 & 9 & 10 & 9 & 9 & 10 \\
19 & 10 & 9 & 10 & 9 & 9 & 10 \\
20 & 10 & 9 & 10 & 9 & 9 & 10 \\
21 & 10 & 9 & 10 & 9 & 9 & 10 \\
22 & 10 & 9 & 10 & 9 & 9 & 10 \\
23\rule[-1em]{0cm}{1em} & $10^{2}$ & 9 & 10 & 9 & 9 & 10 \\ \hline
\end{tabular}
\end{center}
\end{table}
A superscript ``x'' on a table entry indicates that Theorem~A.x of 
Appendix~\ref{AppA} applies, and so these tree Ramsey numbers were 
known prior to this work. The reader can verify that our numerical results 
are in agreement with the theorems of Appendix~\ref{AppA}. Recall from 
Section~\ref{sec4_1} that for graphs with order $N\geq 12$, exhaustive 
search over non-isomorphic graphs was not feasible. For such graphs we used 
the heuristic algorithm Tabu search to look for minima of the tree Ramsey
number objective function. As noted there, Tabu search only yields a lower 
bound for a tree Ramsey number. However, because of Theorem~A.3, we know 
that for $r(\calT_{8}^{1},\calT_{6}^{5})$, the value $12$ is in fact the 
\textit{actual\/} value of this Ramsey number. Thus no lower bounds appear in 
Table~\ref{tableB_9}---all values are actual Ramsey number values. Of the 
$138$ tree Ramsey numbers appearing in Table~\ref{tableB_9},  (to the best of
our knowledge) $132$ are new.

\subsubsection{$r(\calT_{8}^{i},\calT_{7}^{j})$}
\label{AppB_2_2}

The numerical results for the tree Ramsey numbers $r(\calT_{8}^{i},
\calT_{7}^{j})$ for $1\leq i\leq 23$ and $1\leq j\leq 11$ are shown in 
Table~\ref{tableB_10}.
\begin{table}
\caption{\label{tableB_10}Numerical results for tree Ramsey numbers 
$r(\calT_{8}^{i},\calT_{7}^{j})$ with $1\leq i \leq 23$ and $1\leq j \leq 11$. 
Table rows (columns) are labelled by $i$ ($j$). A superscript ``x'' on a table 
entry indicates that Theorem~A.x of Appendix~\ref{AppA} applies, and so 
these tree Ramsey numbers were known prior to this work. The reader can 
verify that our numerical results are in agreement with the theorems of 
Appendix~\ref{AppA}. Numbers marked by an asterisk indicate that these numbers
are  lower bounds on the corresponding tree Ramsey number (see text). The 
remaining $241$ tree Ramsey numbers (to the best of our knowledge) are new.}
\begin{center}
\begin{tabular}{|c|ccccccccccc|} \hline
$\mathbf{r(\calT_{8}^{i},\calT_{7}^{j})}$ & 
   \rule{0cm}{1.5em}  & & & & & & & & & & \\
$i\; \backslash \; j$ & 1 & 2 & 3 & 4 & 5 & 6 & 7 & 8 & 9 & 10 & 11 \\ \hline
1\rule{0cm}{1.5em} & $13^{\ast}$ &  $13^{\ast}$ & $13^{\ast}$ &
    $13^{\ast}$ & $13^{\ast}$ & 
      $13^{\ast}$ & $13^{\ast}$ & $13^{\ast}$ & $13^{\ast}$ & $13^{3}$ & 
          $13^{\ast}$  \\
2 & 11  & 11 & 11 & 11 & 11 & 11 & 11 & 11 & 11 & 11 & 11  \\
3 & 11  & 11 & 11 & 11 & 11 & 11 & 11 & 11 & 11 & 11 & 11  \\
4 & 10  & 10 & 10 & 10 & 9 & 9 & 10 & 10 & 10 & 11 & 10  \\
5 & 11  & 11 & 11 & 11 & 11 & 11 & 11 & 11 & 11 & 11 & 11  \\
6 & 10  & 10 & 10 & 10 & 10 & 10 & 10 & 10 & 10 & 11 & 10  \\
7 & 10  & 10 & 10 & 10 & 9 & 9 & 10 & 10 & 10 & 11 & 10  \\
8 & 10  & 10 & 10 & 10 & 9 & 9 & 10 & 10 & 9 & 11 & 10  \\
9 & 10  & 10 & 10 & 10 & 9 & 9 & 10 & 10 & 9 & 11 & 10  \\
10 & 10  & 10 & 10 & 10 & 9 & 9 & 10 & 10 & 9 & 11 & 10  \\
11 & 10  & 10 & 10 & 10 & 10 & 10 & 10 & 10 & 10 & 11 & 10  \\
12 & 10  & 10 & 10 & 10 & 10 & 10 & 10 & 10 & 10 & 11 & 10  \\
13 & 10  & 10 & 10 & 10 & 9 & 9 & 10 & 10 & 9 & 11 & 10  \\
14 & 10  & 10 & 10 & 10 & 9 & 9 & 10 & 10 & 9 & 11 & 10  \\
15 & 10  & 10 & 10 & 10 & 9 & 9 & 10 & 10 & 9 & 11 & 10  \\
16 & 10  & 10 & 10 & 10 & 10 & 10 & 10 & 10 & 10 & 11 & 10  \\
17 & 10  & 10 & 10 & 10 & 9 & 9 & 10 & 10 & 9 & 11 & 10  \\
18 & 10  & 10 & 10 & 10 & 10 & 10 & 10 & 10 & 10 & 11 & 10  \\
19 & 10  & 10 & 10 & 10 & 10 & 10 & 10 & 10 & 10 & 11 & 10  \\
20 & 10  & 10 & 10 & 10 & 10 & 10 & 10 & 10 & 10 & 11 & 10  \\
21 & 10  & 10 & 10 & 10 & 10 & 10 & 10 & 10 & 10 & 11 & 10  \\
22 & 10  & 10 & 10 & 10 & 9 & 9 & 10 & 10 & 9 & 11 & 10  \\
23\rule[-1em]{0cm}{1em} & $10^{2}$  & 10 & 10 & 10 & 10 & 10 & 10 & 10 & 10 & 
           11 & 10  \\ \hline
\end{tabular}
\end{center}
\end{table}
A superscript ``x'' on a table entry indicates that Theorem~A.x of 
Appendix~\ref{AppA} applies, and so these tree Ramsey numbers were known 
prior to this work. The reader can verify that our numerical results are in 
agreement with the theorems of Appendix~\ref{AppA}. Recall  from 
Section~\ref{sec4_1} that for graphs with order $N\geq 12$, exhaustive
search over non-isomorphic graphs was not feasible. For such graphs we used 
the heuristic algorithm Tabu search to look for minima of the tree Ramsey 
number objective function. As noted there, Tabu search only yields a lower 
bound for a tree Ramsey number. Numbers in Table~\ref{tableB_10} marked 
with an asterisk correspond to lower bounds on the associated tree Ramsey 
number. Of the $253$ numbers appearing in this Table, (to the best of our
knowledge) $241$ are new tree Ramsey numbers, $10$ are lower bounds, and $2$ 
were previously known. 

Before moving on we note that it has been conjectured \cite{fss} that 
$r(\calT_{m},\calT_{n})\leq n + m - 2$ for all trees $\calT_{m}$ and 
$\calT_{n}$. For the trees in Table~\ref{tableB_10} the conjecture gives the
upper bound $r(\calT_{8},\calT_{7})\leq 13$. The conjecture is seen to be 
consistent with all entries in Table~\ref{tableB_10}. 
Furthermore, if the conjecture is true, then all lower bounds in row~$1$ 
become exact values since we would have $13\leq r(\calT_{8}^{1},
\calT_{7}^{j}) \leq 13$.

\subsubsection{$r(\calT_{8}^{i},\calT_{8}^{j})$}
\label{AppB_2_3}

The numerical results for the tree Ramsey numbers $r(\calT_{8}^{i},
\calT_{8}^{j})$ for $1\leq i,j\leq 23$ are shown in Table~\ref{tableB_11}.
\begin{table}
\caption{\label{tableB_11}Numerical results for tree Ramsey numbers 
$r(\calT_{8}^{i},\calT_{8}^{j})$ with $1\leq i,j \leq 23$. Table rows (columns) 
are labelled by $i$ ($j$). Only the upper triangular table entries are shown as 
the lower triangular entries follow from $r(\calT_{8}^{j},\calT_{8}^{i}) =
r(\calT_{8}^{i},\calT_{8}^{j})$. A superscript ``x'' on a table entry indicates 
that Theorem~A.x of Appendix~\ref{AppA} applies, and so these tree Ramsey 
numbers were known prior to this work. The reader can verify that our 
numerical results are in agreement with the theorems of Appendix~\ref{AppA}. 
Numbers marked by an asterisk indicate that these numbers are  lower bounds on 
the corresponding tree Ramsey number (see text). Of the $529$ numbers 
appearing in this Table, (to the best of our knowledge) $479$ are new tree 
Ramsey numbers, $10$ are lower bounds, and $40$ were previously known.}
\begin{center}
\begin{tabular}{|c|cccccccccccc|} \hline
$\mathbf{r(\calT_{8}^{i},\calT_{8}^{j})}$ & \rule{0cm}{1.5em} & & & & & & & & 
                                                                       & & & \\
$i\; \backslash \; j$ & 1 & 2 & 3 & 4 & 5 & 6 & 7 & 8 & 9 & 10 & 11 
                                            & 12 \\ \hline
1\rule{0cm}{1.5em}  &   $14^{\,3}$  & $13^{\,4}$ & $13^{\,4}$ & $13^{\,\ast}$ & 
    $13^{\,\ast}$ &  $13^{\,\ast}$ & $13^{\,4}$ & $13^{\,4}$ & $13^{\,4}$  & 
       $13^{\,4}$ &  $13^{\,4}$ & $13^{\,\ast}$\\ 
2  & & $11^{\,7}$ & 11 & 11 & 11 & 11 & 11 & 11 & 11 & $11$ & 11 & 11 \\
3  & & & 11 & 11 & 11 & 11 & 11 & 11 & 11 & 11 & 11 & 11\\ 
4 & & & & $10$ & 11 & 11 & 10 & 10 & 10 & 10 & 11 & 11\\
5 & & & & & 10 & 11 & 11 & 11 & 11 & 11 & 11 & 11\\
6 & & & & & & $11^{\,7}$ & 11 & 11 & 11 & 11 & 11 & 11\\  
7 & & & & & & & 10 & 10 & 10 & 10 & 11 & 11\\
8 & & & & & & & & $10$ & 10 & 10 & 11 & 11\\ 
9 & & & & & & & & & $10^{6}$ & 10 & 11 & 11 \\ 
10 & & & & & & & & & & 10 & 11 & 11 \\
11 & & & & & & & & & &  & 11 & 11\\ 
12 & \rule[-1em]{0cm}{1em} & & & & & & & & & & & $11^{\,6}$ \\ \hline
\end{tabular}
\begin{tabular}{|c|ccccccccccc|} \hline
$\mathbf{r(\calT_{8}^{i},\calT_{8}^{j})}$ & \rule{0cm}{1.5em} & & & & & & & & 
                                                                       & & \\
$i\; \backslash \; j$ & 13 & 14 & 15 & 16 & 17 & 18 & 19 & 20 & 21 & 22 & 
                         23 \\ \hline
1\rule{0cm}{1.5em}  & $13^{\,4}$ & $13^{\,4}$ & $13^{\,4}$ & $13^{\,4}$ & 
     $13^{\,\ast}$  &   $13^{\,4}$ & $13^{\,4}$ & $13^{\,4}$ & $13^{\,4}$ & 
            $13^{\,4}$ &   $13^{\,4}$  \\ 
2  &11 & 11 & 11 & 11  & 11 & $11$ & 11 & 11 & 11 & 11 & 11  \\
3 &11 & 11 & 11 & 11 & 11 & 11 & 11 & 11 & 11 & 11 & 11  \\ 
4 & 10 & 10 & 10 & 11 & 10 & 11 & 11 & 11 & 11 & 10 & 11  \\
5 &11 & 11 & 11 & 11 & 11 & 11 & 11 & 11 & 11 & 11 & 11 \\
6 &11 & 11 & 11 & 11 & 11 & 11 & 11 & 11 & 11 & 11 & 11 \\  
7 &10 & 10 & 10 & 11 & 10 & 11 & 11 & 11 & 11 & 10 & 11 \\
8 &10 & 10 & 10 & $11$  & 10 & 11 & 11 & 11 & 11 & 10 & 11 \\
9 &10 & 10 & 10 & 11 & 10 & 11 & 11 & 11 & 11 & 10 & 11 \\ 
10 & 10 & 10 & 10 & 11 & 10 & 11 & 11 & 11 & 11 & 10 & 11 \\
11 &11 & 11 & 11 & 11 & 11 & 11 & 11 & 11 & 11 & 11 & 11 \\
12 &11 & 11 & 11 & 11 & $11$  & 11 & 11 & 11 & 11 & 11 & 11 \\
13 &10 & 10 & 10 & 11 & 10 & $11$ & 11 & 11 & 11 & 10 & 11 \\
14 &  & 10 & 10 & 11 & 10 & 11 & 11 & 11 & 11 & 10 & 11 \\ 
15 &  &  & 10 & 11 & 10 & 11 & 11 & 11 & 11 & 10 & 11 \\
16 &&&& 11 & 11 & 11 & 11 & 11 & 11 & 11 & 11 \\  
17 &&&& & 10 & 11 & 11 & 11 & 11 & 10 & 11 \\  
18 &&&& & & 11 & 11 & 11 & 11 & 11 & 11 \\
19 &&&& &&& 11 & 11 & 11 & 11 & 11 \\
20 &&&& &&&& 11 & 11 & 11 & 11 \\ 
21 &&&& &&&&& 11 & 11 & 11 \\
22 &&&& &&&&&& 10 & 11 \\
23 & \rule[-1em]{0cm}{1em}  &&& &&&&&&& $11^{\,2}$ \\ \hline
\end{tabular}
\end{center}
\end{table}
Only the upper triangular table entries are shown as the lower triangular 
entries follow from $r(\calT_{8}^{j},\calT_{8}^{i}) =r(\calT_{8}^{i},
\calT_{8}^{j})$. A superscript ``x'' on a table entry indicates that Theorem~A.x
in Appendix~\ref{AppA} applies, and so these tree Ramsey numbers were known
prior to this work. The reader can verify that our numerical results are in 
agreement with the theorems of Appendix~\ref{AppA}. Recall  from 
Section~\ref{sec4_1} that for graphs with order $N\geq 12$, exhaustive
search over non-isomorphic graphs was not feasible. For such graphs we used 
the heuristic algorithm Tabu search to look for minima of the tree Ramsey 
number objective function. As noted there, Tabu search only yields a lower 
bound for a tree Ramsey number. Numbers in Table~\ref{tableB_11} marked 
with an asterisk correspond to lower bounds on the associated tree Ramsey 
number. Of the $529$ numbers appearing in this Table, (to the best of our
knowledge) $479$ are new tree Ramsey numbers, $10$ are lower bounds, 
and $40$ were previously known. The conjectured upper bound \cite{fss}, 
$r(\calT_{m},\calT_{n})\leq n + m - 2$, evaluates to $r(\calT_{8},
\calT_{8})\leq 14$ and is seen to be consistent with all entries in 
Table~\ref{tableB_11}.


\begin{thebibliography}{99} 
\bibitem{GRS} Graham, R. L., Rothschild, B. L., Spencer, J. H.: Ramsey Theory. 
Wiley, New York (1990)
\bibitem{JNes} Ne\u{s}et\u{r}il, J.: Ramsey theory. In: Graham, R. L., 
Gr\"{o}tschel, M., Lov\'{a}sz, L. (eds.) Handbook of Combinatorics,
pp.~1331-1403. Elsevier, New York (1995), Vol.~$2$
\bibitem{Boll} Bollob\'{a}s, B.: Modern Graph Theory Springer, New York (1998)
\bibitem{radzi} Radziszowski, S. P., \textit{Small Ramsey numbers\/}, Electronic 
J. Combin., Dynamical Survey \# DS1 (2014)
\bibitem{C+H1} Chv\'{a}tal, V., Harary, F.: Generalized Ramsey theory
for graphs II, small diagonal numbers. Proc.\ Am.\ Math.\ Soc.\ 32, 389-394
(1972)
\bibitem{C+H2} Chv\'{a}tal, V., Harary, F.: Generalized Ramsey theory
for graphs III, small off-diagonal numbers. Pacif.\ J. Math.\ 41, 335-345
(1972)
\bibitem{Clan} Clancy, M.: Some small Ramsey numbers. J. Graph Theory 
1, 89-91 (1977)
\bibitem{Hend} Hendry, G. R. T.: Ramsey numbers for graphs with five
vertices. J. Graph Theory 13, 245-248 (1989)
\bibitem{us1} Gaitan, F., Clark, L.: Ramsey numbers and adiabatic 
quantum computing. Phys.\ Rev.\ Lett.\ 108, 010501: 1-4 (2012)
\bibitem{us2} Bian, Z., Chudak, F., Macready, W. G., Clark, L., Gaitan, F.:
Experimental determination of Ramsey numbers. Phys.\ Rev.\ Lett.\ 
111, 130505: 1-6 (2013)
\bibitem{aqo} Farhi, E., Goldstone, J., Gutmann, S., Sipser, M.: Quantum
computation by adiabatic evolution. arXiv.org:quant-ph/0001106v1 (2000)
\bibitem{probM} Spencer, J.: Ten Lectures on the Probabilistic Method.
SIAM, Philadelphia, PA, (1994), 2nd ed.
\bibitem{fg1} Gaitan, F: Simulation of quantum adiabatic search in the
presence of noise. Int.\ J. Quantum Info.\ 4, 843-870 (2006)
\bibitem{fg2} Gaitan, F: Noise-induced sampling of alternative hamiltonian 
paths in quantum adiabatic search. Complexity 14(6), 21-27 (2009)
\bibitem{OEIS} On-line Encyclopedia of Integer Sequences (https://oeis.org),
sequences A000088 and A006125.
\bibitem{nauty} Mackay, B. D., Piperno, A.: Practical graph 
isomorphism, II. J. Sym.\ Comp.\ 60, 94-112 (2014)
\bibitem{tabu} Glover, F.: Tabu search, part I. ORSA J. Comp.\ 1, 190-206 
(1989)
\bibitem{harary} Harary, F.: Graph Theory. Addison-Wesley, Reading, MA (1969)
\bibitem{ger&gya} Gerencs\'{e}r, L., Gy\'{a}rf\'{a}s, A.: On Ramsey-type 
problems. Annales Universitatis Scientiarum Budapestinensis,
E\"{o}tv\"{o}s Sect.\ Math.\ 10, 167-170 (1967)
\bibitem{hara} Harary, F.: Recent results on generalized Ramsey theory 
for graphs. In: Y. Alavi et al. (eds.) Graph Theory and Applications, 
pp.~125-138. Springer, Berlin (1972) 
\bibitem{cocka} Cockayne, E.: Some tree-star Ramsey numbers. J.
Combinatorial Theory Series B 17, no.~2, 183-187 (1974)
\bibitem{Bur&Erd} Burr, S., Erd\"{o}s, P.: Extremal Ramsey theory for 
graphs. Utilitas Mathematica 9, 247-258 (1976)
\bibitem{Gros&Hara&Klaw} Grossman, J., Harary, F., Klawe, M.: Generalized 
Ramsey theory for graphs, X: Double stars. Discrete Mathematics 28, 247-254 
(1979)
\bibitem{fss} Faudree, R. J., Schelp, R.H., Simonovits, M.: On some
Ramsey type problems connected with paths, cycles, and trees. Ars 
Combinatorica 29A, 97-106 (1990)

\end{thebibliography}
\end{document}